\documentclass[preprint]{aastex}
\usepackage{natbib}
\usepackage{rjwmath}
\usepackage{rotating}
\shorttitle{The Distribution of YSOs in the Orion Molecular Clouds}
\shortauthors{Megeath et al.}

\bibpunct{(}{)}{,}{a}{}{;}
\bibpunct{(}{)}{,}{a}{}{;}

\begin{document}


\title{The {\it Spitzer Space Telescope} Survey of the Orion A and B Molecular Clouds II:  the Spatial Distribution and Demographics of Dusty Young Stellar Objects}


\author{S. T. Megeath\altaffilmark{1,11}, R. Gutermuth\altaffilmark{2}, J. Muzerolle\altaffilmark{3}, E. Kryukova\altaffilmark{1}, J. L. Hora\altaffilmark{4}, L. E. Allen\altaffilmark{5},  K. Flaherty\altaffilmark{6}, L. Hartmann\altaffilmark{7},  P. C. Myers\altaffilmark{4}, J. L. Pipher\altaffilmark{8},  J. Stauffer\altaffilmark{9}, E. T. Young\altaffilmark{10},  G.G. Fazio\altaffilmark{4}}

\altaffiltext{1}{Ritter Astrophsical Research Center, Department of Physics and Astronomy, University of Toledo, Toledo, OH 43606 (megeath@physics.utoledo.edu)}

\altaffiltext{2}{Department of Astronomy, University of Massachusetts, Amherst, MA 01003, USA}

\altaffiltext{3}{Space Telescope Science Institute, Baltimore, MD 21218, USA}

\altaffiltext{4}{Harvard Smithsonian Center for Astrophysics,  Cambridge, MA 02138, USA}

\altaffiltext{5}{National Optical Astronomical Observatory, Tucson, AZ 85719, USA}

\altaffiltext{6}{Steward Observatory, University of Arizona, Tucson, AZ 85721, USA }

\altaffiltext{7}{Department of Astronomy, University of Michigan, Ann Arbor, MI 48109, USA}

\altaffiltext{8}{Department of Physics and Astronomy, University of  Rochester, Rochester NY 14627, USA}

\altaffiltext{9}{Spitzer Science Center, California Institute of Technology,  Pasadena, CA 91125, USA}

\altaffiltext{10}{SOFIA-Universities Space Research Association, NASA Ames Research Center, Moffett Field, CA 94035, USA}

\altaffiltext{11}{Completed while on sabbatical at Max-Planck-Institut f\"ur Radioastronomie, Auf dem H\"ugel 69, D-53121 Bonn, Germany}



\begin{abstract}
We analyze the spatial distribution of dusty young stellar objects (YSOs) identified in the {\it Spitzer} Survey of the Orion Molecular clouds, augmenting these data with {\it Chandra} X-ray observations to correct for incompleteness in dense clustered regions. We also devise a scheme to correct for spatially varying incompleteness when X-ray data are not available. The local surface densities of the YSOs range from 1~pc$^{-2}$ to over 10,000~pc$^{-2}$, with protostars tending to be in higher density regions. This range of densities is similar to other surveyed molecular clouds with clusters, but broader than clouds without  clusters. By identifying clusters and groups as continuous regions with surface densities $\ge10$~pc$^{-2}$, we find that 59\% of the YSOs  are in the largest cluster, the Orion Nebular Cluster (ONC), while 13\% of the YSOs are found in a distributed population. A lower fraction of protostars in the distributed population is evidence that it is somewhat older than the groups and clusters.  An examination of the structural properties of the clusters and groups show that the peak surface densities of the clusters increase approximately linearly with the number of members.  Furthermore, all clusters with more than 70 members exhibit asymmetric and/or highly elongated structures.  The ONC becomes azimuthally symmetric in the inner 0.1 pc, suggesting that the cluster is only $\sim 2$~Myr in age. We find the star formation efficiency (SFE) of the Orion B cloud is unusually low, and that the SFEs of individual groups and clusters are an order of magnitude higher than those of the clouds.  Finally, we discuss the relationship between the young low mass stars in the Orion clouds and the Orion OB 1 association, and we determine upper limits to the fraction of disks that may be affected by UV radiation from OB stars or by dynamical interactions in dense, clustered regions.  
\end{abstract}



\section{Introduction}

Since the identification of T Tauri stars as low-mass stars undergoing pre-main sequence evolution \citep{1949ApJ...110..424J,1962AdA&A...1...47H}, young ($< 2$~Myr), low-mass stars have been found in almost every surveyed molecular cloud \citep[e.g.][]{2000ApJ...540..255P,2001ApJ...562..852H,2009AJ....137.4072M}.  Low mass star formation has been observed to occur in a rich diversity of environments, from isolated, cold globules containing only a few young stars to rich clusters in giant molecular clouds heated by OB associations \citep[e.g.][]{2000AJ....120.3139C,2005ApJS..160..379F,2010A&A...518L..87S}. The surface densities of the young stellar objects (YSOs) vary dramatically between these diverse environments, from a few stars per square parsec in the Taurus cloud to hundreds of stars per square parsec in rich clusters \citep{2007A&A...468..425T,2009ApJS..181..321E,2009ApJS..184...18G,2009AJ....137.4072M,2012AJ....144..192M,2010ApJS..186..259R}.

The broad range of stellar densities at which young, low-mass stars are found has led to the division of star formation into distributed and clustered modes, where the distributed mode is characterized by the low stellar densities typical of the Taurus dark clouds and the clustered mode is characterized by the high stellar densities of the cluster found in the Orion Nebula \citep{1962AdA&A...1...47H}.  In the 50 years since this distinction was made, a primary goal of infrared molecular cloud surveys has been to measure the relative number of stars forming in distributed and clustered environments.  In the 1980s, surveys with near-IR cameras revealed the importance of the clustered mode. In a $K$-band survey of the Orion~B cloud, \citet{1992ApJ...393L..25L} found four clusters with a total of 627 stars, and they estimated that the stars in the clusters represented 58-82\% of the total population of young stars in this cloud.  The large uncertainty in this fraction results from the unknown number of foreground and background stars contaminating the survey; young stars in molecular clouds cannot be distinguished by field stars in the line of sight by their $K$-band magnitudes alone. This left open the possibility that almost half of the stars were formed in relative isolation as part of the distributed population. More definitive evidence for a distributed population was found in the Orion A cloud using the 2MASS 2nd incremental release data.  After subtracting out the estimated surface density of contaminating field stars, \citet{2000AJ....120.3139C} found an excess of 2MASS point sources in the Orion A cloud which was interpreted as a distributed population of young stars.  In contrast, there was not a significant excess of point sources in the Orion B cloud; however, the 2nd incremental release of 2MASS only covered half of the Orion B cloud.  Furthermore, due to uncertainties in the  surface density of contaminating field stars, the number of stars in the distributed population was highly uncertain.  Nonetheless, the 2MASS results suggest that over $> 50$\% of the stars in molecular clouds are in clusters.

A second goal of infrared surveys has been to establish the relative importance of star formation in clusters relative to that in groups.   \citet{2003ARA&A..41...57L} define young clusters as gravitationally bound assemblages of stars with densities above 1~M$_{\odot}$~pc$^{-2}$ and more than 35 members, the minimum size at which the relaxation time is greater than the crossing time.  In contrast, \citet{2001ApJ...553..744A} defined clusters as having more than 100 stars, the size at which the relaxation time equals the formation time. They further define groups as having 10-100 members.  In their literature compilation of young clusters and groups in the nearest 1~kpc, \citet{2003AJ....126.1916P} found that groups with sizes of 10-100 members are more numerous than clusters, but that the clusters contain 80\% of the YSOs. A similar result was found by \citet{2000AJ....120.3139C}, who found that the largest clusters in a given cloud contained a much larger fraction of members than the more numerous groups.  Thus, the observations in the current literature indicate the more stars are formed in clusters than groups.  In total, these results suggest that the majority of stars form in clusters, although there remains a significant uncertainty in the number of young low mass stars found in the distributed population.  

Despite the importance of clusters, the nature of embedded clusters and their connection to open clusters has not been well established. Both the physical processes that drive the fragmentation of the cluster forming gas and the subsequent dynamics of the ensuing stars are not understood.  Much of the debate has centered on the best studied young cluster, the Orion Nebula Cluster (hereafter: ONC). \citet{2006ApJ...641L.121T} argued that the ONC is in a quasi-equilibrium configuration that has lasted multiple dynamical times.  In contrast, observations and modeling of the radial velocities of young stars in the ONC by \citet{2009ApJ...697.1103T} and \citet{2009ApJ...697.1020P} indicated that the cluster is sub-virial and undergoing global collapse. Understanding the structure and dynamical state of embedded clusters is an important step toward understanding the formation of open clusters. Although it is likely that embedded clusters are the progenitors of open clusters, \citet{2003ARA&A..41...57L} find that only 7\% of the embedded clusters survive gas dispersal.  It is not clear what properties are required for an embedded cluster to survive and form a bound open cluster.  Of further interest is the origin of the stars found in the distributed population; did these form in relative isolation or did they originate  groups and clusters?



The range of environments in which low-mass stars form may  influence planet formation. Protoplanetary disks surrounding young, low mass stars in rich young clusters can be affected by tidal interactions with other cluster members as well as the UV radiation from massive stars within the cluster.   The gravitational tides experienced by disks during  flybys of cluster members may induce structures within the disks as well as strip the outer regions of the disk.  The rate of such encounters depends strongly on the density of stars within a cluster and appears to be rare at the stellar densities observed in the typical clusters observed in the nearest 1 kpc \citep{2005ApJ...632..397G,2006ApJ...641..504A}.  In contrast, the UV radiation from massive stars has a measurable effect on disks in the Orion Nebula.  The most massive star in the Orion Nebula is $\theta^1$~C with a spectral type of  O7 \citep{1994A&A...289..101B}. VLA and Hubble observations of the nebula show that the UV radiation from $\theta^1$~C is photoevaporating the disks around low mass stars near the massive star \citep{1987ApJ...321..516C,1994ApJ...436..194O,1998AJ....116..293B,1998ApJ...499..758J}. Theoretical analyses show that the radiation erodes the outer disks \citep{2004ApJ...611..360A};  the resulting loss in disk mass has been observed for young stars in the Orion Nebula \citep{2009ApJ...694L..36M,2009ApJ...699L..55M,2010ApJ...725..430M,2014ApJ...784...82M}.  In contrast, it is not clear whether the radiation can destroy the disk.  Theoretical models of photoevaporating, viscous disks suggest that total disk destruction can occur \citep{2003ApJ...582..893M}; however, the observational evidence for the destruction of gas rich disks is mixed.  In support of disk destruction, \citet{2008ApJ...688..408B} find a deficiency of disks in the inner regions of the Rosette nebula where the low mass stars with disks are in close proximity to several massive stars.  However, in an analysis of the disk fraction in the Cep OB3b cluster, \citet{2012ApJ...750..125A} show evidence that observed variations in the disk fraction come from a mixture of ages, and not photoevaporation by  the O7V star in the cluster.   Despite the growing evidence that UV radiation can at least erode the outer regions of disks, the amount of erosion depends sensitively on the distance of the low mass stars orbiting in a cluster from the massive stars \citep{2004ApJ...611..360A,2006ApJ...641..504A}.   Thus, more work is needed to assess the  typical radiation exposure experienced by young low mass stars.  

Surveys of the distribution and density of young stars in molecular clouds have important ramification for all the above topics: the demographics of young stars, the structure of embedded clusters, and the role of environment in planet formation.  Most of the studies described above used maps of the surface density of stars to trace embedded populations; however, this approach is not sensitive to more distributed populations of young stars where the surface density of young stars is similar to that of background stars \citep[see number counts method in][]{2007prpl.conf..361A}.   Cloud surveys with the {\it Spitzer} Space Telescope have now provided an alternative means for mapping the distribution of young stars with dusty disks and infalling envelopes \citep{2004ApJS..154..363A,2007prpl.conf..361A}.  Such dusty young stellar objects (hereafter: dusty YSOs) can be identified even in relative isolation and a map of the distribution of young stars can be obtained down to very low stellar densities \citep{2004ApJS..154..367M,2009ApJS..181..321E}.  

 
The {\it Spitzer} Orion Survey  covered  9~sq.~deg.~of the Orion~A and Orion~B molecular clouds with the IRAC and MIPS instrument onboard {\it Spitzer}. \citet[][hereafter Paper I]{2012AJ....144..192M} combined 2MASS, IRAC, and MIPS 24~$\mu$m phototometry from the survey and published a catalog of 3479 dusty YSOs in the Orion clouds.  In Paper I, we presented the catalog of dusty YSOs, briefly examined the spatial distribution of the YSOs, and then studied the variability of the YSOs between two epochs. In this second paper on the {\it Spitzer} Orion survey, we use the  catalog of dusty YSOs to address the questions posed above.  First, we will examine the spatially varying incompleteness over the surveyed regions and correct for this incompleteness using newly developed methods combining both existing X-ray surveys of Orion clusters and artificial YSO tests.  These corrections allow us to assess the demographics of dusty YSOs in the Orion clouds and present the statistical distributions 
of YSO densities and the relative fraction of stars in clusters, groups and isolation for all YSOs and for protostars alone.  To place the Orion clouds in context, we also compare the demographics of YSOs in Orion to those in other nearby clouds.  Next, we study the structure of the embedded clusters in the Orion survey and examine the relationship of those clusters to the surrounding Orion OB1 association.  Finally, we study the population of young stars with disks, estimating both the fraction of stars with disks and the projected distances of the typical disks from the OB stars in the Orion OB1 association.  

\section{The Completeness of the YSO Catalog}
\label{sec:ysocomp}

In Paper I, we described a method to identify and classify protostars using the eight band 1.2-24~$\mu$m photometry from 2MASS and {\it Spitzer}. We then presented a catalog of YSOs identified in this manner.  In this contribution, we use the same YSOs catalog, with 12 extra dusty YSOs added from one minor modification (Appendix~A).  In addition, seventeen pre-main stars have been reclassified as  protostars, and two protostars have been reclassified as pre-main sequence stars with disks  (Appendix~A).  In this section, we examine the completeness of that catalog. 

\begin{figure}
\plotone{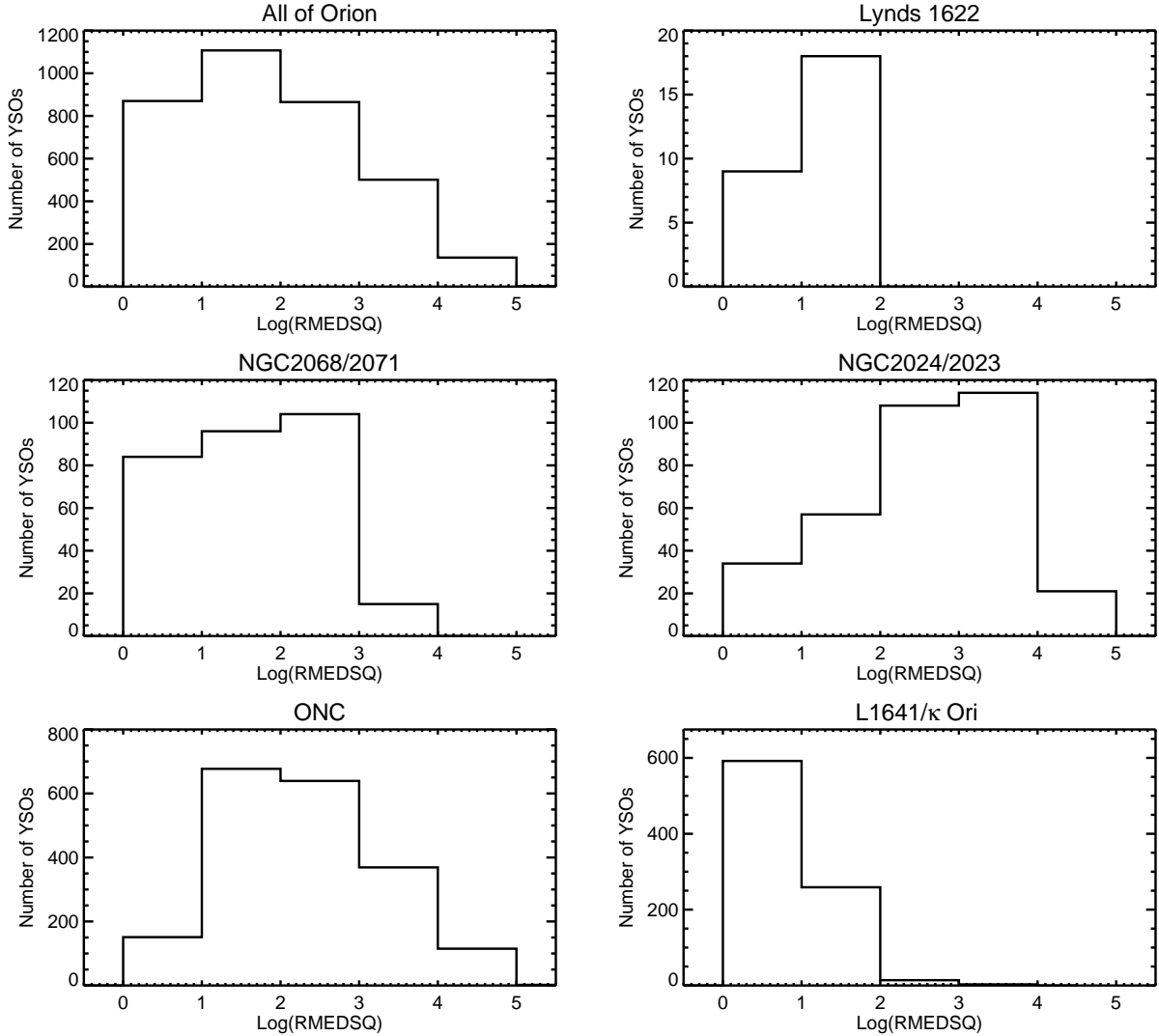}
\caption{The distribution of the 8~$\mu$m $log{\rm (RMEDSQ)}$ values toward dusty YSOs in the entire {\it Spitzer} Orion Survey and for individual fields in the survey  (see Paper I for the definition of the different fields). In the fields containing bright nebulosity due to HII regions (the Orion Nebula and NGC 2024) or reflection nebulae (NGC~1977, NGC~2023, NGC~2068 and NGC~2071), the RMEDSQ varies by three orders of magnitude. }
\label{fig:dist_rmedsq}
\end{figure}

\begin{figure}
\plottwo{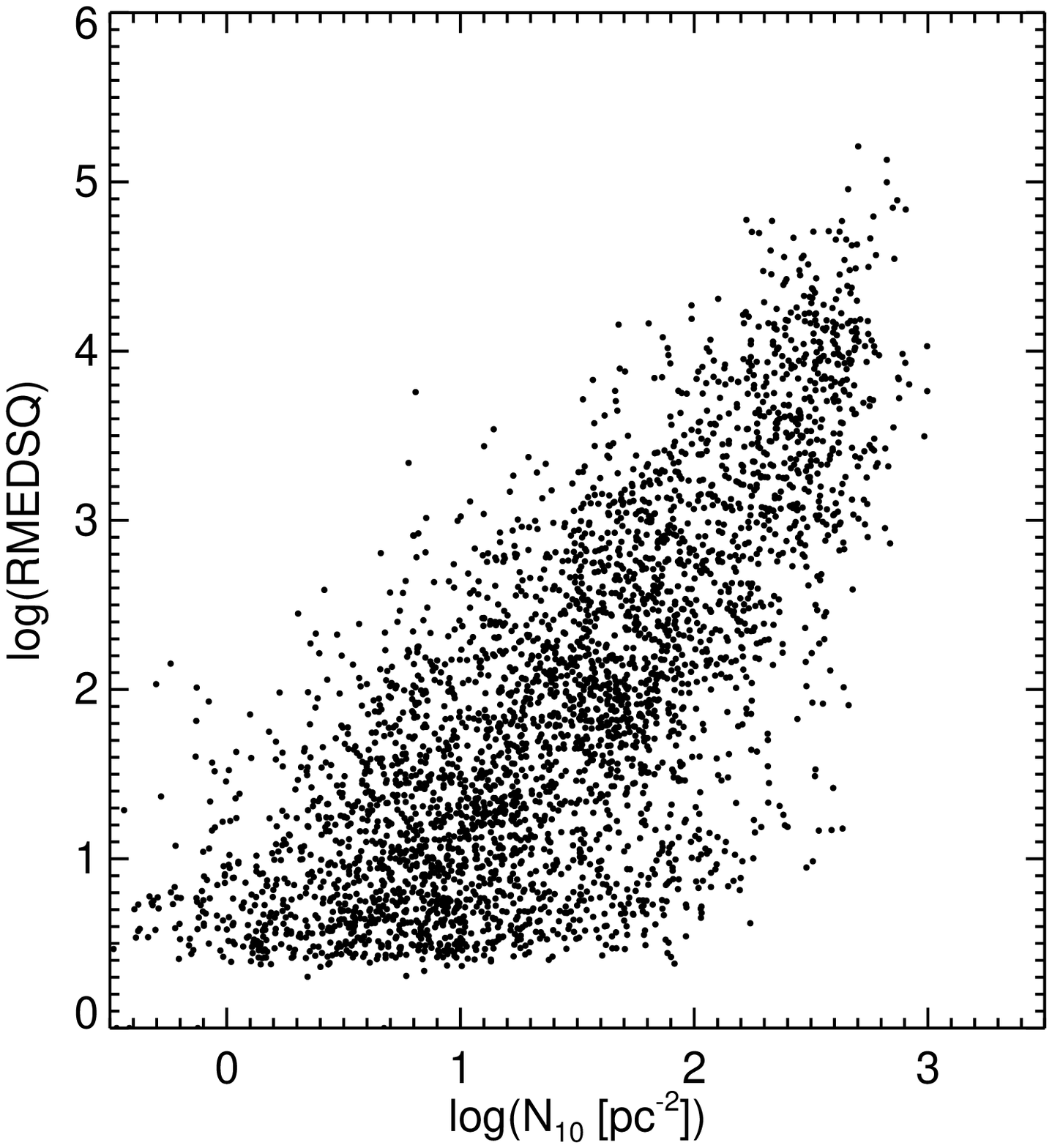}{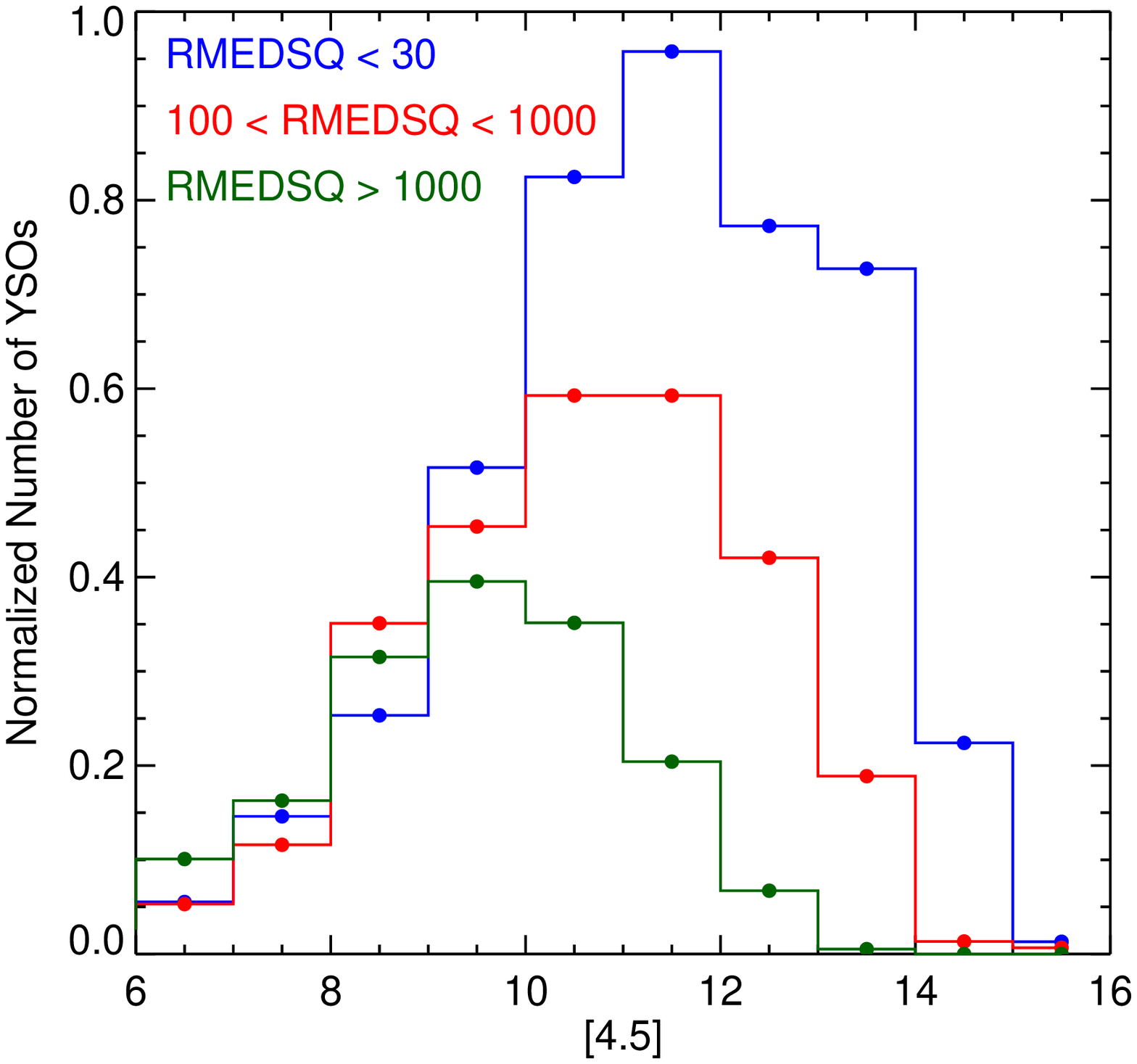}
\caption{{\bf Left:} the log of the nearest neighbor density, $N_{10} =9/(\pi r_{10}^2)$ where $r_{10}$ is the distance to the 10th nearest neighbor, vs.~the log of the RMEDSQ in the 8~$\mu$m band.  Note that the dense, clustered regions show systematically higher RMEDSQ  values. {\bf Right:} histograms of the 4.5~$\mu$m magnitude for young stellar   objects sorted by their RMEDSQ.  For increasing values of RMEDSQ, the  faint end of the histograms become increasingly truncated: this is due  to lower rates of detection in these faint magnitudes bins.  Together, these plots demonstrate lower detection rates of YSOs in crowded regions.  This is mostly due to the bright nebulosity found in clusters which is tracked by the  8~$\mu$m RMEDSQ.}
\label{fig:medvhist4p5}
\end{figure}

A study of the spatial distribution of dusty YSOs requires a correction for the spatially varying completeness found in mid-IR surveys of star forming regions.  Paper I analyzed the completeness of the{\it Spitzer} point source catalog in the four IRAC bands and the MIPS 24~$\mu$m band using artificial star tests. To account for spatial variations,  the completeness in each band was determined as a function of magnitude and the root median square deviations (see Appendix~B for definition of RMEDSQ). This analysis showed that the completeness of the catalog is a strong function of the RMEDSQ and that the completeness varied between the bands.  This spatially varying completeness can bias comparative studies of crowded and sparse regions. In Figure~\ref{fig:medvhist4p5}, we plot the value of the $log({\rm RMEDSQ})$ in the 8~$\mu$m bandpass as a function of stellar density for the dusty YSOs in the Orion survey; this plot shows that the RMEDSQ increases significantly with the stellar density.  This is due to the bright nebulosity, which is particularly apparent at 8~$\mu$m, being strongly enhanced in all the {\it Spitzer} wavelength bands toward clustered regions.    We also show histograms of  4.5~$\mu$m magnitudes for the identified Orion YSOs as a function of their RMEDSQ. As the RMEDSQ values increase, the faint end of the distribution is progressively eroded until only the brightest  stars are left.   In Figure~\ref{fig:dist_rmedsq}, we show histograms of the $log({\rm RMEDSQ})$ values for YSOs in each of the fields displayed in Figures~10-16 of Paper I.  The histograms show that the fields with bright clusters, such as the ONC and the NGC2024/2023 regions, have systematically higher values of RMEDSQ, and hence systematically higher incompleteness.  This leads to a bias in {\it Spitzer} surveys in which bright nebulosity found toward embedded clusters preferentially reduces the number and density of YSOs in those regions.

Our study focuses on the population of YSOs toward the Orion A and B molecular clouds. It is well known that the population of young, low mass stars in Orion extends beyond the molecular clouds and into the older OB1c, OB1b and OB1a subgroups of the Orion OB1 association  \citep[e.g. ][]{2008hsf1.book..838B}.  By design, our survey is spatially incomplete to the somewhat older stars of these subgroups.  Since our focus is on the population of stars associated with the molecular clouds, we focus on the incompleteness to faint YSOs within the spatial boundaries of our survey. 

In the remainder of this section, we address the completeness in two ways.  First, we add artificial YSOs to the IRAC mosaics to estimate the fraction of YSOs recovered as a function of the RMEDSQ. Second,  we use {\it Chandra} X-ray observations of the two clusters with the brightest IR nebulosity, those found in the Orion and NGC 2024 nebulae, to correct for the undetected sources.

\subsection{Correcting the YSO sample for Incompleteness on the Basis of the RMEDSQ}
\label{sec:correct_yso}

To establish the completeness of the YSO catalog, artificial YSOs were added to the mosaics; this process is described in Appendix~B. The goal was to compare the completeness of crowded regions with bright nebulosity to that of sparse regions with comparatively faint nebulosity.   As in the single band completeness analysis in Paper 1, we characterize the amount of fluctuations in the region surrounding a YSO by the RMEDSQ (Eqn.~\ref{eqn:remedsq} in Appendix~B). The spatial variations in the background measured by  the RMEDSQ can be due to both stars and nebulosity; however,  the bright nebulosity typically dominates the background fluctuations at all wavelengths longward of 3~$\mu$m.  We use the 8~$\mu$m RMEDSQ  to characterize the background fluctuations; this band is most dominated by the nebulosity. To ensure that the artificial YSOs had realistic properties, a fiducial sample of dusty Orion YSOs was extracted from regions of our mosaics with low values of RMEDSQ. The colors and magnitudes of the artificial YSOs were then randomly chosen from those of the fiducial YSO sample. The artificial YSOs were then extracted from the mosaics using the methods described in Paper~I.

The resulting fraction of recovered YSOs gives the completeness relative to that found in the low RMEDSQ region of the fiducial sample. The results of the artificial YSO analysis are shown in Figure~\ref{fig:frac}, where we display the fraction of recovered YSOs as a function of the 8~$\mu$m RMEDSQ.   We find a strong dependence of completeness on the RMEDSQ. In Appendix~B, we assess the dependence of the fraction on the chosen fiducial sample.  To determine a functional relationship between the fraction of recovered YSOs and RMEDSQ, we adopted the approach of Paper~I and fit a modified error function. In contrast to the single-band curves in Paper~I, there is no dependence of the fraction of recovered YSOs on magnitude since we added a representative sample of YSOs spanning a range of magnitudes.


 We use the fit displayed in Figure~\ref{fig:frac} to correct for incompleteness in our analyses of YSO surface densities, YSO demographics, and embedded cluster properties. A weight is assigned to every detected YSO which accounts for the expected number of YSOs that were not detected.  A weighted YSO will then be counted as more than one object if it is found in a region with a high RMEDSQ.  To determine the weighting,  we use the best fit function in Figure~\ref{fig:frac} to calculate the fraction of recovered YSOs as a function of RMEDSQ.  For a given YSO, the weighting factor, $w$ and its uncertainty, $\sigma_w$ are then determined by the equations

\begin{equation}
w = \frac{1}{f},~\sigma_w= \frac{\sigma_f}{f^2},
\end{equation}

\noindent
where $f$ is calculated by Eqn.~\ref{eqn:logerf} for the RMEDSQ measured around the YSO.  The value of $\sigma_f$ is the formal uncertainty in $f$ calculated using the coefficients and uncertainties in Appendix~B. We weight all YSOs with a given value of RMEDSQ equally. For example, if we expect 20\% of the YSOs to be missing for a given RMEDSQ, then we apply a weight of 1.25 to all stars with that value of RMEDSQ.  Thus, the YSOs added to correct for the incompleteness follow the spatial distribution of observed YSOs.  Accordingly, YSOs are not added to regions with high RMEDSQ values but without any detected YSOs: we presume that these regions are empty.  Since the distribution of YSOs in molecular clouds is highly non-uniform, the assumption that  the missing YSOs follow the distribution of the observed YSOs is more realistic than adopting an uniform distribution.  

The uncertainty in the weight goes up as the fraction of detected stars decreases; hence, in regions where the completeness is low, an alternative method  is preferable for determining the number of missing stars. In the next two sections, we use X-ray data from the {\it Chandra} observatory to correct for the incompleteness in the ONC and NGC~2024 nebulae; these regions show the brightest nebulosity in the Orion survey and the highest level of incompleteness.

\subsection{Comparison with the COUP Survey of the ONC}
\label{sec:coup}

The {\it Chandra} Orion Ultradeep Project (hereafter: COUP) obtained a nearly continuous 9.7 day exposure of the ONC over a $17' \times 17'$ field of view with the ACIS-I instrument onboard {\it Chandra}; these are the deepest existing X-ray observations of a star forming region \citep{2005ApJS..160..319G,2005ApJS..160..379F}.   Young, low mass stars often exhibit elevated yet highly variable X-ray emission;  consequently, X-ray surveys provide a means to identify young stars both with  and without IR-excesses \citep[e.g.][]{2007prpl.conf..313F,2007ApJ...669..493W}.  Furthermore, X-ray  observations can detect deeply embedded sources and are not  limited in sensitivity by the bright nebulosity typically found in mid-IR observations toward young clusters. 

To assess the incompleteness of the {\it Spitzer} data toward the Orion Nebula, we have examined the radial dependence of the number of X-ray sources, IR-excess sources, and X-ray detected IR-excess sources in the ONC.  To minimize contamination of the COUP from background AGN, we use only COUP sources with detected near-IR analogs taken from Table 10  of \citet{2005ApJS..160..319G}, which tabulated X-ray sources detected in near-IR 2MASS, NTT and VLT imaging of the ONC.  The requirement of an IR detection should eliminate all but a few ($< 10$) extragalactic sources \citep{2005ApJS..160..353G}. The nebulosity in the near-IR images is much weaker than that found in the mid-IR images; consequently, the completeness of the near-IR photometry is much less affected by nebulosity than the {\it Spitzer} mid-IR imaging. In this analysis, we only include {\it Spitzer} sources that are located within the COUP field.

Figure~\ref{fig:coup_radial} shows the azimuthally averaged surface density of X-ray sources, IR-excess sources, and X-ray detected IR-excess sources binned by radial distance.  We also show the surface densities normalized to the surface density of young stars detected in the COUP survey.  The radial distances were calculated relative to a central position defined by the median right ascension and declination of the COUP sources and the densities were then determined for concentric annuli of constant width. We divided the X-ray sources into those that do  have and do not have detections in a sufficient number of {\it Spitzer} bands to test for an IR-excess (as described in Paper I, the detections of IR-excesses require detections in at least two {\it Spitzer} bands, and most criteria for identifying IR-excess sources require detections in 3-4 {\it Spitzer} bands or detections in two {\it Spitzer} bands and two 2MASS bands).   A distinctive peak in the density of X-ray sources is apparent toward the center of the ONC.  This peak is not present in the {\it Spitzer} YSO catalog, demonstrating that there is a very compact clustering of young stars in the center of the ONC where our {\it Spitzer} census of IR-excess sources is incomplete due to the bright mid-IR nebulosity. The center of the peak is dominated by the COUP sources with IR-counterparts that do not have {\it Spitzer} photometry in a sufficient number of bands to be identified as IR-excess sources (hereafter: COUP-only sources).  This confirms that the peak is not apparent in the {\it Spitzer} data because of the lack of detections in the {\it Spitzer} bands.

The COUP data can be used to estimate the number of dusty YSOs missed by {\it Spitzer} in the center of the ONC. 
However, in addition to the dusty YSOs found by {\it Spitzer}, the COUP survey detects diskless young stars without IR-excesses that cannot be identified with the {\it Spitzer} data alone. To correct for this, we estimate the fraction of sources with IR-excesses using two methods.  First, we find the fraction  of X-ray sources with IR-excesses by taking the ratio of the number of X-ray  sources with IR-excesses to the number of X-ray sources which have sufficient infrared photometry to apply the color criteria necessary to detect IR-excesses.  We calculate this ratio for stars between radii of 0.1 and 0.13 degrees, outside the bright center of the Orion Nebula where the {\it Spitzer} data are highly incomplete.  A total of 98/179 X-ray sources, or $0.55 \pm 0.06$, show IR-excesses. This value is comparable to the disk fractions found for the X-ray selected samples of other embedded clusters studied by {\it Spitzer} and {\it Chandra} \citep{2007ApJ...669..493W,2010AJ....140..266W}.  Second, we take the ratio of {\it all} IR-excess sources over the total number of YSOs identified by either IR-excess or X-ray emission.  In this case, the fraction of sources with excesses is 136/193, or $0.70 \pm 0.06$.  



The second method for determining the disk fraction results in a significantly higher disk fraction.  The reason is that there are IR-excess sources in the COUP field that are not detected by {\it Chandra}. The sources lacking {\it Chandra} detections are typically faint, as shown in the $J$ vs. $J-H$ diagram of young stars in the COUP field (Figure~\ref{fig:coup_cm}) .  We find an increasing number of {\it Spitzer}-only sources for  $J > 12$; for a 1 Myr population of stars, this corresponds to masses $< 0.25$~M$_{\odot}$.  The  X-ray luminosity drops with mass, and the lowest mass M-stars and substellar objects can  typically be detected only during flares \citep{2005ApJS..160..401P}; the resulting fraction of stars with {\it Chandra} detections depends on both the duration of  the observations and the rate of flaring.

To augment our catalog of YSOs in the ONC,  we use the COUP catalog to correct for the incompleteness in the central regions of the Orion Nebula by  including stars that lack  {\it Spitzer} detections in enough bands for the identification of an IR-excess. There are two factors which complicate this approach.  First, since we only include stars with IR-excesses in the {\it Spitzer} survey, the fact that COUP detects stars both with and without disks could result in an overestimate in the number of YSOs.  Second, the lack of X-ray detections for very low mass stars could result in an underestimate  of the number of YSOs. To account for both these factors, we set the weight of every COUP-only source to a single value: the density of {\it Spitzer} identified IR-ex sources (whether or not they are detected by COUP) divided by the density of X-ray sources with sufficient IR photometry to determine whether they have an IR-excess. However, as shown in Figure~\ref{fig:coup_radial}, this ratio can change significantly with radius.  We find that the ratio of the IR-excess sources to X-ray sources with sufficient IR photometry varies from 0.78 to 0.62 over radii of $0.0060^{\circ}$ to $0.14^{\circ}$.  Since the annulus at radius $= 0.1275^{\circ}$ is less affected by incompleteness than the annuli at smaller radii, and since the sensitivity of {\it Chandra} decreases at larger radii, we adopt the ratio of $0.75$ found at this radius as the weight for the X-ray sources.  

Recently, \citet{2014ApJ...787..108G} argued that the stars in the center of the ONC are younger and have a higher disk fraction.  To assess the effect of a higher disk fraction in the inner nebula, we estimated the disk fraction for X-ray detected sources in the inner 0.06 pc.  If we only consider stars with sufficient {\it Spitzer} photometry to determine whether they have infrared excesses or not, we find a disk fraction in the inner 0.06 pc of $0.71 \pm  0.05$.  Although this suggests that the disk fraction may increase in the central region of the cluster, this high disk fraction should be considered an upper limit. The sample of stars with {\it Spitzer} photometry is expected to be biased to sources with IR excesses in the inner cluster since  sources with IR-excesses are brighter and easier to detect in the mid-IR.  If we instead use the total number of excess sources divided by the total number of COUP identified YSOs in the central cluster, the disk fraction drops to $0.30 \pm 0.02$ due to the lack of detections in the IR.  Given the large uncertainties in determining disk fraction in the inner cluster, we will assume a constant disk fraction. If the disk fraction does rise up to $0.71$, the ratio of IR-excess sources to {\it Chandra} sources may increase from $0.75$ to as high as $0.97$.  Consequently, if there is an increase in the disk fraction in the inner cluster, we are  underestimating the number of dusty YSOs in the inner ONC by as much as 22\%.   Since we consider the increased disk fraction in the inner cluster to be an upper limit, the 22\% value should be considered an upper limit. 
  

In the following sections, we will analyze the spatial distribution of YSOs using three different variants of the Orion Survey sample: the uncorrected {\it Spitzer} YSO point source catalog, the YSO point source catalog augmented by the {\it Chandra} X-ray sources, and the point source catalog augmented by the X-ray data {\it and} corrected by the weighting factors determined from the RMEDSQ analysis.  In the first case, the X-ray data are not used and the weights of all the {\it Spitzer} sources are set to 1.  In the second case,  the weight of $0.75$ for the COUP-only sources is used and  the weights of all the {\it Spitzer} identified IR-excess sources are set to 1.  In the third case, we include both the COUP-only sources and we apply an RMEDSQ correction outside the COUP field. To ensure consistency in the third case, we need to minimize any discontinuity in the density of sources between the COUP field and the surrounding regions.  In Figure~\ref{fig:coup_radial}, we show the radial density and normalized radial density for IR-excess sources corrected by the RMEDSQ weighting scheme described in the previous sub-section.  At radii of $0.14^{\circ}$ to $0.17^{\circ}$, we find the average RMEDSQ corrected weight is 1.72.  To ensure that the weighting factors throughout the {\it Spitzer} survey are consistent with those within the COUP field, we adopt a weight of 1.72 for all {\it Spitzer} identified IR-excess sources in the COUP field. Furthermore, we will assign the COUP-only sources a weight of $0.75 \times 1.72 = 1.29$.   We note that when we apply this weighting scheme, the ratio of the corrected density of YSOs to the density of COUP detected young stars is remarkably constant with radius (Figure~\ref{fig:coup_radial}).  

\subsection{Comparison with {\it Chandra} Observations of NGC 2024}
\label{sec:chandra}

The NGC~2024 nebula was imaged by {\it Chandra} in a 76 ks exposure \citep{2003ApJ...598..375S}.  Although the sensitivity was much lower than that of the COUP survey, the {\it Chandra} data for NGC 2024 does provide the spatial distribution of young stars unbiased by confusion with the bright nebulosity.  In Figure~\ref{fig:n2024_xray_radial},  we show the radial plots of the azimuthally averaged surface density of sources and  the surface density of sources normalized by the density of young stars in the {\it Chandra} data.  They are shown as a function of radius from the central density peak of the cluster.  For the NGC~2024 data, we count an X-ray source as a young star if it is detected in at least one band by {\it Spitzer} or 2MASS. As was the case for the ONC, a substantial number of sources are detected in the center of the nebula at X-ray and IR wavelengths that lack sufficient IR photometry to be identified as dusty YSOs by {\it Spitzer}.  

To account for these sources, we adopt the same methodology we applied to the COUP survey of the ONC.  The first step is to find the ratio of IR-excess sources to X-ray sources detected in a sufficient number of {\it Spitzer} bands to identify IR-excesses.     The densities of IR-excess sources, X-ray detected IR-excess sources, and sources with sufficient IR photometry are shown in Figure~\ref{fig:n2024_xray_radial}.    At radii $> 0.08^{\circ}$, we find the normalized number of IR-excess sources fluctuates due to the smaller number of sources in these outer regions.  Thus, for the weighting of the X-ray sources, we use the  average ratio of all IR-excess sources to X-ray sources with sufficient {\it Spitzer} photometry between radii of $0.0675^{\circ}$ and $0.0825^{\circ}$; the average ratio equals $1.16$. This weight value is much higher than that found in the ONC. We note that the disk fraction determined by the ratio of the number of X-ray detected IR-excess sources to the number of X-ray sources with sufficient IR photometry is  $0.58 \pm  0.10$, consistent with that found in the ONC. 
 Thus, the higher value of $1.16$ is due to the lower sensitivity of the {\it Chandra} observations toward NGC~2024 and the resulting lower detection rate for the X-ray sources.   
 
 \citet{2014ApJ...787..108G} also found evidence that the typical ages of the stars decrease and the disk fraction increases in the center of the NGC~2024 cluster.  Using the sample of X-ray detected sources with sufficient IR photometry, we find that the disk fraction increases to $0.64 \pm 0.07$ in the inner 0.06~pc of the cluster.  However, these data suffer from the same biases we discussed in the previous section, and we consider this number an upper limit.  If this increase is real, we are underestimating the number of dusty YSOs in the center of NGC~2024 by up to 12\%.

As we described previously, we analyze the spatial distribution of YSOs using separately the uncorrected Spitzer YSO point source catalog,   the YSO point source catalog augmented by the {\it Chandra} X-ray sources, and the point source catalog augmented by the X-ray data {\it and} corrected by the RMEDSQ weighting factors.  For the second case, we use the same approach adopted for the ONC, and we give each X-ray source without sufficient IR-photometry a weight of $1.16$ and each {\it Spitzer} identified IR-excess sources a weight of $1$. For the third case, we must once more minimize the discontinuity between the densities outside the {\it Chandra} field, which are corrected by the RMEDSQ factor, and the densities within the {\it Chandra} field.  To do this, we adopt the typical RMEDSQ weight given to YSOs in the outer radius of the NGC~2024 region: between radii of $0.1275^{\circ}$ and $0.1425^{\circ}$ we find an average weight of 1.44.  This value is much lower than the weight found in the outer regions of the ONC due to sharp decrease in the nebulosity in the outer regions of the NGC~2024 nebula.  Thus, when the incompleteness is corrected across the entire survey, we assign every {\it Spitzer} identified IR-excess source in the NGC~2024 {\it Chandra} field a weight of 1.44, and every {\it Chandra}-only source a weight of $1.44 \times 1.16 = 1.67$.   As was the case in the ONC, the ratio of the corrected YSO density to {\it Chandra} detected young star density is remarkably constant, particularly for radii $< 0.9^{\circ}$ where the density of sources is high.  

\subsection{The Number of Dusty YSOs Before and After Completeness Correction}

The corrections for incompleteness substantially change the number of dusty YSOs in the Orion molecular clouds.  From the Spitzer data, we identify 3481 YSOs in the Orion clouds; 2821 in Orion A and 660 in Orion B. 
After adding in the X-ray sources without sufficient Spitzer photometry to identify IR-excesses, we obtain a total of 3889 dusty YSOs; 698 in Orion~B and 3191 in Orion~A.  In this case, every COUP X-ray source is weighted by 0.75 and every {\it Chandra} source in NGC~2024 is weighted by $1.16$.  Finally, with the full correction, we estimate that there are 5104 dusty YSOs in the Orion clouds: 905 in Orion~B and 4199 in Orion~A. Note that these numbers do not include a correction for the number of YSOs which have already dissipated their dusty disks and envelopes and do not exhibit IR-excesses.  

\section{The Densities and Spacings of Dusty YSOs in the Orion Clouds}
\label{sec:den}

Although young stars are often divided into high density clusters surrounded by a low density distributed population, observations of molecular clouds show a continuum of densities.  \citet{2010MNRAS.409L..54B} constructed the distribution of stellar densities for nearby star formation regions from catalogs of dusty YSOs from the c2d, Gould Belt and Orion Molecular cloud surveys.  They found a continuous distribution of densities spanning four orders of magnitude with no evidence for a break in the distribution that might suggest the presence of two or more distinct populations of young stars.  \citet{2011ApJ...739...84G}, \citep{2012ApJ...752..127M} and \citep{2014ApJ...794..124R} found a power-law relationship between the surface density of dusty YSOs and the column density of gas spanning three orders of magnitude in YSO stellar density.  In this power-law relationship, the YSO surface density scales as the 1.8-2.7 power of gas column density and the efficiency of star formation increases with the gas column density \citep[see also][]{2010ApJ...723.1019H,2013ApJ...773...48B,2013ApJ...778..133L,2013A&A...559A..90L,2014A&A...566A..45L}.  This relationship results in clustered regions with high star formation efficiency (SFE) surrounded by a distributed population of more isolated stars with a SFE efficiency even though there is a continuum of densities and efficiencies and not distinct modes of star formation.

The spatial distribution of dusty YSOs in Orion are shown in Figure~9 of Paper I. This figure shows that the dusty YSOs in Orion extend throughout the cloud complex. Dense clusters such as the ONC are found to be peaks in the YSO surface density.  Extended regions of relatively low YSO density are also apparent, with the distribution of YSOs in such regions tending to follow filamentary structures punctuated by small density peaks.  In this section, we examine the statistical distribution of YSO densities in the Orion molecular clouds.  Our analysis complements that of \citet{2010MNRAS.409L..54B}, who excluded the ONC from their consideration to minimize biases due to incompleteness.  By taking into account the incompleteness in the rich clusters of Orion, we can extend the distribution of YSO densities to the most active star forming region of the Gould Belt.  

\subsection{The Distribution of YSO Surface Densities}
\label{sec:dist_density}

In Figure~\ref{fig:nnden}, we show the distribution of nearest neighbor surface densities for all the  identified dusty YSOs in the survey.  For each YSO, we calculate the nearest neighbor distance using the equation 

\begin{equation}
N_{n} = \frac{n-1}{\pi r_{n}^2},
\label{eqn:nnden}
\end{equation}

\noindent
where $r_n$ is the distance to the {\it n}th nearest neighbor \citep{1985ApJ...298...80C,2005ApJ...632..397G}.  To include the weights and correct for incompleteness,  we modify the equation to take into account the total weight of all the YSOs up to the nearest neighbor distance, including the {\it n}th nearest neighbor and the central YSO.  The resulting density is

\begin{equation}
N_{n} = \frac{w_{tot}}{\pi r_{n}^2}
\label{eqn:nnweight}
\end{equation}

\noindent
The weight, $w_{tot}$, is the sum of the weights for all $n+1$ YSOs at a radius of $r \le r_n$:

\begin{equation}
w_{tot} = \sum_0^n {w_i -2 -0.5(w_n-1)}.
\label{eqn:weight}
\end{equation}

\noindent
Since we are measuring the density inside the annulus between the central YSO and the outer YSO, we subtract $2$ from the weight: $1$ for the central YSO and $1$ for the outer YSO. For the inner YSO, we add the excess weight above 1, i.e. $w_0-1$; thus, if the weighting for the central YSO is 1.5, then we include a total value of 0.5 for that YSO in our density measurement.  For the outer YSO (the {\it n}th YSO), we assign a value of $0.5(w_n-1)$.  In this case, if the outer YSO has a weight of $1.5$, we assign $0.25$ as the number of YSOs inside the annulus; the remaining $0.25$ are considered to be outside the annulus.  The variance of the nearest neighbor density is given by  

\begin{equation}
\sigma_{n} = \frac{N_{n}}{\sqrt{n-2}}
\end{equation}

\noindent
\citep{1985ApJ...298...80C}.  This is used for both the weighted and unweighted values. For $n=10$ and $n=5$, which are used throughout this paper, the uncertainties are 35\% and 58\%, respectively.  The variance characterizes the range in fluctuations in $N_{n}$ for multiple realizations of a randomly distributed set of stars with an average density of $N_{n}$, and it is not an expression for the uncertainty in the {\it measurement} of $N_{n}$  at a particular location in the Orion clouds. The uncertainty in the measured $N_{n}$ is instead dominated by the uncertainties in the distance to the Orion clouds and the incompleteness in the YSO sample and cannot be simply characterized.  Uncertainties due to the distance, which may be as much as 12\% due to the depth of the Orion cloud complex \citep[e.g.][]{2005A&A...430..523W}, are on the order of 25\%.  Uncertainties due to the incompleteness correction are approximately 10\% given the spread of measured values for the fraction of YSO recovered shown in Figure~\ref{fig:frac}.

The distribution of $N_{n}$ densities is displayed in Figure~\ref{fig:nnden} for three separate cases.  The first case uses the  catalog of YSOs with IR-excesses identified with {\it Spitzer}.  As described previously, this plot may not reproduce the high stellar densities in clusters where the bright nebulosity lowers the completeness.  Consequently, high density regions are incomplete relative to the low density regions (Figure~\ref{fig:medvhist4p5}). To correct for the incompleteness in the Orion and NGC 2024 nebulae, the second case has the additional X-ray sources as described in Secs.~\ref{sec:coup} and \ref{sec:chandra}. The third case employs the RMEDSQ based weighting correction for all sources outside the {\it Chandra} fields, this corrects for incompleteness throughout the entire survey.  This final case is our most aggressive attempt at creating a density distribution unaffected by the spatially varying completeness and should be considered an upper envelope to the $N_{10}$  distribution.   It extends to higher densities than the other cases since the density of the clustered regions have been scaled up by the weighting scheme. 

The uncertainties are derived by a combination of the variance in the $N_{10}$ values and the Poisson statistics for the number of objects in each bin. We do not include systematic uncertainties in the distance or in the incompleteness correction. To take into account the variance in the individual surface densities, we perform 1000 iterations of the curve where we vary the density at each point by a normal distribution where the peak and standard deviation are given by $N_{10}$ and $\sigma_{10}$, respectively. We then calculate the mean and standard deviation, $N$ and $\sigma_{bin}$, of the 1000 iterations for each of the bins.  We add this uncertainty, $\sigma_{bin}$ to the
Poisson uncertainties for each bin:

\begin{equation}
\sigma(bin) = \left( \sigma_{bin}^2 + <w>^2 N \right)^{1/2},
\end{equation}

\noindent
where $<w>$ is the average weight of the stars in a given bin and $N$ is the number of YSOs in the bin.  In the cases where we do not correct for incompleteness by assigning weights to the  YSOs, $<w> = 1$. 

All three cases show a broad peaked distribution where the peak extends between 10 and 100~pc$^{-2}$.   All three also extend to densities above 1000~pc$^{-2}$.  The peak densities approach $10^4$~pc$^{-2}$ when the {\it Chandra} data are used to augment the source catalog in the ONC and NGC~2024, where the highest YSO densities are found. These distributions shows that the observed YSO surface density varies by almost five orders of magnitude.  The low density region below 10~pc$^{-2}$ is  well represented by a log-normal distribution; however the entire distribution is too broad to be represented by a log-normal distribution or the superposition of two log-normal distributions.  In all three versions, the high stellar density side of the curve show a peak between 40-60~pc$^{-2}$ and a wing that extends to densities above 1000~pc$^{-2}$.  

The divergence from a log-normal distribution is not surprising for two reasons.  First, extinction maps of molecular clouds exhibit gas column density distribution that are not log-normal, but instead are best fit by a lognormal function at lower column densities and a power-law tail or a 2nd lognormal function at high column densities \citep{2009A&A...508L..35K,2013A&A...549A..53K}.  Furthermore, if the column density of YSOs appears to scale as 2nd to 3rd power of the gas column density  \citep{2011ApJ...739...84G}, the exponential tails apparent in the  gas column density distribution should be even more prominent in maps of YSO surface density.


\subsection{Mapping the YSO Surface Density}


In Figure~\ref{fig:nn_map}, we show the map of nearest neighbor densities for all the identified YSOs.  For each point in a rectangular grid, we calculate the density using Eqn.~\ref{eqn:nnweight}.  Since there is no longer a central YSO (except in rare chance coincidences when a YSO is located at the grid point), the annulus is now defined by the outer YSO and the central grid point.  Correspondingly, the central YSO does not need to be subtracted out of the weight term and the weight  is defined as:

\begin{equation}
w_{tot} = \sum_0^n {w_i -1-0.5(w_n-1)}
\label{eqn:weightmap}
\end{equation}

\noindent
where $w_n$ is the weight of the {\it n}th star \citep{1985ApJ...298...80C}.   In this map, we use the {\it Chandra} data to augment the density distribution in the ONC and NGC~2024. 

A large range in stellar densities is again evident.  Spatially extended regions of high densities correspond to the previously known embedded clusters and groups. These include  the highly elongated ONC, the NGC~2024 cluster, the small group towards the reflection nebula NGC~2023, the double peaked cluster found toward the reflection nebulae NGC~2068 and 2071, and the numerous small groups found in L~1641 \citep{1992ApJ...393L..25L,1994ApJS...90..149C,1995PhDT..........A,1998ApJ...492..540H}. The clouds also contain large regions of relatively low stellar densities \citep{1993ApJ...412..233S,2000AJ....120.3139C}, these distributed regions dominate the cloud in area. In Sec.~4, we probe the demographics of clustering in the Orion complex, i.e. the fraction of stars in large clusters, in small groups, and in the distributed population.


\subsection{Comparing the Spatial Distributions of Protostars and Pre-main Sequence Stars with Disks}
\label{sec:comparing}
 
In the previous analysis of the spatial densities of YSOs, we ignored the distinction between protostars \citep[ages $< 0.5$~Myr,][]{2009ApJS..181..321E} and the older pre-main sequence stars with disks \citep[hereafter: disks sources; ages $< 5$~Myr,][]{2008ApJ...686.1195H}.  Do protostars and disk sources show distributions of surface densities similar to that for all YSOs, as shown in Fig~\ref{fig:nnden}, or are there systematic differences?  Is there evidence for an evolution in the spatial distribution of YSOs, and in particular, evidence for the migration of the older disk sources from their formation sites? Finally, what do the spacings of protostars imply about the fragmentation process and the potential for subsequent interactions between protostars? We approach these questions through a comparison of the nearest neighbor separations and nearest neighbor densities of protostars and those of the more evolved disk sources.  

We first examine nearest-neighbor separations between protostars and between stars with disks to facilitate comparisons with previous analyses performed with {\it Spitzer} data for the NGC~1333, Serpens Main and AFGL 490 clusters \citep{2007ApJ...669..493W, 2008ApJ...674..336G,2010AJ....140..266W,2012ApJ...752..127M}. In Figure~\ref{fig:orion_nn}, we show the separations of protostars and disks; these have not been corrected for incompleteness. We have included all protostellar candidates: the protostars, faint candidate protostars and red candidate protostars from Paper~I. Regions with the highest YSO densities, particularly the Orion nebula and NGC~2024, are highly incomplete; consequently, the most tightly spaced protostars and disks are not accounted for in the displayed distributions.  However, given the difficulties in correcting the number of protostars when one of the primary bands for identifying protostars, the 24~$\mu$m band, is saturated in the densest clusters, we have chosen not to augment the nearest neighbor distributions like we have done for the distribution of all dusty YSO densities.   Instead, we also show the nearest neighbor separations for the combined L1641 cloud and $\kappa$~Ori region (hereafter: L1641/$\kappa$~Ori region, see Paper I for the definitions of these regions), which contains both a high number of objects, yet due to the lack of  massive stars \citep{2013ApJ...764..114H}, does not contain the bright nebulosity that reduces completeness (see Figure~\ref{fig:dist_rmedsq}).


The cumulative distributions of the nearest neighbor distances (hereafter:~nn2) are shown in Figure~\ref{fig:orion_nn}.  The cumulative distribution for the protostars shows that for any fraction of sources, the spacing of the protostars are larger than that of the more evolved disk sources. The median spacing between protostars, 0.13~pc, is larger than that between disk sources, 0.08~pc.  The difference is significant, a  Kolomgorov-Smirnov (K-S test) gives a probability of the distributions being drawn from the same parent distribution as only $log(P) = -18$.  However, much of the difference between these two samples may be due to the incompleteness to protostars in the dense centers of the ONC and NGC~2024 clusters, where the saturation of the 24~$\mu$m band and the lower sensitivity in the lower wavelength IRAC-bands limits our ability to detect and identify protostars.  To reduce this bias, we perform the same analysis for the L1641/$\kappa$~Ori region.  In L1641/$\kappa$~Ori, the median separation between protostars is 0.17~pc while the median separation between disk sources is 0.13~pc.  The K-S test give a probability of $log(P) = -4.1$ that the two distributions are drawn from the same parent distribution.  In contrast, the protostars in the  the Serpens and AFGL~490 clusters haves smaller median separations than the disk sources \citep{2007ApJ...669..493W, 2010AJ....140..266W,2012ApJ...752..127M}, and in the NGC~1333 cluster, the separations of protostars and disks are indistinguishable \citep{2008ApJ...674..336G}. 

Part of the reason for the longer separations between the protostellar sources is their short lifetimes, and hence rarity, of protostars.  In other words, disks sources have shorter separations simply because there are more of them.  To remove this bias, we follow the analysis  of \citet{2009ApJS..184...18G} and examine the separations between protostars to the 5th nearest dusty YSO and disk sources to the 5th nearest dusty YSO.  By choosing the separation to the 5th nearest neighbor (hereafter:~nn6), we reduce the effect of random fluctuations on the nearest neighbor distance. The cumulative distributions for the nn6 distances are plotted for protostars and disks sources in Figure~\ref{fig:orion_nn}; as was done for the nn2 analysis, we consider both the entire Orion sample and the sample of objects in the L1641/$\kappa$~Ori region.  For the entire Orion sample, there is not a clear difference between the protostars and disk sources.  The median nn6 distance between protostars to YSOs is 0.20~pc while the median nn6 distance for disk sources to YSOs is 0.22~pc, and the K-S probability that they are drawn from the same sample is $log(P) = -0.9$.  However, if we restrict the sample to L1641/$\kappa$~Ori , we find that the protostars have systematically smaller nn6 distances: the median nn6 is 0.27~pc for protostars and 0.37~pc for disk sources.   The probability that the distributions for protostars and disks are from the same parent distribution, as given by the K-S test, is $log(P) = -3.7$.  

For comparison with the results in Sec.~\ref{sec:dist_density}, we also plot the nearest neighbor density histograms for the protostars and disk sources in Figure~\ref{fig:orion_proto_disk_nnden}.  The nearest neighbor density is calculated using  Eqn.~\ref{eqn:nnden} with no weights applied to the sources.  The calculation of the uncertainties are described in Sec.~\ref{sec:dist_density}. These results reinforce those of the nn6 distance analysis.  We see no difference between the protostars and disk sources in the cumulative distribution for the entire survey. If we limit our analysis to the L1641/$\kappa$~Ori region, however, we find that the protostars tend to be located in denser environments than the more evolved disk sources.  Since we have not corrected for incompleteness, the comparison between the protostars and disk sources in the entire ONC survey will be affected by the higher level of incompleteness for the protostars in the dense clustered regions.   However, if we limit our analysis to the L1641/$\kappa$~Ori region, the distributions of both the nn6 separations and nearest neighbor densities show than the protostars are found in systematically denser regions than the disk sources.  This  is in agreement with the result of \citet{2009ApJS..184...18G}, who found that the protostars in their sample of 36 embedded clusters also have systematically smaller nn6 distances than the disk sources.  Thus, we conclude that the weight of the evidence favors a tendency for protostars to be found in higher density environments than more evolved pre-main sequence stars.  We will discuss the implications of this result  in Sec.~\ref{sec:demoproto}.



Previous analyses of the nearest neighbor separations of protostars and YSOs in general suggested a characteristic separation similar to the local Jeans length \citep{2006ApJ...636L..45T,2009ApJS..184...18G}.  In the Orion molecular clouds, we find a wide range of separations, from 0.01~pc to 2.8~pc in L1641/$\kappa$~Ori.  The median separations are  0.13 and 0.17~pc ($2.6$ and $3.4 \times 10^4$~A.U.) for the Orion complex and for the L1641 cloud, respectively.  For, a random orientation, the corresponding 3D separation would be 0.17 and 0.22~pc  ($3.4$ and $4.5 \times 10^4$~AU). In comparison, a Jeans length of 0.2~pc  ($8 \times 10^4$~AU) requires $H_2$ densities of 1.5 and $3 \times 10^4$~cm$^{-3}$  for kinetic temperatures of 20 and 40~K, respectively.  These values are very similar to the kinetic temperatures and volume densities determined for dense cores in the Orion~A cloud by \citet{1999ApJ...525..343W}. Consequently, the median separations are consistent with the length-scale predicted for thermal, Jeans-type fragmentation of the gas. However, the distances that we used were the median separations, and the full distribution of separations spans almost 3 orders of magnitude. Future work should examine whether the range in separations can be explained by Jeans fragmentation in the very inhomogeneous and structured gas of the Orion molecular clouds.

The wide range of separations in Orion suggest that while some protostars form in relative isolation, others may be found in densely packed groups of interacting protostars. Studies of other star forming regions have found dense groups of protostars that could potentially interact \citep{2007ApJ...669..493W}; such interacting groups are also found in some simulations of cluster formation in turbulent clouds \citep[e.g][]{2012MNRAS.419.3115B}.  To assess the importance of interactions between protostars, we estimate the fraction of protostars where the projected separations to the nearest neighbor protostar is small enough that interactions may occur. In the entire sample, 11\% of the protostars have projected separations $\le 0.024$~pc, or 5000~AU; this percentage decreases to 7\% for the L1641/$\kappa$~Ori region.  An average projected separation of 5000 AU would correspond to a 3D separation of 6400~AU if the separation of the protostars were constant and the orientation of pairs of protostars with respect to the observer were random.  This separation is close to the size of molecular cores; \citet{2008ApJ...684.1240E} found deconvolved core diameters of $59''$ in the Ophiuchus cloud, corresponding to 7080 AU for the cloud distance of 120 pc   \citep{2008ApJ...675L..29L,2008A&A...480..785L}.  
Thus, with the caveat that cores have a range of diameters and could be systematically different in size within the Orion clouds, $\sim11\%$ of protostars of Orion could be part of interacting pairs/groups of protostars. 

Since the 3D distances are not known, the percentage of interacting protostars should be considered an upper limit to the actual percentage.  For this reason,  the interactions can at most affect 11\% of the Orion protostars, and the observed protostars typically are not close enough to directly interact.  We conclude that the protostars in Orion are found in a range of environments, ranging from small, dense, potentially interactive groups to protostars in relative isolation.  However, most protostars, even in the more clustered regions of Orion, are spaced at distance which make interactions unlikely and from this perspective, can be considered essentially isolated. 


\section{The Demographics of Dusty YSOs}
\label{sec:demo}

In the previous section, we found that the column density of YSOs  varies by more than three orders of magnitude within the Orion clouds, with the areas of high density organized into contiguous regions with varying sizes and morphologies (Figure~\ref{fig:nn_map}).  In this section and the following section, we focus on the contiguous regions of high YSO column density and their properties.  We refer to the large contiguous regions containing hundreds of dusty YSOs as clusters, while smaller regions with ten to a hundred YSOs we refer to as groups.\footnote{\citet{2003ARA&A..41...57L} define young clusters as assemblages with  more than 35 stars, the minimum size at which the relaxation time is greater than the crossing time.    Since gas dispersal can lead to a decrease in the cluster membership due to the ejection of stars, 35 members should be considered a lower limit. In order to use logarithmic binning, we define clusters as containing $\ge 100$ dusty YSOs.} The remaining YSOs that are not found in clusters and groups with 10 or more members are referred to as the distributed population.

It has been often suggested in the literature that clustered and isolated star formation are two distinct modes of star formation, potentially driven by separate physical processes; however, recent observation suggest that the clusters and distributed populations may be part of a continuum of densities and star formation efficiencies with no clear break between the two \citep{2007prpl.conf..361A,2010MNRAS.409L..54B,2011ApJ...739...84G}. Nevertheless, an analysis of the demographics of star formation - i.e. the fraction of YSOs that are found in clusters, groups or relative isolation - as well as the the properties of the individual groups and clusters can provide a unique characterization of how the observed YSOs are aggregated together. This characterization is needed to address some of the key problems posed in the introduction.  An analysis of the number of YSOs (and consequently the total stellar mass) found in  embedded clusters and the diameters, densities and morphologies of those clusters may provide clues into which clusters survive gas dispersal and the resulting distribution of open cluster masses.  Furthermore,  the fraction of YSOs in groups and clusters and the properties of those assemblages can provide a better understanding of the environments in which stars and planets form.

In this section, we develop a methodology for identifying clusters and groups in the Orion molecular clouds above a threshold density.  We then use this methodology to characterize the demographics of the Orion molecular clouds and we discuss how the adopted threshold density affects the demographics.  Finally, we compare the Orion clouds to other molecular clouds within 500 pc of the Sun.


\subsection{The Demographics of the Orion A and B Clouds}
\label{sec:demo_cloud}

An analysis of the demographics requires a methodology for isolating clusters.  Previous methods have relied on surface densities or projected spacings between stars \citep{1991ApJ...371..171L,2000AJ....120.3139C,2007prpl.conf..361A,2008ApJ...682..445C,2008ApJ...688.1142K,2009ApJS..184...18G}; we choose a similar strategy for the following analysis and search for contiguous regions above a threshold YSO surface density.   The primary parameter in this analysis is the threshold density.  Figures~\ref{fig:nnden} and  \ref{fig:nn_map} show no clear break between clustered and distributed  populations.  Except possibly in the case of NGC~2024, the clusters appear as peaks in more extended distributions of stars \citep{2007prpl.conf..361A}.  Hence, there is no apparent critical YSO separation or density that can be used to separate clustered and distributed YSOs. \citet{2009ApJS..184...18G} defined the boundaries of  clusters by searching for an increase in the gradient of the YSO surface density and the corresponding decreasing of branch lengths in minimum spanning trees.  Although this works well for individual clusters, it is more difficult on cloud scales where nested hierarchical structures and  variations in the YSO density over the length of a molecular clouds make this approach difficult to apply uniformly in a single cloud.  

An alternative approach is to compare the spatial distribution in the Orion clouds to that in other nearby molecular clouds. Although the average of the solar neighborhood  gives a continuous surface density with an approximately log-normal distribution  \citep{2010MNRAS.409L..54B}, individual clouds can exhibit density distributions which  can diverge significantly from the average.  This is shown in Figure~\ref{fig:nnden_nearby}, where we display the density distribution for clouds in the Orion survey, from the c2d survey, and from the Taurus molecular cloud.   In these maps, we have not corrected the Orion data for incompleteness.  Of particular importance is the distinction between the nearby dark clouds, Taurus, Lupus and Chameleon 2, and the molecular clouds with clusters, Perseus, Serpens, Ophiuchus, Orion~A and Orion~B.  In each of the plots, we have indicated the density 10~pc$^{-2}$, the distributions for the nearby dark clouds peak below this density while the distributions for molecular clouds containing embedded clusters peak at densities above this value.  For this reason,we initially pick 10~pc$^{-2}$ as our threshold density.  A comparison of different stellar density thresholds used to identify clusters is found in \citet{2010MNRAS.409L..54B}.

Next, we group together  sources found in contiguous regions where $N_{10} \ge 10$~pc$^{-2}$.  To identify contiguous regions, we use a friend of a friend method.   For a given YSO, the 10 nearest YSOs which also show $N_{10} \ge10$~pc$^{-2}$ are friends.  Each friend of a friend is a friend. In Figure~\ref{fig:clusterid10}, we show the results of this method for the Orion A and B clouds.  In this figure, each cluster identified by our friend of a friend technique is given a distinct color. Stars that are not assigned to a group or cluster with 10 or more members are given the color black; these are the distributed population.

In Figure~\ref{fig:demo}, we show the number of members as a function of the size of the group and/or cluster.  We have binned our sources in logarithmic intervals: large clusters with 10,000-1000 members, clusters with 1000-100 members, groups with 100-10 members, and the remaining objects are in the distributed population. We display this for each of the three cases, the {\it Spitzer} sources alone, the {\it Spitzer} sources augmented by {\it Chandra}, and the fully corrected sample; these three cases are also found in Table~\ref{tab:demographics}. In all three cases we get the same result as found by  \citet{2000AJ....120.3139C}: that 50\% to 80\% of the members are in the largest clusters. In the Orion~A cloud, most of the members are in the ONC, and in the Orion~B cloud, most of the members are in the NGC~2024 cluster, and the NGC~2068/2071 cluster.  In the combined distribution of the Orion A and B clouds, around 50\% of the members are in the ONC. The fraction of YSOs found in the distributed population depends strongly on the completeness correction, decreasing from 21\% in the uncorrected sample to 13\% in the fully corrected sample, with Orion~B exhibiting a slightly smaller fraction of distributed YSOs than Orion~A.  

How sensitive is this result to the chosen threshold density?  Figure~\ref{fig:demo_trend} shows how the fraction of members in groups and clusters and the number of groups and clusters as a function of the threshold density. As we increase the threshold density, there is a tradeoff between the distributed stars and the clusters; 47\% of the YSOs are found in the distributed population if we raise the threshold  to 100~pc$^{-2}$. Over the entire range of densities, the  ONC still has more members than all the small clusters and groups combined. The fraction of members in the  ONC drops continuously with increasing threshold density.  At 75~pc$^{-2}$, the ONC is broken into two clusters, which is seen as a jump in the number of small clusters in the displayed trends. The number of small clusters (100-1000) members and the fraction of YSOs in these clusters drops until 50~pc$^{-2}$, at which point only the NGC~2024 cluster is left. This cluster has a remarkably high average density which makes it relatively insensitive to the threshold density (Sec~\ref{sec:prop}).  In contrast, the NGC~2068/2071 and ONC clusters have lower average densities and more complicated internal structures; their properties are more strongly dependent  on the threshold density.  Interestingly, the number of groups (10-100 members) and the fraction of members in these groups is relatively insensitive to the threshold density.

In summary, we find that the ratio of YSOs in large clusters to distributed YSOs depends on the chosen threshold. However, the result that the larger clusters contain more than YSOs than the smaller clusters or groups seems to be insensitive to the adopted threshold.  Thus, our analysis is in agreement with that of \citet{2000AJ....120.3139C}, who found that in the Perseus, Orion and Mon R2 clouds that the large clusters contain more stars than the more numerous groups or small clusters.

This result is inconsistent with studies of demographics integrated over large regions of our galaxy and of clusters in other galaxies, in which the  number of stars per logarithmic interval of cluster  membership is found to be constant  \citep{2003ARA&A..41...57L,2007AJ....133.1067W,2010ApJ...719..966C}.  However, such studies include clusters formed from many different molecular clouds which presumably span a range of cloud masses.  We speculate that if the size of the largest cluster increases with the total cloud mass, the flat distribution of cluster sizes in galaxies may result from the distribution of cloud masses that produced the clusters. Future analyses of Spitzer molecular cloud surveys that adopt the same methodology applied to Orion are needed to determine whether the Orion cloud is unusual, or whether the  largest cluster(s) typically dominate the demographics of molecular clouds.  


\subsection{Comparison of the Orion A and B Clouds to Other Nearby Clouds}
\label{sec:comp_other_clouds}

How do the demographics of the Orion~A and B clouds compare to other nearby clouds?  We contrast the properties of the nearby clouds discussed in Sec.~\ref{sec:demo_cloud} using a simplified analysis of clustering based on two diagnostics: the fraction of YSOs with $N_{10} \ge10$~pc$^{-2}$ and the median value of $N_{10}$ (Figure~\ref{fig:nearby_summary}). We plot these two values against the total number of YSOs in the clouds.  For each cloud, we include the number of {\it Spitzer} identified YSOs, and have not performed a correction for incompleteness. The sample includes five clouds with clusters (the Orion A and B, Perseus, Serpens and Ophiuchus clouds) and three nearby dark clouds that do not have clusters (the Chameleon, Lupus and Taurus clouds).  The  $N_{10} \ge10$~pc$^{-2}$ density appears to bifurcate the sample, with the  clouds with clusters having fractions above 0.6 and the three nearby dark clouds having fractions below 0.4.  Furthermore, the clouds with clusters have median densities  above 25~pc$^{-2}$ while the nearby dark clouds have median densities below 10~pc$^{-2}$.

The fraction of YSOs in clusters and the median density do not show a clear dependence on cloud mass.  For example, Ophiuchus and Taurus have similar cloud masses \citep{2008A&A...489..143L,2010A&A...512A..67L} and a similar number of YSOs \citep{2008ApJ...672.1013P,2010ApJS..186..259R}.   Despite these similarities, most of the YSOs in Ophiuchus are clustered in the central Lynds~1688 core while the YSOs in the Taurus cloud are distributed throughout extended filaments \citep{2008hsf2.book..351W,2010ApJS..186..111L}. This comparison demonstrates that the relative number of YSOs in clusters is not simply a function of the size  and mass of a cloud and its embedded population, but is also a function of the structure of the gas.  Furthermore, for the clouds with clusters, the fraction of YSOs in clusters and the median YSO density do not seem to depend strongly on the number of YSOs or the mass of the cloud.   

One apparent trend is that we find no YSO rich ($> 500$ YSO), massive clouds ($10^5$~M$_{\odot}$) that do not have clusters.  An open question is whether there exist clouds with gas masses and YSO numbers similar to the Orion clouds that do not contain a significant number of YSOs in clusters.  One possible example is G216-2.5 or Maddalena's cloud, a giant molecular cloud ($> 10^5$~M$_{\odot}$) which harbors only a low density, Taurus-like star forming region \citep{1994ApJ...432..167L,1996ApJ...472..275L,2009AJ....137.4072M}.  However,  G216-2.5 may contain only  $\sim 100$ YSOs and it is characterized by low gas column densities; in this respect it appears to be much more similar to Taurus than Orion despite its large mass of molecular gas \citep{2009AJ....137.4072M, 2015ApJ...803...38I}.

\subsection{The Demographics of Protostars}
\label{sec:demoproto}

The demographics presented above are for the total sample of all dusty YSO protostars identified by {\it Spitzer}.  The question arises, do the protostars follow the same demographics? In  Figure~\ref{fig:cluster_proto}, we show the number of protostars logarithmically binned by the number of members in their parent assemblage. We also display the ratio of the number of protostar to number of disk sources in each logarithmic bin.  For the reasons discussed in  Sec.~\ref{sec:comparing}, we do not correct the number of objects  for incompleteness.  To address the issue of incompleteness, we adopt the approach of Sec.~\ref{sec:comparing} and perform the analysis for the entire Orion sample and again for the L1641/$\kappa$~Ori region alone.
   
Of particular interest is the ratio of protostars to disk sources. Since the ratio of protostars to disks decreases monotonically with time for a steady star formation rate, this ratio is a proxy for the the age of a star forming region. For the entire Orion sample, Figure~\ref{fig:cluster_proto} shows the protostars/disks ratio decreasing as we ascend in the number of members and go from groups, to clusters and then to large clusters.  In contrast, the fraction is relatively  constant between groups and clusters in the L1641/$\kappa$~Ori sample. This suggests that the decrease observed in the entire Orion sample may be the result of incompleteness, with the protostars affected more by incompleteness than disk sources.  Protostars may be more incomplete since their identification depends on having either 24~$\mu$m data (which is strongly affected by nebulosity and is saturated toward the ONC  and NGC~2024 clusters), a 5.8~$\mu$m detection (which is also  strongly affected by nebulosity), or a $H$ and $K$ detection (which are strongly affected by extinction for protostars). We conclude that the fraction appears to be constant and that groups and clusters have similar ages for the L1641/$\kappa$~Ori sample, and potentially for the entire cloud sample.  

In contrast, we see a significant drop in the protostars/disks fraction in the distributed population relative to groups and clusters in both the full cloud and L1641/$\kappa$~Ori samples.  The protostar/disk ratio of the distributed population increases if we limit  the analysis of L1641/$\kappa$~Ori to  regions where $A_V > 4$ using the extinction map of \citet[][ also see Figure 1 in Paper I]{2011ApJ...739...84G}, and thereby concentrate on regions with high  gas column densities,  but it remains lower than the protostar/disk ratio for clusters and groups. The low protostar/disk ratio for the more isolated stars is consistent with the results of Sec.~\ref{sec:comparing}, where we found that protostars tend be found in denser regions than pre-main sequence stars with disks.    

 If we adopt 0.5~Myr as the typical duration of the protostellar phase \citep{2014prpl.conf..195D}, then the variation in the protostar/disk ratios can result from systematically different ages between the stars in groups and clusters and the stars found in the distributed population.  Assuming a  constant star formation rate  for $t \le {\rm age}$, the clusters and groups  have an age of $\sim~2$~Myr, while the distributed population has an age of $\sim 3$~Myr for an $A_V > 4$.  (The age is given by  $(1+n_d/n_p) \times 0.5$~Myr where $n_p/n_d$ is the protostar/disk fraction.) This suggests that either the distributed population started to form before the formation of groups or clusters, or that distributed population contains stars that formed in clusters and groups which have since dispersed. 

The broad range of ages of the distributed population may also result from a mixture of reasons, as  this populations appears to have multiple origin environments.  A total of 26 protostars, 14 of which are toward regions of the molecular cloud where $A_V \ge 3$, are found in the distributed population. This shows that some of distributed population formed in isolation and this process may contribute many of the youngest stars. In addition, the halo of distributed YSOs surrounding the ONC appears to have resulted from the migration of pre-main sequence stars from the filaments in which they formed \citep[see Figure~16 in Paper I][]{2013ApJ...768...99P}.  Thus, stars that have migrated from existing clusters and groups may provide some of the intermediate age stars.  Finally,   groups and clusters  that have  already dispersed may provide some of the older stars.

\section{The Structure of the Orion Embedded Clusters and Groups}
\label{sec:prop}

We now turn to the structural properties of the clusters and groups found in the Orion molecular clouds.  For simplicity, we define the clusters and groups by adopting the methodology described in Sec.~\ref{sec:demo} with the 10~pc$^{-2}$ surface density threshold. As discussed in that section, this threshold density distinguishes between the crowded clusters of the Orion clouds from the more dispersed population of YSOs found in nearby dark clouds such as Taurus.  By adopting this single threshold, we can compare the properties of the clusters and groups in a uniform manner.  

\subsection{The Global Properties of the Orion Embedded Clusters and Groups}
\label{sec:global_prop}

To compare the global structural properties of the Orion groups and clusters, we use the properties established in \citet{2009ApJS..184...18G}. The global properties of the extracted groups and clusters are given in Table~2. They are plotted as a function of the number of cluster members in Figure~\ref{fig:cluster_radius_size}, where we show the properties derived from both the corrected and uncorrected samples to illustrate their sensitivity  to the corrections for incompleteness.  For the cases of the ONC and NGC~2024, we also show the cluster statistics corrected by the {\it Chandra} data, but without the weighting correction. Finally, Figure~\ref{fig:cluster_prop} displays histograms for the fully corrected cluster properties.

The radii, $R_{hull}$, are given by $\sqrt{A/\pi}$, where $A$ is the area of the convex hull surrounding a group or cluster.  The values of $R_{hull}$ are $< 1$ pc for the groups and between 1 and 4~pc for the clusters (Figure~\ref{fig:cluster_prop}).  This range of values is similar to the $R_{hull}$ values found in the embedded cluster survey of \citep{2009ApJS..184...18G}, although the ONC has the largest $R_{hull}$ of the combined sample of clusters from that paper and Orion.


To quantify the deviations from circular symmetry, \citet{2005ApJ...632..397G} defined the Azimuthal Asymmetry Parameter, or AAP. This parameter uses the number of members in 16 equal-area, Nyquist-sampled wedges filling a circle centered on the cluster. The parameter  measures the deviations in the number of members relative to the deviations expected for Poisson statistics in a uniform population.  A value of AAP $> 1.5$ implies an asymmetry at the 3$\sigma$ level.  In addition, we measure the aspect ratio of the convex hull surrounding each clusters, as defined by the ratio of the circular area over the convex hull area. \citep[The circular area is that of the smallest circle that can encompass the group or cluster,][]{2005ApJ...632..397G}. Since this parameter depends on the boundaries of the clusters and not the total number of stars, it is not strongly dependent on the completeness correction. 

The clusters and groups show significant departures from azimuthal symmetry.  The AAP for all the groups and clusters over 70 members exceeds 1.5, except for one cluster in L1641 (number 12 in Table~2)  where the AAP is reduced from 1.66 to 1.37 by the weighting correction.  The plot shows a clear correlation of  the AAP with the number of cluster members; however, this trend may result from dependence of the AAP on the numbers of members.  Specifically,  higher values of the AAP result from the greater numbers of sources in the wedges dividing up larger clusters since the uncertainty in the number of stars in each wedge is determined by Poisson statistics. Further evidence for asymmetry is found in the aspect ratios, which exceed 1.25 for all the clusters and groups larger than 70 members.  In summary, we find that all the clusters and large groups with 70 or more members show clear evidence for azimuthal asymmetry.  For the smaller groups, low number statistics make it impossible to draw a conclusion.

The average YSO density is the number of members divided by the area of the convex hull surrounding the members.  The values are between 10-60~pc$^{-2}$; this value is relatively independent of the number of members although the lowest densities are found primarily in small groups. This implies that most of the clusters and groups have average densities that are not much higher than the threshold density used to identify clusters. This is not surprising.  Since the clusters are density peaks in an extended distribution of young stars,  the average density is determined in part by the choice of threshold density.  NGC~2024 has the highest average density in a cluster: 49~pc$^{-2}$ for the uncorrected, {\it Spitzer} only, sample and  62~pc$^{-2}$ when augmented by the {\it Chandra} X-ray observations and corrected by the weighting.  The high average density and the steep surface density gradient surrounding this cluster are the reasons why the number of stars in the NGC~2024 does not depend strongly on the density threshold used to isolate the cluster (Figure~\ref{fig:demo_trend}).  The ONC also has a comparatively high density, ranging from 29~pc$^{-2}$ in the {\it Spitzer} only sample to 50~pc$^{-2}$ in the fully corrected sample.  In this case, the average density is being raised by the high density in the center of the cluster. The outer boundary of the ONC cluster is less distinct than those of NGC 2024; hence, changing the threshold density of the ONC can make large changes in the number of members (Figure~\ref{fig:demo_trend}).

We define the peak YSO density as the maximum  $N_{10}$  value found in a cluster \citep[in comparison,][uses $N_{5}$ for their peak densities]{2009ApJS..184...18G}; unlike the average YSO density, the peak density is not affected by the chosen threshold.   We find a strong correlation between the peak density and the number of members.  The peak densities and number of members both increase by two orders of magnitude between the smallest groups and the largest clusters. There is an approximately linear dependence of the peak density on the number of members which is well fit by $N_{10}(peak) \propto n_{YSO}^{1.2 \pm 0.1}$, where $n_{YSO}$ is the number of members.

We include in our number vs. peak density plot the spectral types of the most massive known stars in the four clusters \citep[these are taken from][]{1968AJ.....73..233R,1994A&A...289..101B,2003A&A...404..249B,2008hsf1.book..621A}.  This illustrates that the mass of these stars increase with both the size and peak density of the clusters.  Although this trend by itself does not demonstrate a shift in the IMF;  \citet{2012ApJ...752...59H} found evidence that the L1641 cloud, in which the largest cluster has $\sim 100$ members (L1641 contains the group and clusters numbers 9-20 in Table~2), is deficient in O and early B stars relative to the ONC.   Consequently,  it is unlikely that the mass functions of the ONC and L1641 are drawn from the same parent IMF \citep{2013ApJ...764..114H}.  This hints at a possible connection between peak  stellar densities and the masses of the most massive stars in clusters \citep[also see][]{2012AJ....144...31K, 2013MNRAS.434...84W}.


\subsection{The Internal Structure of the Three Largest Clusters}
\label{sec:anatomy}

The three largest clusters in the Orion clouds are the ONC (number~7 in Table~2), the NGC~2024 cluster (number~4), and the cluster that encompasses the NGC~2068 and NGC~2071 nebulae (number~3). In this section, we examine the internal structure of these three clusters. We perform this analysis in two ways. First, we generate $N_{10}$ maps of the clusters.  Second, we select all the YSOs that have nearest neighbor densities above a varying threshold density and construct the convex hulls for those YSOs.  We thereby create a nested series of convex hulls that encompass the YSOs found at increasing levels of surface density.  We then determine the number of members, $R_{hull}$, the average density and the AAP from the YSOs above a given density threshold. Using this analysis, we can examine how the properties of a cluster vary with $R_{hull}$. In comparison to analyses of cluster properties along azimuthally or elliptically averaged radial bins \citep[e.g.][]{2009AJ....138...33D}, this approach is better adapted to  irregularly shaped clusters because it does not rely on choosing a cluster center nor does it impose an azimuthal symmetry upon a cluster.  
 
We examine the ONC for both the weighting corrected, {\it Chandra} augmented {\it Spitzer} sample and for the X-ray sample from the COUP survey alone. The ONC is distinguished by its highly elongated morphology and the high stellar densities toward its center. In Figures~\ref{fig:onc_map} and \ref{fig:onc_prop}, we display the internal structure using a variety of diagnostics. The cluster elongation is clearly evident in the surface density map and in the nested convex hulls in Figure~\ref{fig:onc_map}. This elongation was also found by \citet{1998ApJ...492..540H}.  Furthermore,  the peak density is offset from the center of the outer convex hull, indicating that there is also a north--south asymmetry in the cluster.  Fig~\ref{fig:onc_prop} shows that the cluster is strongly centrally condensed, with the density increasing from 70~pc$^{-2}$ to 10,000~pc$^{-2}$,  more than two order of magnitude, from the outermost to innermost $R_{hull}$. This figure also  shows the AAP increases rapidly with cluster size. We find that the $AAP \ge 2$ for all $R_{hull}$ greater than 0.1~pc. The asymmetry only disappears in the very inner region of the cluster; at $R_{hull} \le 0.1$~pc the AAP decreases to 1.5 for both the combined and COUP survey samples. 

The NGC~2024 cluster is also shown for the weighting corrected,  {\it Chandra} augmented {\it Spitzer} sample and for the  {\it Chandra} X-ray sample alone (Figures~\ref{fig:n2024_map} and \ref{fig:n2024_prop}). The cluster shows a strongly peaked surface density profile, with the mean density within the convex hulls growing from 78 to 2664~pc$^{-2}$.  Similar to the ONC, the surface density map and convex hulls show a significant elongation, with one end of the cluster narrower than the other. The AAP values exceed 1.5 for $R_{hull} > 0.3$~pc  for the combined {\it Spitzer} and X-ray sample, but do not exceed 1.5 for the X-ray sample alone.  This may be due to the lower  number of sources and weaker statistics in the X-ray sample.

The cluster associated with the NGC~2068 and 2071 nebulae exhibits a double density peak and would be classified as a hierarchical cluster by \citet {2003ARA&A..41...57L}.   This cluster was also studied with deeper near-IR data by \citet{2015A&A...581A.140S}; we identify more members since they require detections in all IRAC bands and the MIPS 24~$\mu$m bands to identify dusty YSOs.  In Figure~\ref{fig:n2068_map}, we display $N_{10}$ maps and convex hulls for the two density peaks separately, and in Figure~\ref{fig:n2068_prop}, we give the properties for each density peak.  For this cluster, there is no X-ray data.  Again, significant density peaks are found, but the range in average density goes only from 33 to 236~pc$^{-2}$ for the NGC~2068 sub-cluster and 51 to 232~pc$^{-2}$ for the NGC~2071 sub-cluster.  The two cluster peaks are asymmetric: they are elongated and are not centered on the massive B members that heat the NGC~2068 and NGC~2071 nebula. The AAP values typically exceed 1.5  for $R_{hull} \ge 0.25$~pc in both sub-clusters.  

\subsection{Constraints on the Ages of the Clusters from their Structure}
\label{sec:const_ages}

The lack of azimuthal symmetry suggests that the clusters in Orion have not undergone relaxation, except potentially in their inner regions.  We can therefore put a limit on the age of a cluster by using the radius at which the cluster is no longer azimuthally symmetric.  In the ONC, that radius is 0.2~pc.  To calculate the relaxation time, we use the standard equation \citep{orion_lite_cluster_v11}

\begin{equation}
t_{relax} = \frac{N}{6 log(N/2)} t_{cross},
\end{equation}

\noindent
where $N$ is the number of stars and $t_{cross}$ is the crossing time.  To determine the crossing time, we assume
that  the cluster within $R_{hull}$ is virialized.  We derive the potential using a constant density sphere with radius $R_{hull}$ and total mass $M(r \le R_{hull})$:

\begin{equation}
t_{cross} = 2 R_{hull} \sqrt{\frac{10 R_{hull}}{3 G  M(r \le R_{hull})}}.
\end{equation}

\noindent
Initially, we set the mass within $R_{hull}$ to $ N \times 0.5$~M$_{\odot}$.  We note that we are assuming most of the mass is in the stars.  This is a reasonable assumption since most of the Orion stars in the center of the cluster are optically visible and have a low extinction \citep{2000ApJ...540..236H}.  In this case, the relaxation time within 0.2~pc is 1 Myr; and the resulting maximum age for the ONC is 1 Myr. We note that if we assume instead that the  stars are only 20\% of the mass, then the relaxation time and maximum age increase to to  2.2~Myr.  Thus, the asymmetry of the interior cluster indicates an age of the cluster of $\le 2.2$~Myr.  This is somewhat shorter than the mean isochronal age for the ONC of 2.5~Myr \citep{2010ApJ...722.1092D,2011MNRAS.418.1948J}.  It is not clear whether these can be reconciled.  The uncertainties in the pre-main sequence tracks and in the stellar birth line make isochronal ages uncertain \citep[e.g.][]{1997ApJ...475..770H,2012ApJ...756..118B}.  On the other hand, the assumption of a virialized cluster may underestimate the velocities after gas dispersal and thereby underestimate the relaxation time.  Finally, this analysis assumes that the cluster is coeval; however, the stars in the inner core of the ONC may be systematically younger \citep{2014ApJ...787..109G} or older \citep{2013A&A...549A.132P} than the outer region of this cluster.

We can also perform a similar analysis for the other three clusters.  For NGC~2024, we find a maximum age of 3.7~Myr, for NGC 2068, a maximumage of 2.3~Myr and NGC~2071, a maximum age of 2.1 Myr.  These are consistent with the isochronal ages found for these regions \citep{2006ApJ...646.1215L,2008AJ....135..966F}.  



\subsection{The Structure and Dynamical State of the ONC}
\label{sec:struct_dyn_onc}

The ONC is one the three largest young ($< 5$~Myr) clusters within 1~kpc of the Sun \citep[the others being the Cep~OB3b cluster and the NGC~2264 cluster, see][]{2003ARA&A..41...57L,2003AJ....126.1916P,2012ApJ...750..125A} and thus of particular interest in understanding the evolution of embedded clusters and their potential to form bound clusters \citep{2001MNRAS.321..699K}.  Prior to this paper, previous authors have demonstrated the elongated nature of this cluster. \citet{1998ApJ...492..540H} used data from a combination of visible light and near-IR imaging to show that the ONC cluster has an elongated structure aligned with the molecular filament from which it is forming.  By fitting ellipses to isodensity contours of the surface density of stars, they found that ellipses were centered near the $\theta^1$ Ori C and aligned with the cloud.  The eccentricities of the ellipses ranged from 0.29 for the innermost ellipse to 0.54 for the outermost ellipse.  (The aspect ratio we find for the cluster corresponds to a eccentricity of 0.8, but our value is determined over a much larger region than that considered by \citet{1998ApJ...492..540H}).  Furthermore, they found the outermost contour shows the highest eccentricity, in agreement with our analysis of the AAP.  They also find a peak surface density similar to ours of $\sim 10,000$ stars pc$^{-2}$.  Recently, \citet{2014ApJ...787..107K} fit isothermal ellipsoids to the distribution of stars in the ONC. They also found that that the structure is fit by elongated ellipsoids aligned with the molecular filament.  Three separate ellipsoids had to be fit to the central cluster to reproduce a core-halo morphology.  The ellipsoid representing the inner core has a smaller ellipticity than the ellipsoid representing the outer halo,  in agreement with our results and those of \citet{1998ApJ...492..540H}. 


 \citet{2008ApJ...676.1109F} and \citet{2009ApJ...697.1103T} measured the radial velocities of  1613 young stars toward the ONC. The most prominent dynamical feature is a  gradient in velocity with declination observed in the OMC~2/3 region north of the Orion nebula.  This gradient is also apparent in the gas and can be seen in the CS ($2 \rightarrow 1$) map shown in \citet{2008hsf1.book..590P}.  \citet{2009ApJ...697.1103T} interpreted the gradient as infall onto the massive cores in the Orion Nebula region.  The interpretation of the gradient as infall is not unique; the velocity gradient can be explained as either expansion, contraction, or  rotation.  However, the magnitude of the gradient is consistent with infall.  N-body simulations by \citet{2009ApJ...697.1020P}  showed that the gradient can be explained by the initial collapse of a sub-virial and elongated cluster embedded in a molecular clump. The magnitude  of the velocity shift in \citet{2008hsf1.book..590P} is 1.5~km~s$^{-1}$ over a region extending from 0.36~pc to 1.44~pc from the central, massive clump.  Adopting a mass for the central clump of 627~M$_{\odot}$ \citep{2010ApJ...710.1247S}, gravitational acceleration would result in a velocity shift of 3.4~km~s$^{-1}$.  Considering the unknown inclination of the filament,  the magnitude of the observed velocity gradient is consistent with gravitational infall.


Another conspicuous feature in the gas velocity structure of the ONC are the  multiple velocity components apparent toward the southern half the cluster \citep{1987ApJ...312L..45B}. At the same location, the Spitzer image of the Orion Nebula  shows a large bubble extending  to the southwest of the nebula (Figure~14 in Paper I);  this region, known as the Extended Orion Nebula, is filled with hot X-ray emitting  gas \citep{2008Sci...319..309G}.  The multiple velocity components appear to be due to the interaction of the molecular gas with the OB stars in the Orion Nebula.  Figure~14 in Paper~I also shows that the  bubble is filled with YSOs which appear to trace the inner region of the bubble.  An interpretation of this morphology is that the  stars formed in the walls of the bubble as it was blown out by the winds and UV radiation of the OB stars. Since the gas is continually accelerated by the massive star winds and  by the photoevaporation of gas from the surfaces of the bubble exposed to the UV radiation, the  bubble walls will eventually expand faster than the stars that formed within them. In this manner, the dusty YSOs  will be left behind to fill the interior of the bubble as it expands. \citet{2008ApJ...676.1109F} found evidence that the stars toward the bubble are slightly blue shifted relative to the main molecular cloud filament,  consistent with this interpretation (see their Figure~9).

There is also evidence for non-coevality in the ONC and its surroundings.  North of the ONC is a  small cluster  of pre-main sequence stars toward NGC~1981; this cluster shows a much higher ratio of Class III to Class II objects than the ONC, suggesting that the first star formation in the vicinity of the ONC was at the northern tip of the cluster \citep{2013ApJ...768...99P}. There is also a halo of stars that extends laterally beyond the filament (Fig~\ref{fig:onc_map}, Fig~9 in Paper I).  An older age for this halo is suggested by XMM and {\it Chandra} observations that show a lower disk fraction and a systematically lower J-band luminosity for their X-ray luminosity \citep{2013ApJ...768...99P,2014ApJ...787..109G}.  These may be stars that have either migrated from the embedded cores of the cluster or stars that formed in the outer regions of the filament as it contracted \citep{2013ApJ...768...99P,2014ApJ...787..109G}.

The highly elongated and irregular structure of the ONC and its parental molecular cloud indicate a dynamic, evolving, non-relaxed structure  (Figures~\ref{fig:orion_nn}, \ref{fig:onc_map} and Figure~14 in Paper~I). Although the center of the ONC shows evidence of dynamical relaxation (Sec.~\ref{sec:anatomy}, Figure~\ref{fig:onc_prop}), this only occurs in the inner 0.1 pc of the cluster.  On larger radii, the cluster shows evidence of the infall of an actively star forming filament onto the massive center of the cluster and a star-forming bubble driven outward by the OB stars. It also shows a bubble on the northernmost tip of the cluster driven by a B1V star \citep[the NGC~1977 nebula, ][]{2008hsf1.book..590P}.

The non-equilibrium and evolving nature of the ONC is also suggested through the comparison of the various timescales of the cluster.  The relaxation time for the inner, circularly symmetric region is about 1 Myr, which we suggest might be the typical age of the cluster (Sec.~\ref{sec:anatomy}).  This relaxation time increases with increasing radius, and can be tens of millions of years for the outer region of the cluster. In comparison, if we calculate a crossing time of the longest dimension of the cluster using the velocity dispersion of the gas, $t_{cross} = 2 R_{hull}/\sqrt{3}\sigma$ where $\sigma$ is the 1-D average velocity dispersion in the $^{13}$CO emission, the value of the crossing time is $t_{cross} = 1.8$~Myr.  We can also estimate a free fall time for the cluster by computing the average density in a sphere of radius $R_{hull}$, 

\begin{equation}
\rho = \frac{(0.5 M_{\odot} N_{star}/0.75 + 5000 M_{\odot})}{\frac{4}{3} \pi R_{hull}^3}, t_{ff} = \sqrt{\frac{3 \pi}{32 G \rho}};
\end{equation}

\noindent 
where $5000$~M$_{\odot}$ is the mass of the integral shaped filament in which the ONC is embedded \citep{1987ApJ...312L..45B}, 0.5~M$_{\odot}$ is the mean stellar mass of the YSOs, 0.75 is the fraction of young stars with IR-excesses (see Sec.~\ref{sec:irex}), and  $N_{star}$ is the number of dusty YSOs.  Adopting $N_{\star} = 5000$, the resulting free fall time is $t_{ff} = 2.2$~Myr.  The  gas dispersal timescale is difficult to ascertain, but is expected to be on the order of a few Myr \citep{2007prpl.conf..361A}.  

Thus, we find that the crossing time, free fall time, and gas dispersal time are comparable, and they are similar to the time needed for the observed circularization of the inner cluster.  
In total, this suggests that star formation in the ONC is not occurring within virialized molecular clumps and equilibrium structures, but within rapidly evolving gas structures where the timescales for star formation, cloud destruction through feedback, collapse, and crossing are comparable yet are long enough to allow for the relaxation of the dense central core of the cluster.   Because of this likely non-equilibrium nature, it is difficult to predict whether the ONC will produce a bound cluster.

\section{The SFE of the Orion Clouds and Clusters}
\label{sec:sfe}

The star formation efficiency, $SFE$, is defined as $M_{\star}/(M_{\star}+M_{\star})$ where $M_{\star}$ is the mass in stars that have formed from the gas and $M_{gas}$ is the mass of the remaining gas.  In the standard picture that has emerged from surveys of molecular clouds, the SFE of entire molecular cloud complexes is low while the SFE in clusters is comparatively high \citep[e.g.][]{1992ApJ...393L..25L}.  To test this picture, we perform a simple analysis of the SFE of the entire clouds and the clusters. Masses of the Orion~A and B clouds have been determined from the CO survey of \citet{2005A&A...430..523W} and the wide field NIR extinction map of \citet{2011A&A...535A..16L}.  These maps are sensitive to the low density molecular gas that comprises a significant fraction of the total mass of the cloud. Scaling the masses to a common distance of 414~pc for comparison, we find that the Orion A mass is 86000 and 83000~M$_{\odot}$ from the CO and extinction maps, respectively.  For the Orion B cloud, we find masses 67000 and 74000~M$_{\odot}$ from the CO and extinction data, respectively.  Given the similarity of these values, we use the average of the two.  To calculate the mass of young stars, we adopt an average mass of 0.5~M$_{\odot}$ and an IR-excess fraction of 0.75. We use both the number of detected dusty YSOs augmented by the X-ray surveys in NGC 2024 and ONC and the number of dusty YSO estimated using the weighting correction plus {\it Chandra} augmentation. 

The resulting SFEs are  2 to 3\% and 0.6-0.8\% for the Orion A and B clouds, respectively.  The higher numbers are those that include the weighting correction.   \citet{2009ApJS..181..321E} uses the extinction at $A_K \ge 0.2$ to determine the SFEs of clouds in the c2d survey.  To compare the SFEs of the Orion clouds  with those of the c2d clouds, we use the masses for $A_K \ge 0.2$  from \citet{2011A&A...535A..16L}; after correcting to a distance of 414 pc these are 56000 and 39000~M$_{\odot}$ for Orion A and Orion B, respectively. The resulting SFEs are 3-5\% for Orion A and 1.1-1.5\% for Orion~B.   In comparison,  \citet{2009ApJS..181..321E} find SFEs in the c2d clouds of 3-6\%. These numbers show that the Orion~A cloud has a similar SFE to that of the c2d molecular clouds while the SFE for the Orion~B cloud is unusually low.

 To measure the SFE toward individual groups and clusters, we used the $A_V$ map of \citet{2011ApJ...739...84G} and the $^{13}$CO ($1 \rightarrow 0$) maps from \citet{2013MNRAS.431.1296R}.  For each cluster, we identify all the pixels of those maps within $100''$ of one of the cluster members. We then add up the column densities in those pixels and derive a M$_{gas}$ for an adopted distance of 414~pc.   Once again, we calculate the efficiencies using both the X-ray augmented sample  and the weighting corrected sample of YSOs.  Although there is a significant amount of scatter in the resulting SFEs, as shown in Figure~\ref{fig:sfe}, the groups and clusters have SFEs that are approximately 3-30 times higher than the cloud efficiencies, consistent with the standard picture.  There is a weak trend of the SFE increasing with the number of cluster members, particularly for those with more than 100 members.   This trend may result from the approximately linear increase in the SFE with gas column density observed in molecular clouds \citep{2011ApJ...739...84G,2013ApJ...778..133L,2014A&A...570A..15L}.  This trend must be confirmed with future maps of the column density of the Orion clouds that more accurately determine the densities in regions of high gas column density and high stellar surface density \citep[e.g. ][]{2014A&A...566A..45L}.
 
\section{The Relationship between the Orion Embedded Clusters and the Orion OB1 Association}
\label{sec:ob1}

The Orion molecular clouds are located in the Orion OB1 association, which contains nine O stars, five B0 stars, and 67 B1-B3 stars \citep{1994A&A...289..101B}.   In Figure~\ref{fig:ob}, we show the spatial relationship of the clouds to the OB association; the positions of the OB stars are from \citet{1994A&A...289..101B}.   We find that the overall spatial extent of the young stars within the molecular clouds and the OB stars are similar: i.e. both extend $\sim  100$~pc along their longest dimension.   As was known previously,  most of the OB stars in the association are not coincident with the molecular clouds and were created in previous episodes of star formation within the Orion region.  The OB stars in the association have been divided into four distinct subgroups \citep{1964ARA&A...2..213B,1978ApJS...36..497W}.   Only the youngest subgroup of the association,  the OB1d subgroup,  contains massive stars still associated with their natal clouds including the massive stars in the ONC \citep{1978ApJS...36..497W,2008hsf1.book..590P}.   Although they are not shown in Fig~\ref{fig:ob}, a large population of low mass stars coexists with the OB subgroups outside the cloud \citep{2008hsf1.book..838B}.  The ages of the low mass stars found in the subgroups range from 2-3 for the OB1c subgroup, 4-6~Myr for the subgroup OB1b and 7-10~Myr for the subgroup OB1a \citep{2005AJ....129..907B, 2008AJ....135.1616S}.  

There are multiple lines of evidence that the OB association is interacting with the Orion A and B clouds.  The bright PAH emission observed toward the surface of the Orion B cloud near NGC2024 and NGC2023 is clear evidence that the cloud is illuminated by the OB1 association (Figure~13 in Paper~I).  This surface hosts the Horsehead nebula, a bright rimmed cloud that is the most dramatic example of the interaction between the OB association and the molecular cloud  \citep{2006sf2a.conf..247P,2006MNRAS.369.1201W,2009AJ....137.3685B}. The two richest clusters, the ONC and the NGC~2024 Cluster, are in the regions of the molecular clouds closest to this concentration of older OB stars; suggesting that the star formation in these region may have been enhanced by the compression of the gas by the OB stars \citep[e.g.][]{2008hsf1.book..459B}.   

The diameter of the association is similar to the length of the molecular clouds, suggesting that the overall size of the association is set by the distribution of star formation sites and not the expansion of the association members. It is not clear whether the Orion cloud complex is bound since the gravitational and kinetic energies appear comparable to within a factor of a few \citep{2005A&A...430..523W}.  An uncertainty in the total  mass is the number of low mass stars in the association.  Although we have a good census of young low mass stars associated with the Orion A and B clouds, in the Ori OB1 association subgroups outside the molecular clouds, we only have a complete census of the massive stars.  If we use the total number of intermediate to massive stars in the OB1a, b and c subgroups from \citet{1994A&A...289..101B} and scale by the IMF \citep{2001MNRAS.322..231K,2003PASP..115..763C}, we estimate there are 6300 to 9100 stars in these subgroups.  If we add that to the 5000 dusty YSOs and adopt a 75\% disk fraction, then the number of young stars is between 13000 and 16000.  With an average stellar mass of around 0.5~M$_{\odot}$, this is far less than the molecular gas mass in the molecular clouds of $\sim 200,000$~M$_{\odot}$ \citep{2005A&A...430..523W,2011A&A...535A..16L}.  Furthermore, if we assume that most of the association is filled with a low density atomic or ionized gas with a density of 1 cm$^{-2}$, that would increase the mass by  only 10,000~M$_{\odot}$.  Thus, the molecular clouds dominate the mass of the region. 

If we then assume a diameter of 100~pc, and approximately viral velocities, the crossing time of the association is $\sim 30$~Myr.  Since OB associations appear to disperse their molecular gas and cease star formation after $\sim 10$~Myr, as inferred from the lack of residual gas in the 5-20 Myr associations and clusters \citep{1989ApJS...70..731L,2002AJ....124..404P,2012ApJ...746..154P}, the gas dispersal time appears to be much less than the crossing time.  \citet{2007MNRAS.380.1589B} find that regions with low star formation efficiencies and rapid gas dispersal times much faster than a crossing time will not form bound clusters; hence we expect that even if the Orion OB1 association is currently gravitationally bound, it will form an unbound system after gas dispersal.

In contrast, the inner core of the Orion Nebula Cluster has a star formation efficiency of 30\% and crossing times which are on the scale of 1-2 Myr, comparable to the age of the ONC and clearly less than the gas dispersal times of this partially embedded cluster \citep{2007MNRAS.381.1169J,2011MNRAS.418.1948J}.  Thus, localized regions within the OB association such as the ONC may undergo relaxation and survive gas dispersal as bound clusters.  \citet{2005MNRAS.359..809C} simulated star formation in a turbulent, gravitationally unbound molecular cloud where the total kinetic energy exceeded the gravitational potential energy. The simulations show the formation of embedded clusters within the unbound cloud which may remain bound after the molecular gas is dispersed. The clusters are not bound to each other and expand away from one another with time.  These simulations provide an attractive framework for understanding the Orion star forming complex, with its unbound association and bound embedded clusters. It should be noted it is unclear whether any of the Orion clusters will survive gas dispersal.  Although the OB stars currently found outside of the molecular clouds must have  formed in embedded clusters, none of these are  known to be surrounded by bound clusters which have survived gas dispersal in the OB association. 

Observations of Orion and other associations paint a picture where stars in the molecular clouds have ages $\le 2$~Myr \citep{2007MNRAS.381.1169J,2008AJ....135..966F},  while stars in the association have ages ranging up to 10~Myr \citep{2008hsf1.book..838B}. This requires some method for creating and destroying clouds on $\sim 2$~Myr timescales, while sustaining star formation within the OB association for a 10~Myr period.  We note that this 10~Myr period is similar to the timescale needed to create molecular clouds by colliding flows of atomic gas  \citep{2008ApJ...689..290H}; thus the 10~Myr period may reflect this timescale if cloud formation is not synchronized across a 100~pc region in a colliding flow.  Alternatively, the long timescale may result from compression and shuffling of molecular gas through feedback from massive stars \citep{1977ApJ...214..725E}. Understanding the processes that sustain star formation in OB associations over these long timescales is a key step towards understanding star formation on galactic scales.

\section{The Fraction of Stars with IR-excesses}
\label{sec:irex}

The {\it Spitzer} Orion survey identifies only young stars  and protostars with IR-excesses from  dust grains in disks and infalling envelopes.  The  IR-excesses from  these disks and envelopes evolve and disappear rapidly: protostellar envelopes  persist for $\sim 0.5$~Myr \citep{2007A&A...468.1009H,2009ApJS..181..321E}, while optically thick disks  around low mass stars typically persist for a few million years \citep{2008ApJ...686.1195H}.  It is thus of great interest to assess the fraction of young stars which can be identified by their IR-excesses. In the Orion molecular clouds, the fraction of  IR-excesses has been previously measured in the ONC and NGC 2024 clusters.  \citet{1998AJ....116.1816H} used the dereddened I-K color and found the fraction of pre-main sequence stars with disks in the ONC is between $61\%$ and 88\%.  \citet{2001ApJ...558L..51M} used a combination of $JHKL$-band photometry toward the central $7' \times 7'$ of the ONC and found an IR-excess fraction of $80\% \pm 7\%$.  In NGC~2024, a IR-excess fraction of $\ge 86\% \pm 8\%$ was determined from $JHKL$-band  observations \citep{2001ApJ...553L.153H}.

To reassess this value in the Orion A region with the {\it Spitzer} data, we require a means to determine the number of young, pre-main sequence stars without IR-excesses.  We do this using three independent  methods.

First, we use the near-IR variability to identify young stars; diskless pre-main sequence stars may be distinguished by variability due to rotating star spots. This methods assumes that the same fraction of young stars with and without IR-excesses are variable. We use the near-IR variability survey of \citet{2001AJ....121.3160C}, which repeatedly observed a $0.^{\circ}84 \times 6^{\circ}$ strip centered on the Orion Nebula with the 2MASS telescope at Cerro Tololo, Chile over a 2 year period.  We apply the criteria adopted by \citet{2000AJ....120.3139C}  and take all sources with near-IR  Stetson indexes $\ge 0.55$ as variables and  YSOs.  In total, 284 of the variables can be placed on the IRAC 4-band,  JH[4.5] or HK[4.5] diagrams needed to identify IR-excesses.  A total of 222 variables,79\% of the sample, show infrared excesses in these diagrams.  To search for a dependence of the fraction of sources with IR-excesses with position, we plot the dependence of the disk fraction on right ascension and declination (Figure~\ref{fig:variex}).  For decl.$ < -4.75^{\circ}$, we find IR-excesses toward $82\% \pm   0.03$ of the variables with sufficient photometry to identify an IR-excess.   For decl.$ > -4.75^{\circ}$, there is a sharp drop off in the IR-excess fraction; this region is the NGC~1981 cluster located north of the Orion~A cloud and the NGC~1977 nebula \citep[see Figure~14 in Paper I, ][]{2013ApJ...768...99P}. There is  no drop in the fraction of all variables with IR-excesses toward the Orion Nebula and the central O7 star $\theta^1$~C, which is at a coordinate of $R.A. =  83.8186^o$, $decl. = -5.38968^o$.  This suggests that photoevaporation by the O-star has not had a significant impact on the frequency of hot inner disks traced by the IRAC-bands.

Second, we use the X-ray data from the COUP and NGC~2024 fields to establish the fraction of X-ray detected YSOs with IR-excesses.  Using the COUP data for the Orion Nebula region displayed in Figure~\ref{fig:coup_radial}, we compare the  ratio of the X-ray detected sources with IR-excesses to the X-ray sources with sufficient IR photometry to detect IR-excesses. We avoid the inner regions of the the clusters due to the incompleteness of the {\it Spitzer} data.  Between $0.0675^o$ and $0.1275^o$ from the cluster center, we find a IR-excess fraction of $0.54 \pm 0.02$.  We repeat the  same analysis using the NGC~2024 data (Figure~\ref{fig:n2024_xray_radial}). In this case, between $0.0675^o$ and  $0.1275^o$ of NGC~2024~IRS~2, we find a IR-excess fraction of $0.58 \pm 0.10$; this is consistent with the value derived for the Orion Nebula. 

Finally, we measure the disk fraction in the combined groups and clusters in the L1641 region using number counts to estimate the total number of young stars.  The L1641 region was chosen as it does not contain bright nebulosity  which creates spatially varying completeness and since it has not been measured in previous studies.  The contamination from background stars is minimized by considering only regions with YSO densities $> 10$~pc$^{-2}$ (Figure~\ref{fig:nn_map}).  These regions typically show high extinction which further reduces the number of background stars.  The total area subtended by these regions is $0.46^{\circ}$.  Toward these regions we find 380 IR-excess sources and 1229 sources which have detections in sufficient infrared bands to determine if they have  an excess, but appear to be pure photospheres without excess emission from disks or envelopes.

To subtract out the background contamination from the pure photospheres, the density of background stars is estimated using the $A_V$ map of the clouds and number counts  vs. $K_s$ magnitude take from nearby  reference fields.  We chose two different methods to  estimate the number of background  stars.  In the first method, we use the observed reference fields near the Orion A cloud, these are the ones  within $2^o$ of the ONC (Figs~1 and 9 in Paper~I). In these fields, we have IRAC data and we only consider sources with sufficient photometry to identify IR-excesses that are not likely extragalactic contamination.  The  second method uses two circular 1~sq.~deg. reference fields centered at  $l = 216^o$ and $l = 209^o$ both with a galactic latitude of $b = -19.3^o$.   For these fields, the source counts were extracted from the 2MASS point source catalog; these two fields have the same galactic latitude as the Orion~A cloud and straddle it in galactic longitude. In these two fields, we consider all sources detected by 2MASS.

In Figure~\ref{fig:disk_l1641}, we show the K-band magnitude histograms for the IR-excess sources, for the estimated background contamination, and for all sources with sufficient photometry subtracted by the expected background contamination.  We consider the disk fraction for $m_K \le 13$~mag.; at fainter magnitudes the number of background stars exceeds the number of members.  We find that the disk fraction for the first and second methods are  $0.72 \pm   0.07$ and $0.81 \pm 0.08$.  These are consistent within the uncertainties.  


We note that there may be biases in the methods which identify members on the basis of variability or X-ray emission. At the mid-IR wavelengths, pre-main sequence stars with disks and protostars show much higher incidence of variability than pre-main sequence stars without disks  \citep[Paper I, ][]{2009AJ....138.1116S,2009ApJ...704L..15M,2011ApJ...733...50M}.  This variability appears to arise in passively heated inner disks, and may result from fluctuations in the luminosity generated by the accretion of gas from the  disks onto the stars and from structures orbiting in the disks  \citep{2011ApJ...733...50M,2011ApJ...732...83F}. This variable disk  emission may also be apparent in the near-IR and thus could increase the incidence of variability toward stars with IR-excesses \citep{2000AJ....120.3139C}.

In contrast, the detection of X-ray emission toward pre-main sequence stars may be biased against stars with disks and protostars. This can result in a IR-excess fraction for the X-ray sample that is systematically lower than those found by other methods.  \citet{2010AJ....140..266W} found the  disk  fraction of X-ray detected sources in the young NGC~1333 cluster  is significantly lower than that found by \citet{2008ApJ...674..336G}  from number counts.  This  bias against detecting pre-main sequence accreting stars with disks and protostars may be due to lower X-ray luminosities and shorter flare durations \citep{2007A&A...468..425T,2008ApJ...688..437G,2010AJ....140..266W}.  Furthermore, they  found that none of the Class 0 protostars and only 23\% of the Class I protostars were detected, compared to 52\% for the Class II objects.  The bias against detecting protostars could further lower the fraction of X-ray sources with IR-excesses.

A more detailed analysis of the disk fraction is deferred to future studies of the Orion clouds which focus on obtaining unbiased samples of young stars with and without disks.  Instead, we focus on how these results may impact our analysis of the distributions of young stars as traced by the dusty YSOs.  First, we note that the fraction of IR-excess sources is approximately $\sim 75\%$; we are tracing 3/4 of the young stars.  Second, we find that toward the cloud, the disk fraction appear to be remarkably invariant; this is evident from the lack of spatial variations in the disk fraction of the variables, the similarity of the L1641 disk fraction with disk fractions in other regions in the clouds, and the consistency of the fraction  of X-ray selected sources with IR-excesses in the Orion nebula and NGC~2024 clusters.  This lack of apparent  variations in the IR-excess fractions supports our operating assumption that the dusty YSOs are tracing the spatial  distribution of all young stars within the cloud.

\section{The Environment of Planet Formation in the Orion Molecular Clouds}
\label{sec:environment}

The intense UV radiation and high densities of young stars found in rich young clusters can alter disks around low mass stars, potentially influencing the process of planet formation.  The clearest evidence for the alteration of disks is found in center of the ONC, where there is  clear evidence for the UV photo-ablation of disks by  the  O-stars $\theta^1$~C and $\theta^2$~A \citep{1987ApJ...321..516C,1994ApJ...436..194O,1998AJ....116..293B}.   Alternatively, close encounters of disks may also truncate disks, although such tidal interactions may not be common even in dense clusters \citep{2005ApJ...632..397G,2006ApJ...641..504A}.  Both the photo-ablation of disks by UV radiation and the tidal interaction of disks will affect primarily the outer regions of disks \citep{1998ApJ...499..758J,2004ApJ...611..360A,2006ApJ...641..504A}; these outer regions are not traced by the infrared excess detected in the {\it Spitzer} 3.6-24~$\mu$m bands. Since our data do not trace the outer regions of disks which are directly affected by interactions and UV radiation, we forgo an examination for direct evidence for the impact of environment.  Instead, we address the following question: what fraction of disks around low-mass stars may plausibly be affected by their environment?  We use the {\it Orion} Spitzer survey to determine the fraction of YSOs found in the extreme environments that can  alter disks.

The determination of the UV radiation at the surface of the disks requires a knowledge of the distance between the disk and hot OB stars and the attenuation of the UV radiation by the intervening gas.  Both the distance and attenuation cannot be directly measured by the current data.  However, we can assess whether the stars are close enough to be affected by UV radiation by using the projected distances between the identified YSOs and known OB stars.  The projected distance provides a lower limit to the actual distance and can be used to determine which YSOs are too distant to be affected.   We ignore the attenuation for two reasons. First, we can only measure the extinction along the line of sight to the YSO; the attenuation along the sightline between the YSO and  OB star could be different.  More importantly, the amount of extinction may drop as the OB stars clear the surrounding  gas or as the low mass stars orbit in a cluster.

We use the catalog of \citet{1994A&A...289..101B} (Figure~\ref{fig:ob}).  As discussed before, the low mass stars are found distributed throughout the regions showing molecular gas.  In comparison,the OB stars are much rarer, and many are found in the neighboring association.   We group the O stars and B0 stars together since stars with these spectral types produce intense extreme UV  radiation fields \citep[EUV, $h\nu >13.6$~eV]{1973AJ.....78..929P}.   Although the stars with strong EUV fields are the most destructive to disks, intense far UV radiation fields produced by the more numerous B0.5-B3 stars (FUV, $6$~eV~$< h\nu < 13.6$~eV) may also photo-ablate  disks \citep{2004ApJ...611..360A}.   We find that the OB stars are spread widely throughout the Orion constellation, with many near to or coincident with the clouds.   For each YSO in our sample, we have found the nearest O-B0 or B0.5-B3 star (as seen in projection), and determined the projected distance from the YSO to that star. The cumulative distributions of the number of YSOs vs. projected distance to the nearest O or B star are displayed in Figure~\ref{fig:dist_nearest_ob}.  In this analysis, we have included the COUP stars and have applied the completeness correction to the exhibited distributions; the correction for incompleteness is particularly important for this analysis since the  regions near OB stars typically exhibit  bright nebulosity.   We have not included the O9 star  $\iota$~Ori, the brightest star in the sword (Figure~\ref{fig:ob}), as it appears to be part of a foreground cluster \citep{2012A&A...547A..97A,2013ApJ...768...99P}.

From the distributions in Figure~\ref{fig:dist_nearest_ob}, we find the median distance of a YSO to an O or B star is 2.6~pc and 1.6~pc, respectively.   In the Orion Nebula, only the stars within 0.5 pc of $\theta^1$~C show evidence for photo-ablation \citep{2005A&A...441..195V}; the others are too distant for the FUV radiation to warm the disk enough to generate a flow of material off the disk \citep{1998ApJ...499..758J}.  For OB stars with later spectral types than $\theta^1$~C, the distances may be even smaller; thus 0.5~pc may be considered an upper limit.  On the other hand, the orbits of the observed YSOs may carry them closer to the O stars during part of their lifetime. With this in mind, we find the percentage of stars that are within 0.5~pc and 1~pc of massive stars.  The former distance represents the upper limit for current photo-ablation, the later distance includes sources that may have passed near the massive stars.  For the sample of O-B0 stars, we find that 24\% and 16\% of the YSOs are within projected distances of 1 and 0.5~pc, respectively.   For the B0.5-B3 stars, the corresponding percentages are 39\% and 20\% of YSOs that are within projected distances of 1 and 0.5~pc, respectively.   These are projected distances, hence the actual distances are larger and the percentage are accordingly upper limits.  In conclusion, less than 24\% of YSO disks are plausibly exposed to intense EUV fields, and less than 39\% are exposed to intense FUV fields.   Even in molecular clouds associated with young massive stars, the majority of low mass stars are too distant from the OB stars to be affected by their UV radiation fields.

Tidal interactions may also affect the outer regions of disks during a close encounter between two YSOs.  The probability of a close approach depends on the velocity dispersion, the spacing between stars, and the radius of interaction.  \citet{2005ApJ...632..397G} show that for a 1~km~s$^{-1}$ velocity dispersion (consistent with the C$^{18}$O line widths toward clusters), a stellar density of $10^4$~pc$^{-3}$, an interaction radius of 100~AU (which may affect a disk with a 50~AU radius disk), and a duration of 1~Myr (after which the cluster will expand due to gas disruption), only 10\% of the disks may be affected.  However, there is a distribution of disk sizes, and the duration of the high density phase of the cluster is uncertain.  Thus, with an interaction radius of 200~AU (corresponding to a 100 AU disk) and a duration of 2.5~Myr, 100\% of the disks at that density may be affected.  The actual number probably lies between these two numbers; a more rigorous analysis demands the adoption of a distribution of disk sizes, stellar masses, and a treatment of the cluster orbits during gas dispersion and is beyond the scope of this paper.  Nevertheless, regions with stellar densities $\ge 10^{3-4}$~pc$^{-3}$ are likely required for more than $> 10\%$ of the disks to be affected.  Assuming a depth of 0.1~pc, this corresponds to a regions with stellar densities in excess of $10^{2-3}$~pc$^{-2}$.  Using the fully corrected distribution in Fig~\ref{fig:nnden}, we find that 15\% of YSOs are found at these column densities.  We conclude that  approximately $\le 13$\% of the YSOs are found in regions where tidal interactions frequently affect disks. If the initial radii of disks are much larger, the fraction of disks that undergo tidal interactions may be higher 
\citep{2015A&A...577A.115V}.

\section{Summary}

We have performed a detailed analysis on the distribution of dusty YSOs identified in the {\it Spitzer} Orion survey (Paper I). The goal of this study is to examine the spatial distribution of YSOs in the Orion A and B cloud, determine the demographics of the clustering in these clouds, study the detailed structure of the clusters, and assess the potential affect of environment on circumstellar disks.  The results of the study are as follows:

\begin{enumerate}

\item{} The catalog of dusty YSOs extracted from the {\it Spitzer} Orion Survey has a spatially varying completeness.  By adding artificial YSOs to the IRAC data, we quantified the level of completeness as a function of the fluctuations in the signal surrounding the YSO at 8~$\mu$m; the level of the fluctuations is largely determined by the level of spatially structured nebulosity.  This analysis shows a steep drop in the completeness in nebulous regions.  Since clusters show the brightest nebulosity, primarily due to the UV heating of PAHs by OB stars, this results in a systematic decrease in the completeness toward clustered regions.  To apply a correction we take a two step approach.  First, we use {\it Chandra} X-ray observation of the ONC (the COUP survey) and the NGC 2024 region to identify YSOs which are not detected in a sufficient number of IR bands to be identified by their IR-excesses.  Second, in regions without X-ray surveys, we use the level of fluctuations in the vicinity  of each source to assign a weighting to each dusty YSO. This weighting corrects for the number of sources not detected due to confusion with the surrounding nebulosity and nearby point sources (Section~\ref{sec:ysocomp}).

\item{}  A total of 3481 dusty YSOs are identified in the data.  This number rises to 3889 by using data from the {\it Chandra} X-ray observatory to augment the number of YSOs in the Orion Nebula and NGC~2024 region. If these numbers are further corrected for the spatially varying nebulosity using the assigned weights, there are 5104 YSOs in the Orion clouds (Section~\ref{sec:ysocomp}).

\item{}  The nearest neighbor surface densities of dusty YSOs range from 1~pc$^{-2}$ to $10^4$~pc$^{-2}$. The distribution of densities is not lognormal.  We compare the distribution of protostars and non-protostars.  In the L1641 region, where incompleteness has the smallest effect on the detection of protostars, we find that the density distribution of protostars is  biased significantly to higher densities relative to more evolved pre-main sequence stars with disks.  The median spacing of the protostars is 0.17~pc.  This is similar to the thermal jeans length, but there is a broad distribution of spacings.   We find that less than 7\% of the L1641 protostars, and less than 11\% for the entire Orion protostar sample, are close enough to likely interact (Section~\ref{sec:den}).  

\item{}  The distributions of nearest neighbor surface densities differ between clouds with clusters - i.e. Perseus, Serpens, Ophiuchus and the Orion Clouds - and the nearby dark clouds without clusters - Taurus, Lupus, Chameleon.  The  clouds with clusters show broad density distributions that peak above 10~pc$^{-2}$ and exhibit median surface densities of YSOs above 10~pc$^{-2}$, while dark clouds without clusters have peak and median densities below 10~pc$^{-2}$.  We adopt 10~pc$^{-2}$ as a threshold for identifying clustered stars  (Sections~\ref{sec:demo_cloud} and \ref{sec:comp_other_clouds}).

\item{}  We find 47-59\% of the stars are in the one cluster with more than 1000 members, the ONC, 14-16\% are found in clusters with 100-1000 members, 18-12\% are found in groups of 10-100 members, and 21-13\% are found in groups of less than 10 YSOs or in relatively isolation. This later category we refer to as the distributed population. The ranges in these percentages are due to the inclusion of completeness corrections which increase the percentages in clusters and decrease the percentages found in groups or relative isolation. In this analysis, we define clusters as contiguous regions with nearest neighbor densities over an adopted threshold of 10~pc$^{-2}$, and the fraction of stars in clusters, groups and isolation varies with the chosen threshold value (Section~\ref{sec:demo_cloud}).

\item{} In the L1641 region, where we are more complete to protostars due to the lack of bright nebulosity, the protostellar/pre-main sequence star ratio is $\sim 30\%$ for clusters and groups, but drops to $\sim 19\%$ for the distributed population.  This indicates that the stars in the distributed population are older ($\sim 3~Myr$) than the stars in groups and clusters ($\sim 2~Myr$).  The implication is that either the distributed population is older or that it contains stars from groups and clusters that have already dispersed.  We suggest that the distributed population comes from a combination of stars formed in relative isolation, stars that have migrated from existing groups and clusters, and stars that formed in groups and clusters which have since dispersed (Section~\ref{sec:demoproto}).

\item{} The radii of the groups and clusters range from 0.34~pc to 4.3~pc, with all the clusters having radii over 1 pc. The peak YSO surface densities of the groups and clusters range over two orders of magnitude and increase approximately linearly with the number of member YSOs. In contrast, the YSO surface density averaged over a group or cluster does not exhibit a clear trend. For the four clusters, the mass of the most massive star increases with the peak density, hinting at a possible connection between the peak density and the mass of the most massive stars.  (Section~\ref{sec:global_prop})

\item{} The clusters show statistically significant deviations from circular symmetry and are typically elongated.  The highly elongated ONC does not appear to be dynamically relaxed, with the  crossing time, duration of star formation in the cluster, and the dispersal times of the parent cloud being comparable. The ONC becomes increasingly circularly symmetric in the inner regions of the cluster, as has been reported by previous authors.  If this is due to relaxation, then this would suggest an age of $\le 2.2$~Myr for the ONC (Sections~\ref{sec:anatomy} - \ref{sec:struct_dyn_onc}).

\item{} The star formation efficiency (SFE) is $2-3\%$ for the Orion~A cloud and $0.6-0.8\%$ for the Orion~B cloud; the low value for the Orion~B cloud is unusual compared to other molecular clouds within 500 pc of the Sun surveyed by {\it Spitzer}. The SFEs of the individual clusters and groups within the clouds are approximately an order of magnitude higher. The small SFEs, large sizes, long crossing times, and rapid rate of gas dispersal of the clouds are consistent with the the formation of an association and not a bound cluster of stars; however, dense clusters within the clouds may potentially survive gas dispersal to form bound clusters within a larger association (Section~\ref{sec:sfe}).

\item{} The Orion molecular clouds  and the population of dusty YSOs associated with the clouds extends over the same linear length as the Orion OB1 association; hence, the size of the OB association is determined in part by the original star formation configuration and does not simply result from the expansion of the stars from a more compact configuration (Section~\ref{sec:ob1}).

\item{} The fraction of sources with IR-excesses indicative of disks and envelopes is estimated in three ways: the IR-excess fraction of near-IR variables in the ONC cluster (79\%), the IR-excess fraction of X-ray detected sources in the ONC and NGC 2024 clusters (54\% and 58\%, respectively), and the IR-excess fraction of young stars in dense groups in L1641 (72-81\%).  Although there are variations in the fractions given by the different methods, we find the typical IR-excess fraction to be  $\sim 75\%$ implying that the dusty YSOs are 3/4 of all the young stars and protostars in the Orion clouds (Section~\ref{sec:irex}).

\item{} We assess the potential impact of the environments found in the Orion clouds on disks around pre-main sequence stars.   Less than $24\%$ of the dusty YSOs are within 1 pc from an O-B0 star and less than $39\%$ are within a projected distance of 1 pc from a B0.5-B3 star.  A majority of disk are more than 1 pc from a massive star and are unlikely to be affected by photoevaporation by the intense UV fields produced by the OB stars.  We also examine the potential for tidal interactions with nearby young stars and find that $\le 13\%$ of young stars are found in environments where tidal interactions are likely to affect disks (Section~\ref{sec:environment}).

\end{enumerate}

\section{Appendix A: The identification of dusty YSOs in the {\it Spitzer} Orion Survey: Modifications from Paper I}

In Paper I, we presented a catalog of dusty YSOs based on a suite of color and magnitude criteria adapted primarily from the work of \citet{2009ApJS..184...18G} and \citet{2012AJ....144...31K}.  In this paper, we have used that catalog with three very minor modifications. In Paper I, the criteria described in Eqn.~3,

\begin{equation}
[3.6]-[4.5] \ge 0.5 + \sigma_{[3.6]-[4.5]},~[4.5]-[5.8] \ge 0.25 + \sigma_{[4.5]-[5.8]},
\end{equation}

\noindent 
are only applied to sources without 24~$\mu$m detections.  This is appropriate in the case of protostars, where we only consider sources identified by their 3.6, 4.5 and 5.8~$\mu$m magnitudes if they are not detected at 24~$\mu$m.  However, requiring a 24~$\mu$m non-detection can also reject bonafide disk sources and faint protostar candidates which  otherwise would been identified. Thus, in this paper, we apply the above criteria to all sources where the 3.6, 4.5 and 5.8~$\mu$m data satisfy our uncertainty limits.  However, we only identify such sources as protostars if they do not have 24~$\mu$m detection and they have colors satisfying Eqn.~8 in Paper I.  This leads to the detection of five additional sources, four pre-main sequence stars with disks and one faint candidate protostar,  and the total number of dusty YSOs increases to 3474.

Furthermore, we discovered a group of sources that were being rejected as potential outflow shocks, even though they had an IR-excess at 24~$\mu$m.  Outflow knots are not expected to show strong emission in the 24~$\mu$m band.  In our modifieid criteria, all sources identified as YSOs on the basis of a 24~$\mu$m excess will no longer be rejected if they have IRAC colors similar to those of outflow shocks.  This results in the identification of seven additional dusty YSOs, bringing up the total number to 3481.  The seven additional sources are all pre-main sequence stars with disks. All twelve new identified YSOs are found in Table~3.

We also reclassified sources due to minor changes in our source classification criteria.  In Paper I, we required sources classified as protostars to be located within the 4.5~$\mu$m mosaic. However, due to the spatial offset between the 3.6/5.8~$\mu$m FOV and the 4.5/8~$\mu$m FOV, there are regions with 3.6, 5.8 and 24~$\mu$m detections without 4.5~$\mu$m data.  This meant that IR-excess sources with 3.6 and 5.8~$\mu$m detections that fell off the 4.5~$\mu$m mosaic would be classified as disk sources even though they showed a rising SED between 5.8 and 24~$\mu$m.  We therefore have removed the requirement that the sources be within the 4.5~$\mu$m mosaic.  

In addition, we had required that protostars detected at 4.5 and 24~$\mu$m be within the 3.6~$\mu$m mosaic.  We now have removed that requirement and use the 2MASS $H$ and $K_s$-band detections to establish an upper limit at 3.6~$\mu$m.  The upper limit is extrapolated from the $H$ and $K_s$-band magnitudes using the equation

\begin{equation}
[3.6]_{limit} = \left(\frac{log(3.6)-log(2.15)}{log(2.15)-log(1.64)}\right) (m_{K_s}-m_H)
\end{equation}

\noindent
if there are $H$ and $K_s$-band detections, and is equated to the $K_s$-band magnitude if there is no $H$-band detection. In all cases, the 3.6~$\mu$m upper limit is required to be $\le 15.5$~mag, and if there is no H or $K_s$-band detection, then $[3.6]_{limit} = 15.5$~mag.

Faint protostar candidates are sources which exhibit the colors of protostars, that are not found to be extragalactic contamination by the criteria of \citet{2009ApJS..184...18G}, yet do not satisfy the criteria $m_{24} \le 7$~mag required for high reliability protostar candidates.  In Paper I, only sources with 24~$\mu$m detections were allowed to have the faint protostar classification.  In this paper, we also classify faint candidate protostars that do not have 24~$\mu$m detections, yet $[4.5]-[24]_{limit} \ge 4.76$~mag where $[24]_{limit}$ is the lower limit to the 24~$\mu$m magnitude described in Paper I.  

Due to these changes,  seventeen disks sources are reclassified: three disk sources are now protostars and fourteen disk sources are considered faint prototstar candidates.  In addition, two protostar are now classified as a disk source.  
The reclassified sources are  found in Table~4.

\section{Appendix B: Measuring the Completeness to YSOs}

The completeness of YSOs as a function of the nebular background and sources confusion is difficult to assess from the single band completeness functions described in Paper 1.  The identification of a source as a YSO depends on the completeness in at least three of the eight 2MASS and {\it Spitzer} wavelength bands (Paper I) and is a much more difficult problem than determining the completeness  in an individual band.  Utilizing the single-band completeness functions would require us to devise a methodology that takes into account the completeness limits in the eight available wavelength bands, the multiple color and flux criteria used to identify YSOs combinations of those bands, and the magnitudes of the YSOs in those bands.

A  simpler approach is to add a representative sample of YSOs directly to the IRAC images and then determine the fraction  of YSOs recovered as a function of the RMEDSQ.  This value, which is introduced in Paper 1, is given by:

\begin{equation}
RMEDSQ(i_0,j_0) = \sqrt{median[(S_{ij}-median[S_{ij}])^2]}
\label{eqn:remedsq}
\end{equation}

\noindent
where $i_0$ and $j_0$ are the pixel coordinates of the source, and $i$, $j$ are the pixels which are found in an annulus  centered on the source.  For the IRAC photometry, the annulus typically extends from 6 to 11 pixels (7\farcs 2 to 13\farcs 2).   For bright stars, fluctuations in the point source response function, or PRF, may dominate the RMEDSQ in this annulus. To ensure that the PRF itself does not  contribute to the RMEDSQ, the radius of the annulus is extended to larger radii until variations in the PRF make a negligible  contribution to the RMEDSQ. The maximum allowed size of an annulus is from 45 to 50 pixels (54\arcsec to 60\arcsec).   Since the units of our mosaics are DN, the resulting RMEDSQ are in units of DN.  As in Paper 1, we use the RMEDSQ measured in the 8~$\mu$m band; this band shows the strongest nebular emission from the Hydrocarbon bands and provides the best measure of the nebular emission that dominates source confusion in the the IRAC data.

%
%
%

\begin{figure}
\plotone{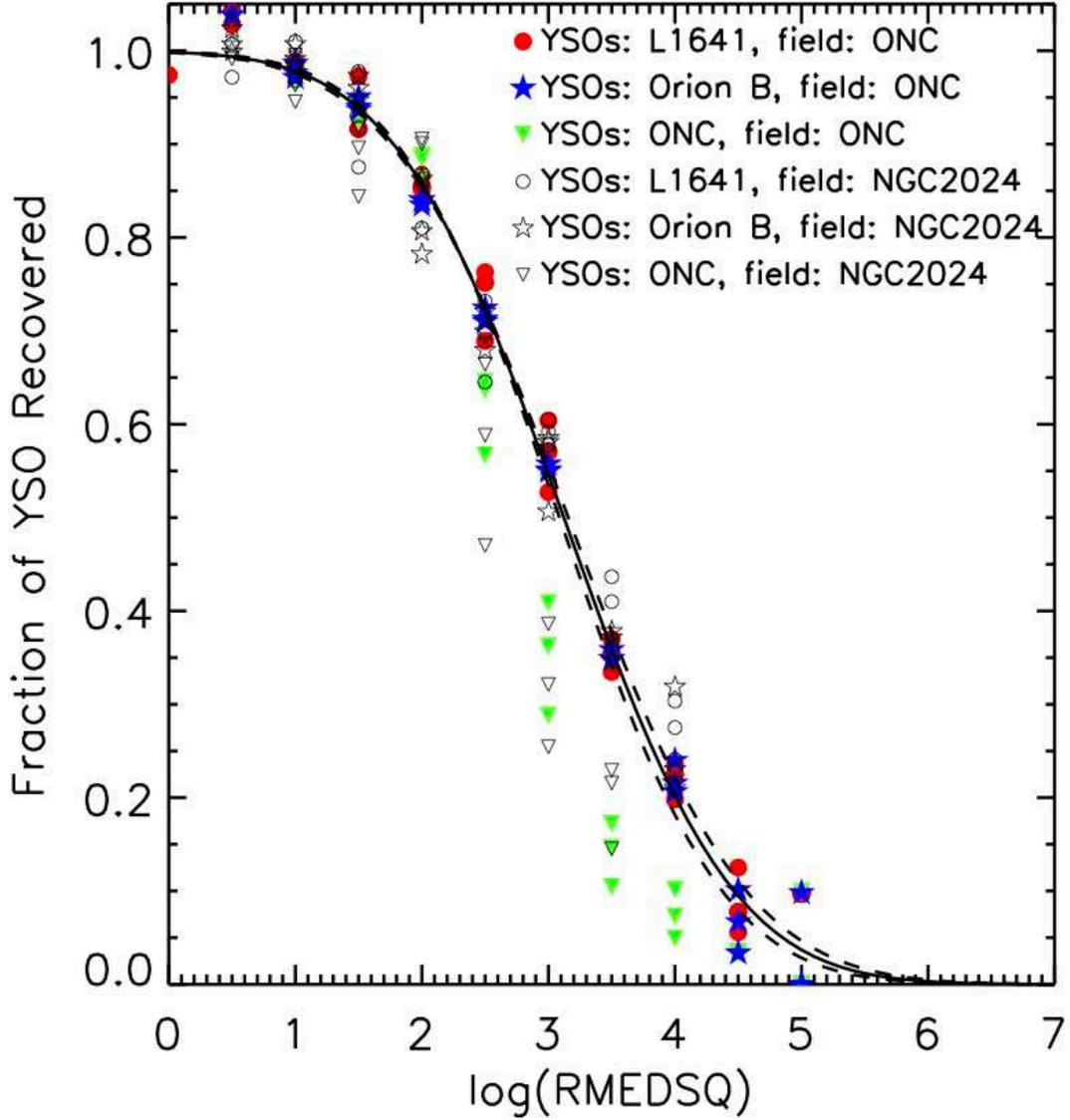}
\caption{The fraction of recovered young stellar objects vs.~$log({\rm RMEDSQ})$ at 8~$\mu$m. For a given value of $log({\rm RMEDSQ})$, we show fractions determined from different fiducial YSOs samples and survey fields. The symbols corresponding to the various combinations of the samples and fields are defined in the key printed within the plot; see Appendix B for a description of these combinations.  The solid line shows the adopted fit and the dashed line shows the $\pm 1$~$\sigma$ uncertainties of the fit. The functional form and adopted parameters for the fit are presented in Appendix B. }
\label{fig:frac}
\end{figure}

\begin{figure}
\plotone{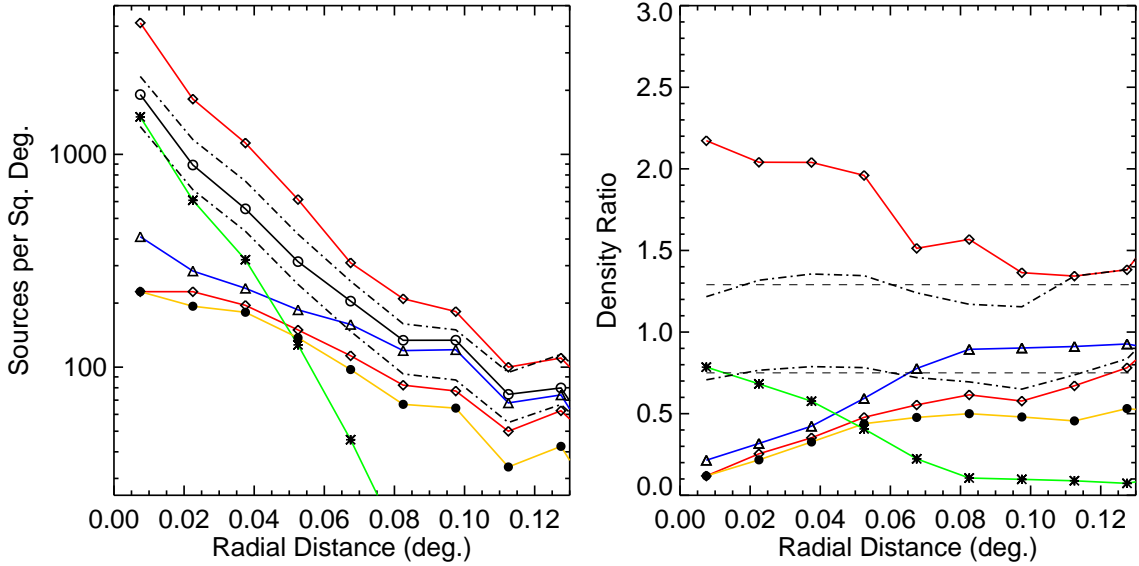}  
\caption{{\bf Left:} the azimuthally averaged surface density of sources as a function of radial distance from 
the median R.A.~and decl.~of all COUP sources (R.A. =  5:35:16.8, decl. = -5:22:60).  The black line/open circles shows the density of all sources.  The color lines show the densities for sources belonging to four different categories.  The green line/asterisks is the density of COUP sources that are not detected in enough IR-bands to identify an IR-excess.   The blue line/triangles shows the density of COUP sources that are detected in enough IR-bands bands to identify an IR-excess, and the orange line/filled circles show the density of those X-ray sources which have IR-excesses.  The red line/diamonds shows the density of all IR-excess sources; the lower curve shows the density uncorrected for incompleteness and the upper curve shows the density corrected by the RMESQ derived weights.  The black dot-dash lines show the density for the adopted weighted combinations of X-ray and IR-excess sources; the lower line is for the case when there is no correction for incompleteness outside the COUP field, while the upper line is corrected by the RMEDSQ weights to  account for the incompleteness of the census at the the outer radii and outside the COUP survey.  {\bf Right:} the density of sources in each of the five categories normalized by the density of young stars identified in the COUP survey. For comparisons, the black dashed lines give the weights of the X-ray sources without (0.75) and with (1.29) the correction for the RMEDSQ weights. Note that the fraction of COUP sources which do not have enough photometry to be identified as an IR-ex sources rises to 75\% in the center of the ONC.}
\label{fig:coup_radial}
\end{figure}

\begin{figure}
\plotone{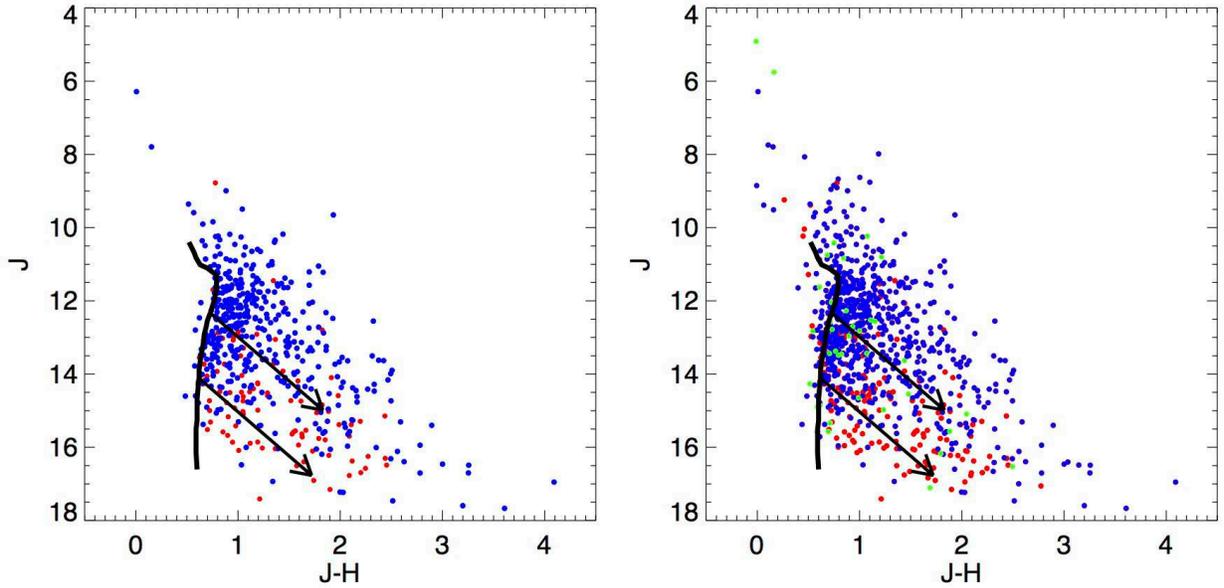}
\caption{$J$ vs. $J-H$ color magnitude diagrams for young stars detected in the COUP field.  On the left we show the IR-excess sources: those detected with COUP are blue while those detected only with {\it Spitzer} are shown in red. On the right we show all young stars. The green dots mark the X-ray detected stars that lack the photometry needed for the detection of an IR-excess, blue dots are the X-ray detected that have sufficient photometry for the detection of IR-excesses, and the red dots are young stars identified by {\it Spitzer} that are not detected in the COUP survey. The black curve is the 1~Myr isochrone from \citet{1998A&A...337..403B} and two lines are extinction vectors extending to 1 $A_K$ for 1 Myr stars with masses of 0.25~M$_{\odot}$ and 0.08$~M_{\odot}$.  These diagrams show that the COUP data are not complete for very low mass stellar and sub-stellar members ($< 0.25~M_{\odot}$).}
\label{fig:coup_cm}
\end{figure}

\begin{figure}
\plotone{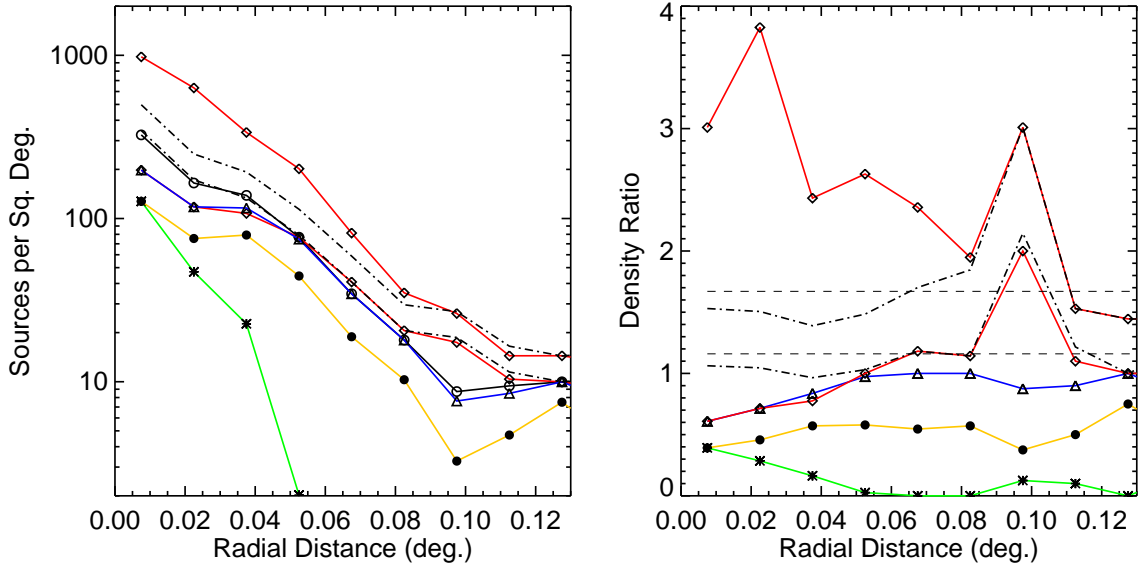} 
\caption{{\bf Left:} the surface density of source within the NGC2024 field as a function of radial distance from a central R.A.~and decl.~(R.A. = 5:41:45.8 and decl. = -01:54:30).  The curves/symbols show the densities for the same five categories of sources as displayed for the ONC. {\bf Right:} the density of sources in the five categories normalized by the density of X-ray identified young stars. For comparisons, the black dashed lines give the weights of the X-ray sources without (1.16) and with (1.67) the correction for the RMEDSQ weights. The fraction of {\it Chandra} identified young stars without sufficient IR  photometry to be identified as an IR-excess source peaks at 40\% in the center of the cluster.}
\label{fig:n2024_xray_radial}
\end{figure}

\begin{figure}
\plottwo{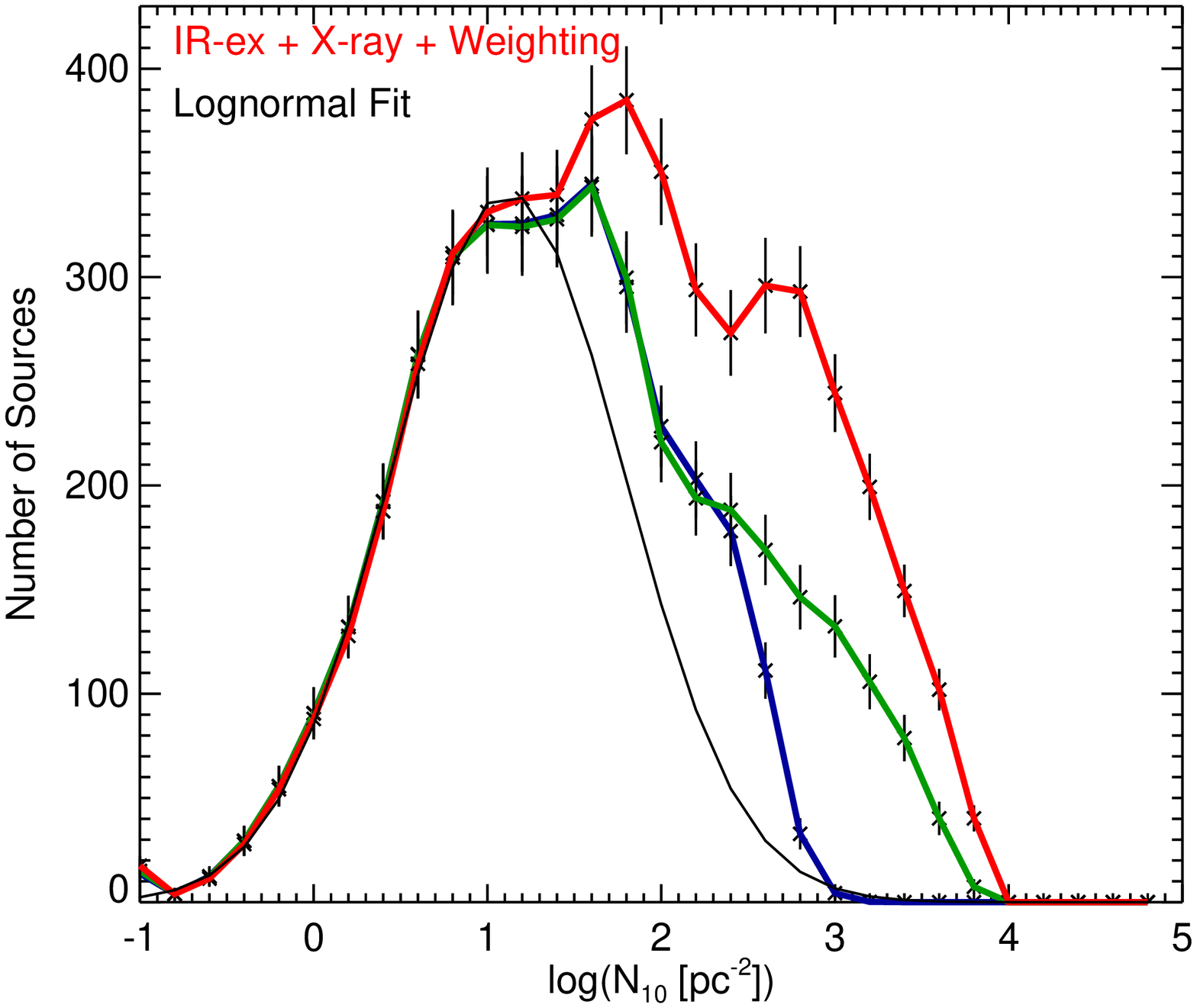}{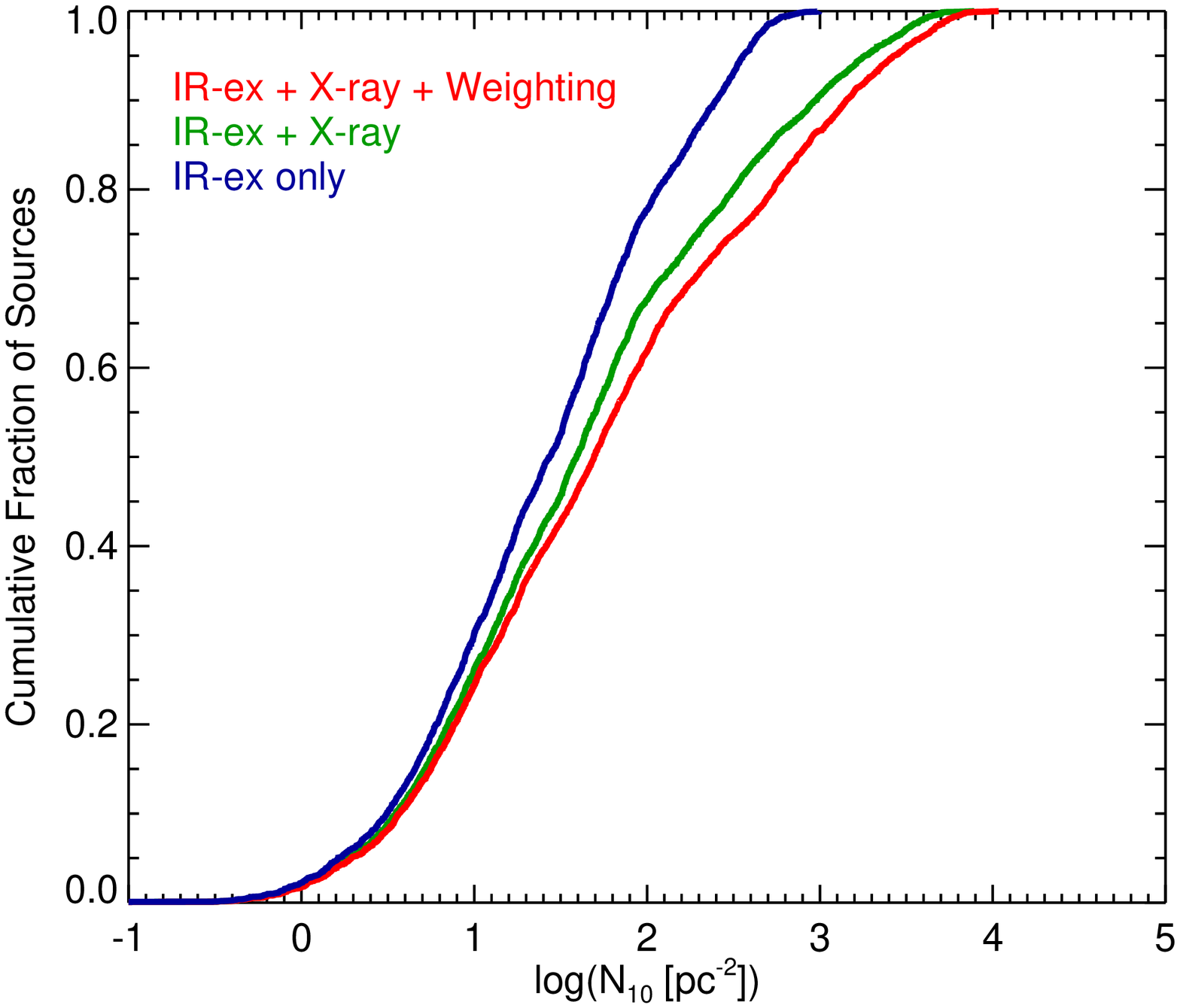}
\caption{{\bf Left:} the distribution of nearest neighbor densities for the dusty YSO sample.  The distance to  the 10th nearest neighbor was used to estimate the local density around each source. We show the distribution calculated for three cases: the dusty YSOs identified by {\it Spitzer}, the dusty YSOs augmented with the {\it Chandra} sources, and the {\it Spitzer} plus {\it Chandra} YSOs corrected for incompleteness.  For comparison, we fit a lognormal function to the low density end of the distribution, the high density end cannot be fit with a lognormal.  {\bf Right:} the normalized cumulative distribution of YSOs for each of the three samples. }
\label{fig:nnden}
\end{figure}

\begin{figure}
\plotone{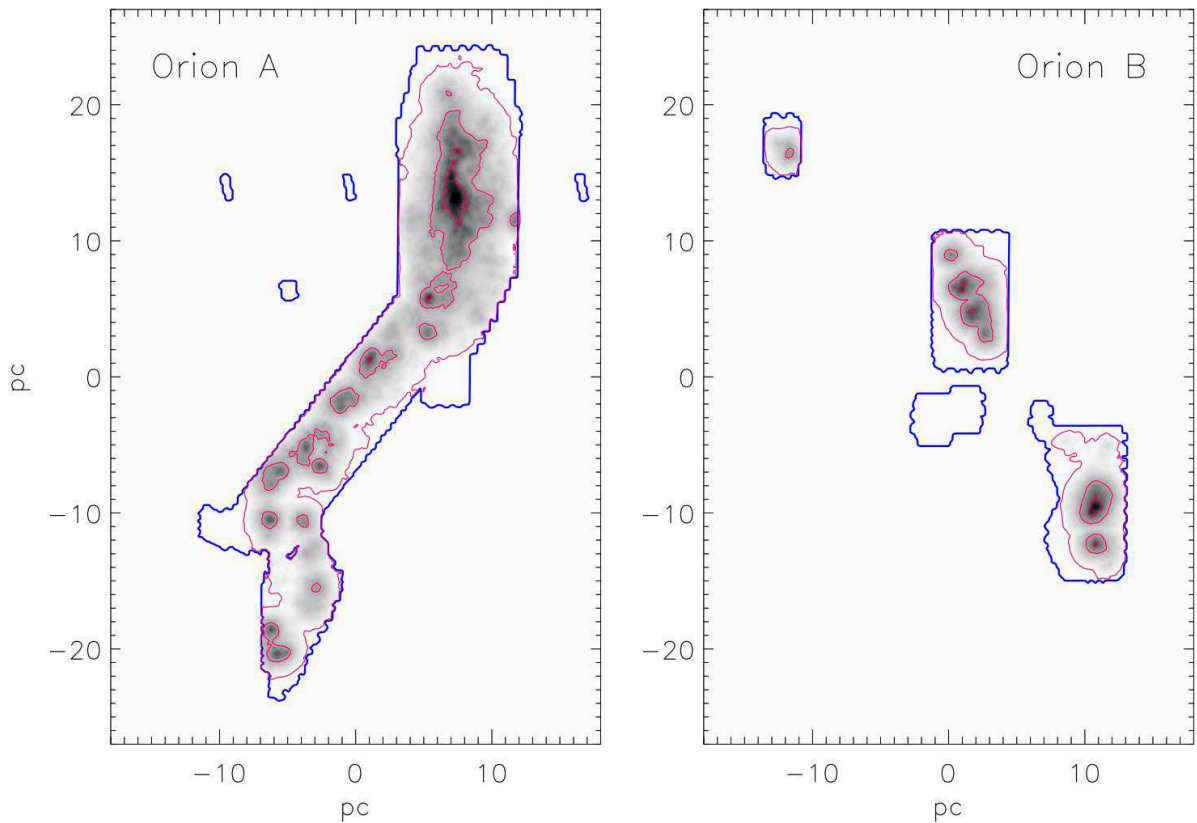}
\caption{Maps of the nearest neighbor surface density in the Orion A and B molecular clouds.  We have used the 10th nearest neighbor and have corrected the densities for incompleteness. The blue contour gives the outline of the IRAC field.  The inverted gray scale images renders the densities with a logarithmic scaling.  The red contours are for 1, 10 and 100 YSOs~pc$^{-2}$.   
The adopted distance is 414 pc. }
\label{fig:nn_map}
\end{figure}

\begin{figure}
\plotone{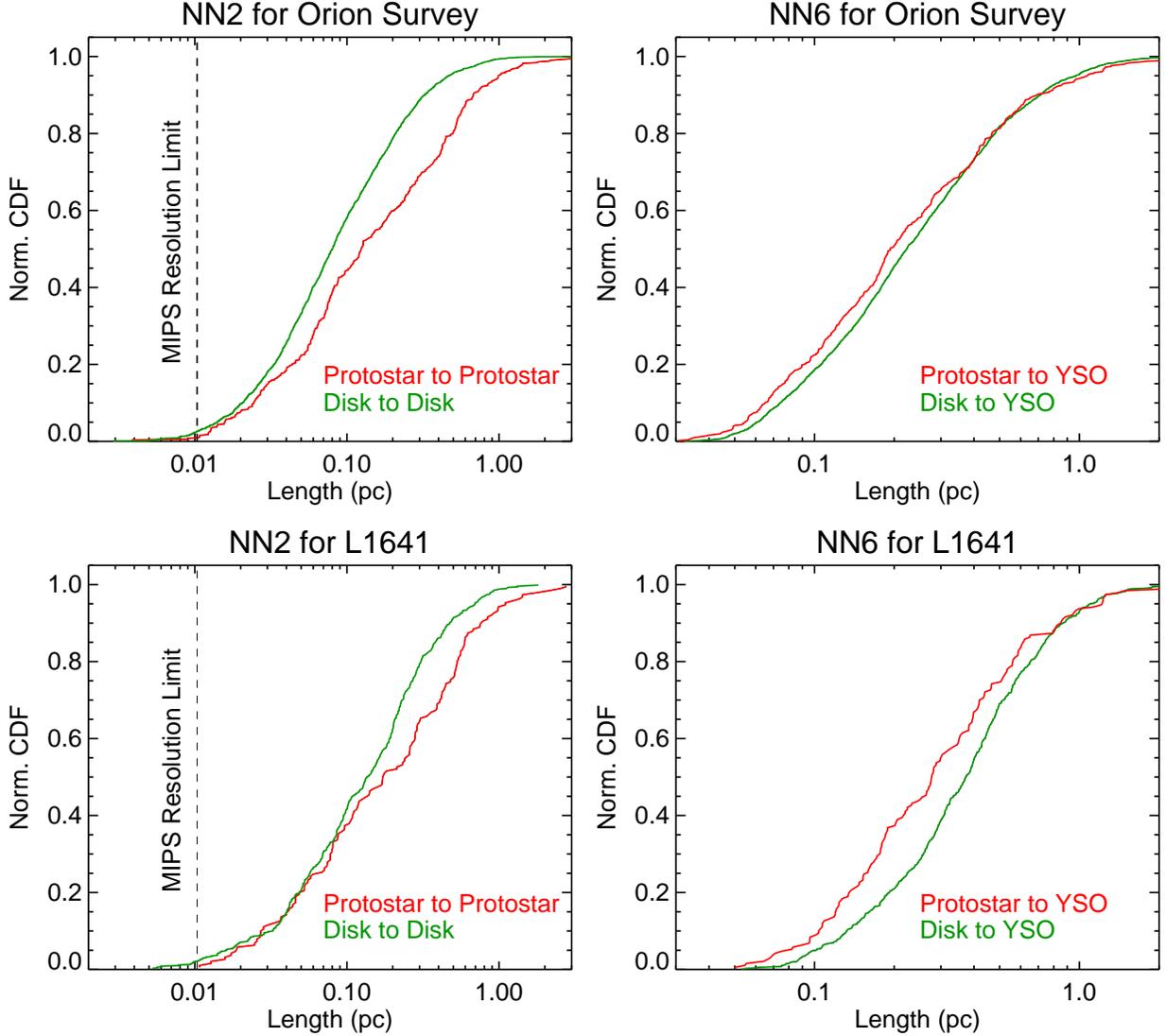}
\caption{{\bf Left panels:} the cumulative distributions of nearest neighbor (nn2) distances  for protostars and pre-main sequence stars with disks.  On the top we show the nn2 distances  for the entire sample while on the bottom we show the distances only for the L1641/$\kappa$~Ori region.  These plots show that nn2 distances are typically smaller for disk sources than for protostars. {\bf Right Panels:} cumulative distribution of 5th nearest distances (nn6) between protostars and all dusty YSOs and between disk sources and all dusty YSOs.  We show the nn6 distances for the entire sample on the top panel and the nn6 distances for the L1641/$\kappa$~Ori region in the bottom panel. These show that in the L1641/$\kappa$~Ori region, that protostars are typically found in denser regions than the disk sources.  Although this is not seen for the entire sample, the comparison between protostars and disk sources in the entire sample is affected by our inability to detect and identify protostars in the bright nebulosity found in the cores of the dense clusters.}
\label{fig:orion_nn}
\end{figure}

\begin{figure}
\plotone{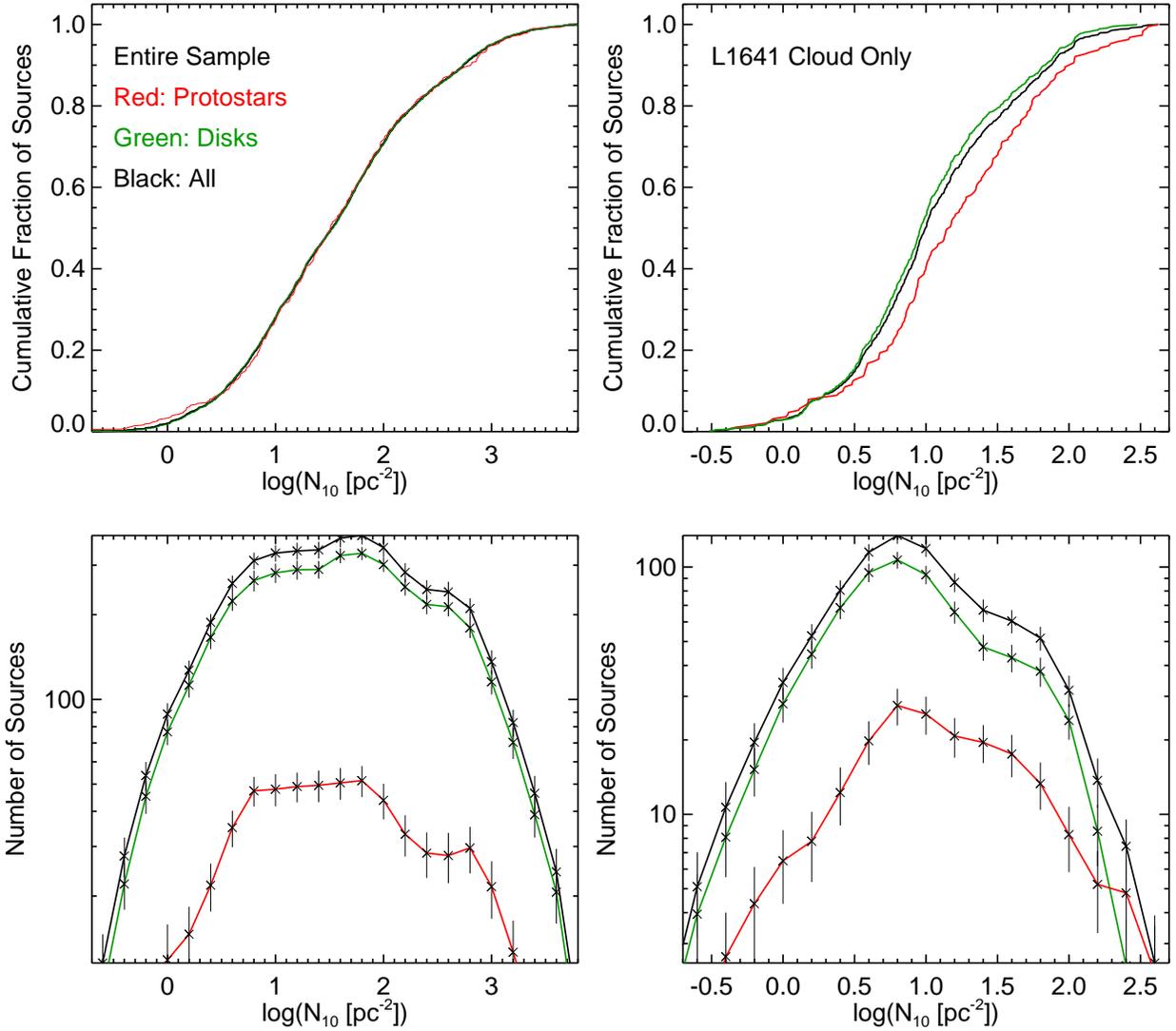}
\caption{The distribution of nearest neighbor densities for the entire clouds and the L1641/$\kappa$ Ori region.  The upper panels give the cumulative distributions while the lower panels give the differential distributions.  The key defines the different colors used in the plots.  The left panels shows the distributions for the entire sample, the right panels show the distributions for the L1641/$\kappa$ Ori region.  Again, the protostars in the L1641/$\kappa$~Ori  are located in denser regions than the disk sources.}
\label{fig:orion_proto_disk_nnden}
\end{figure}

\begin{figure}
\plotone{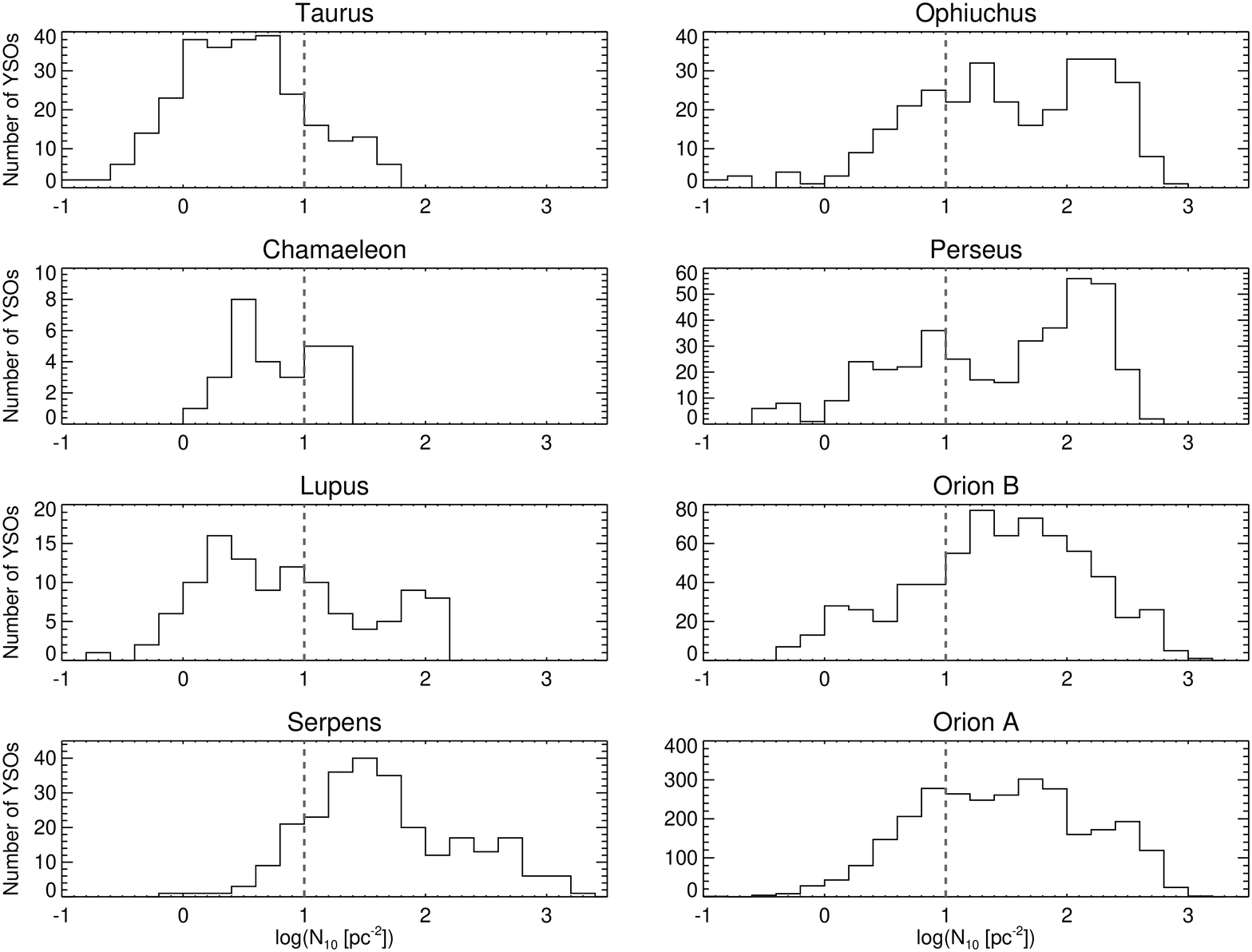}
\caption{A comparison of the distribution of nearest neighbor densities for the uncorrected Orion~A and Orion~B cloud YSO samples from this paper, for the c2d catalogs of the  Chameleon, Lupus, Ophiuchus, Perseus and Serpens clouds and for the Taurus cloud (K. Luhman, p. Com). The nearest neighbor density for the 10th nearest neighbor ($N_{10}$) was used to estimate the local density around each source. The dashed line is drawn at a density of 10 pc$^{-2}$.}
\label{fig:nnden_nearby}
\end{figure}

\begin{figure}
\plotone{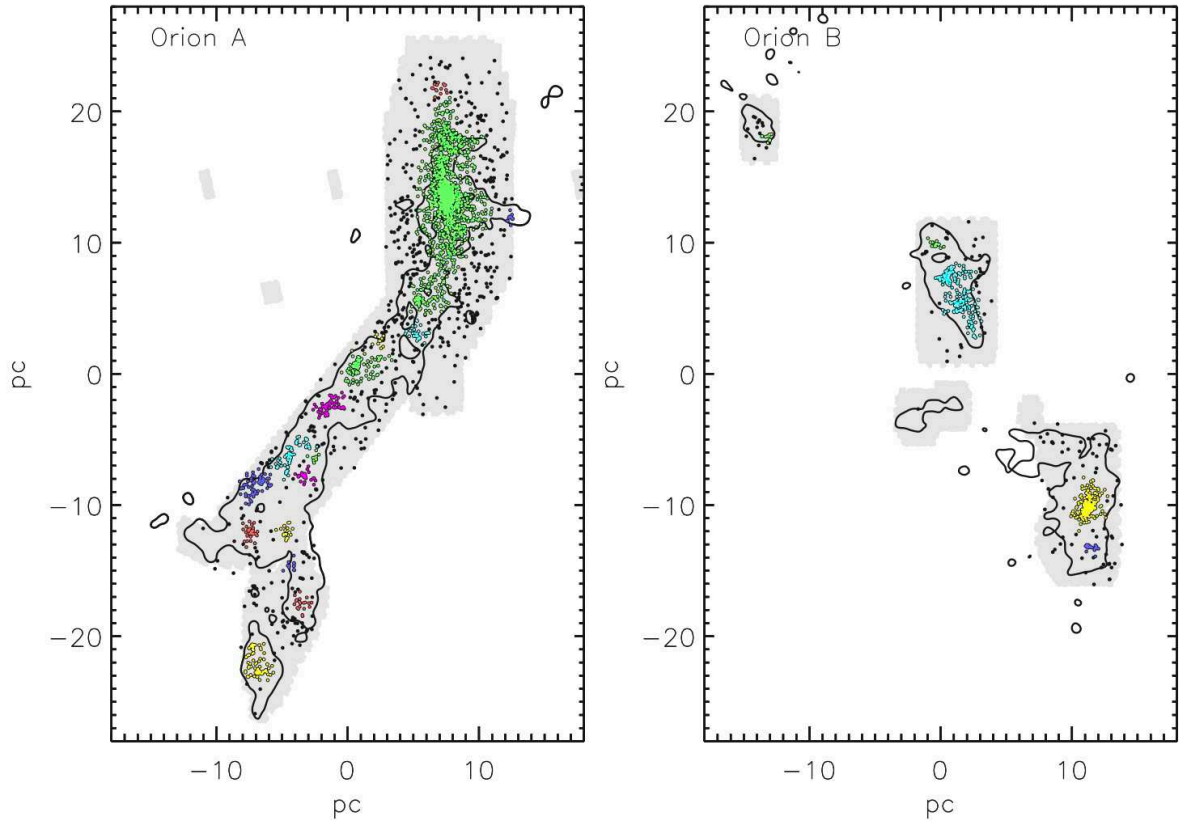}
\caption{The clusters identified in the Orion A and Orion B clouds. The $A_V = 3$ contour of the two clouds are shown. The grey area shows the region surveyed by with all four IRAC bands. The colors show the different groups and clusters identified above a threshold of 10~pc$^{-2}$, the black dots are YSOs which are not included in a group or cluster with 10 or more members.}
\label{fig:clusterid10}
\end{figure}

\begin{figure}
\plotone{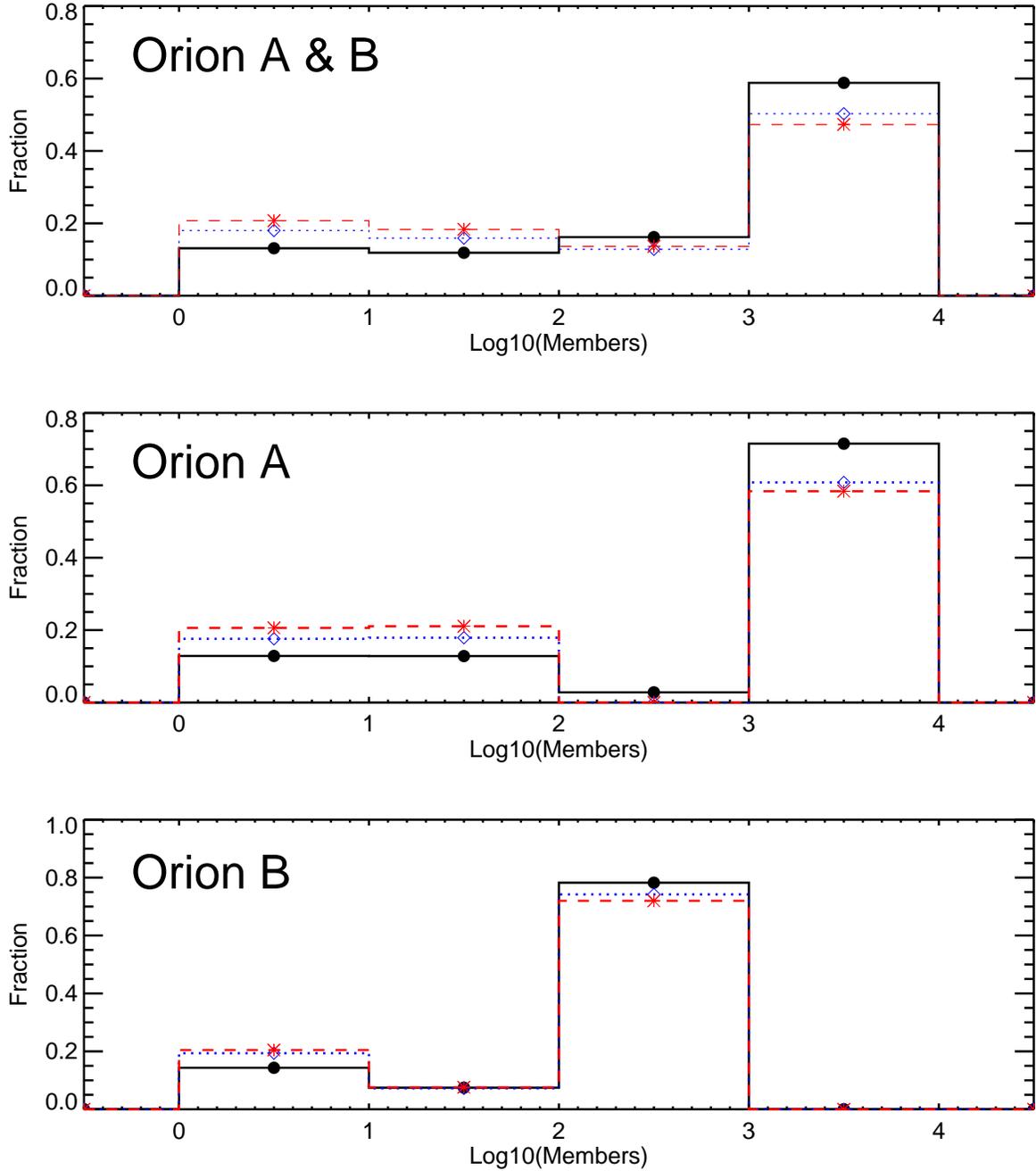}
\caption{The fraction of dusty YSOs in groups and clusters binned in logarithmic intervals of the number of members. Sources which are not found in a group or cluster with 10 or more members are put in the first bin.  We display the fractions for the combined Orion clouds ({\bf Top}), the Orion~A cloud ({\bf middle}) and the Orion~B cloud ({\bf bottom}).  We show this for the fully corrected fractions (black), the {\it Chandra} augmented fraction without the weighting  corrections (red) and the sample without completeness corrections or the addition of the {\it Chandra} sources (blue).}
\label{fig:demo}
\end{figure}

\begin{figure}
\plottwo{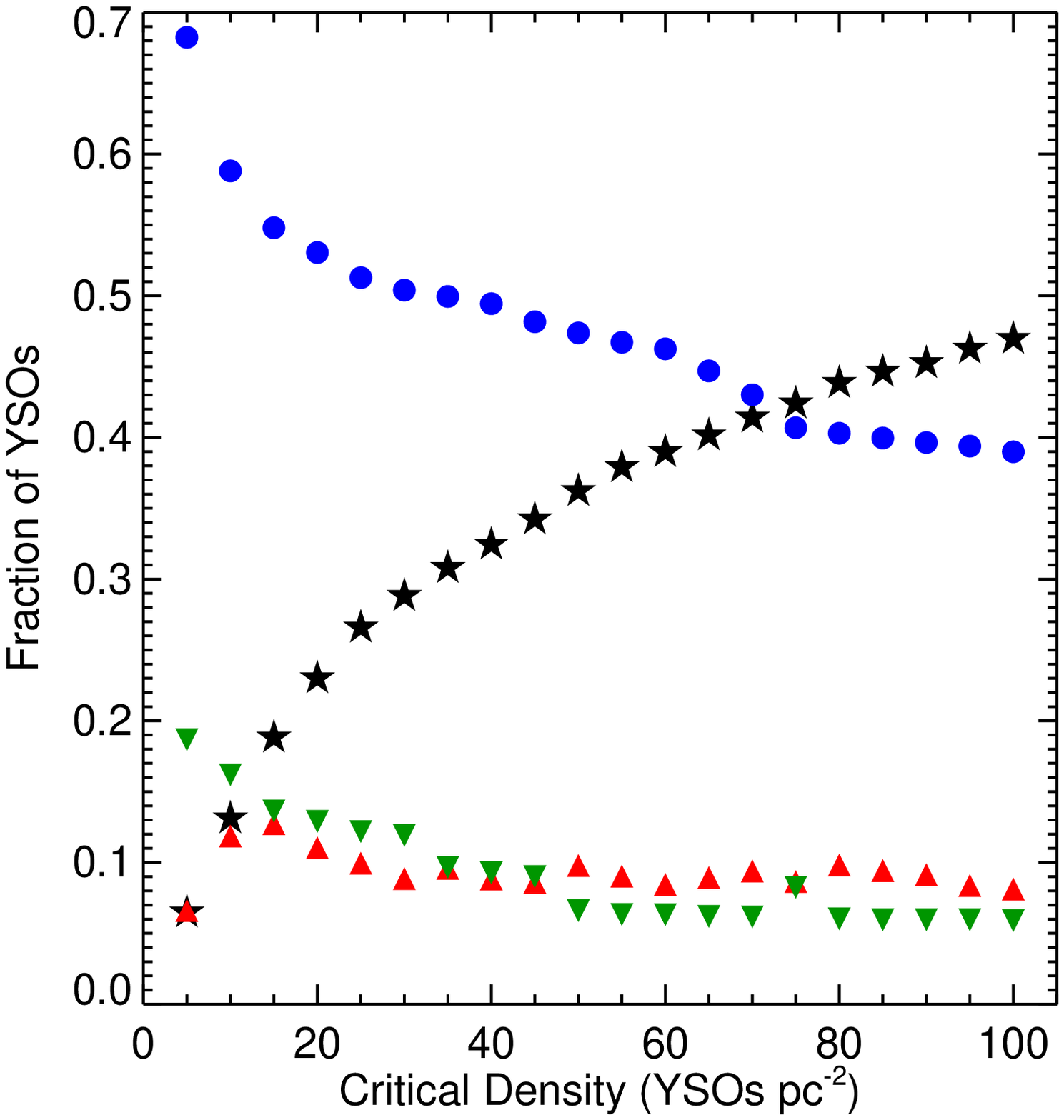}{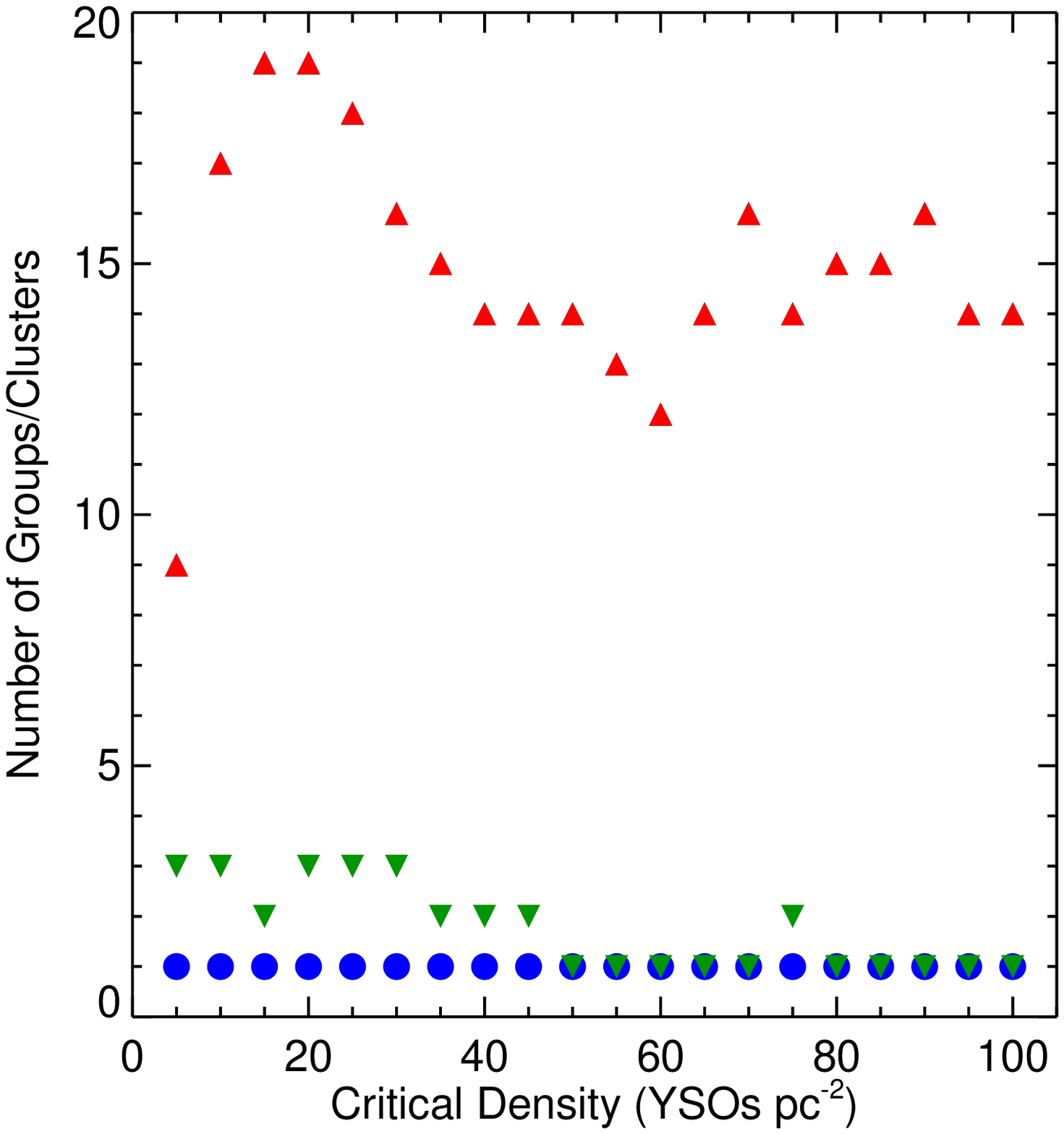}
\caption{{\bf Left:} the fraction of dusty YSOs in groups and clusters as function of the critical threshold density using the incompleteness corrected fractions.  The blue circles are the fraction of dusty YSOs in clusters with $\ge 1000$ members, the green upside down triangles are the fraction of YSOs in clusters with  100 to 1000 members, the red triangles are the fraction in groups with 10 to 100 members, and the black stars are the fraction in the distributed population.  {\bf Right:} the number of groups and clusters as a function of critical density.  The symbols are the same as the right panel.  As we raise the critical density, YSOs switch between the largest clusters and the distributed population; however, the fraction of stars in small groups (10-100 members) and the number of small groups remain relatively  constant.}
\label{fig:demo_trend}
\end{figure}

\begin{figure}
\plotone{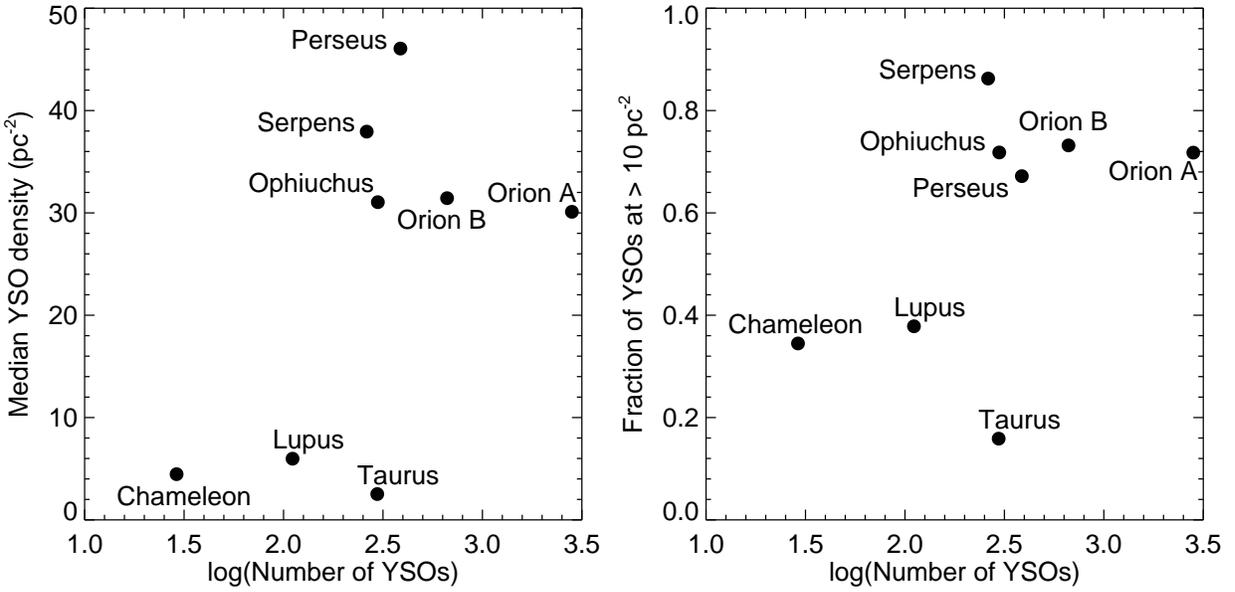}
\caption{A comparison of stellar surface densities in star forming regions within 500 pc. {\bf Left:} the total number of YSOs vs. the median YSO surface density. {\bf Right:} the number of YSOs vs. the fraction of YSOs at densities above 10 pc$^{-2}$. The densities were determined using the 10th nearest neighbor from each YSO in the respective clouds (see histograms in Figure~\ref{fig:nnden_nearby}).}
\label{fig:nearby_summary}
\end{figure}

\begin{figure}
\plottwo{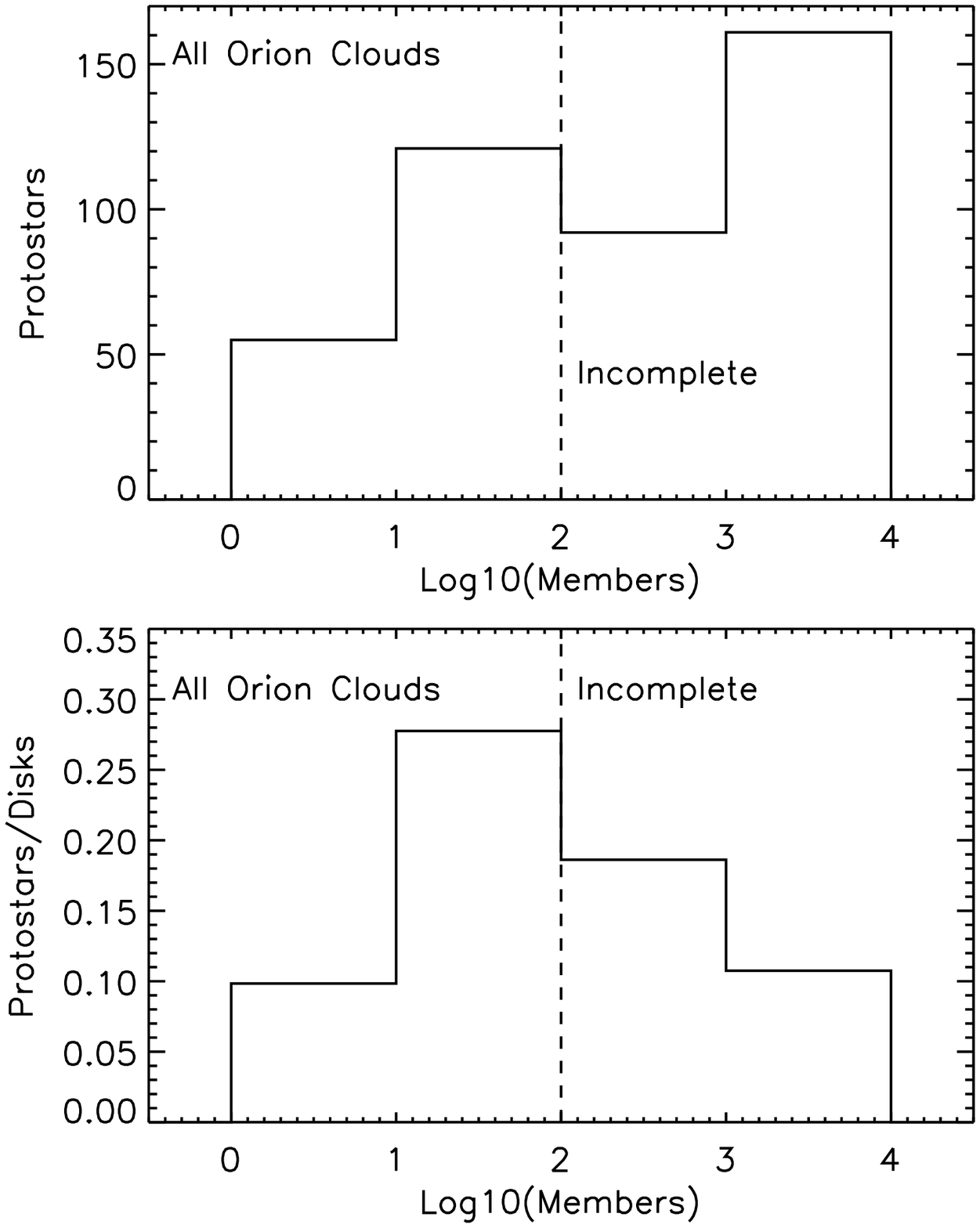}{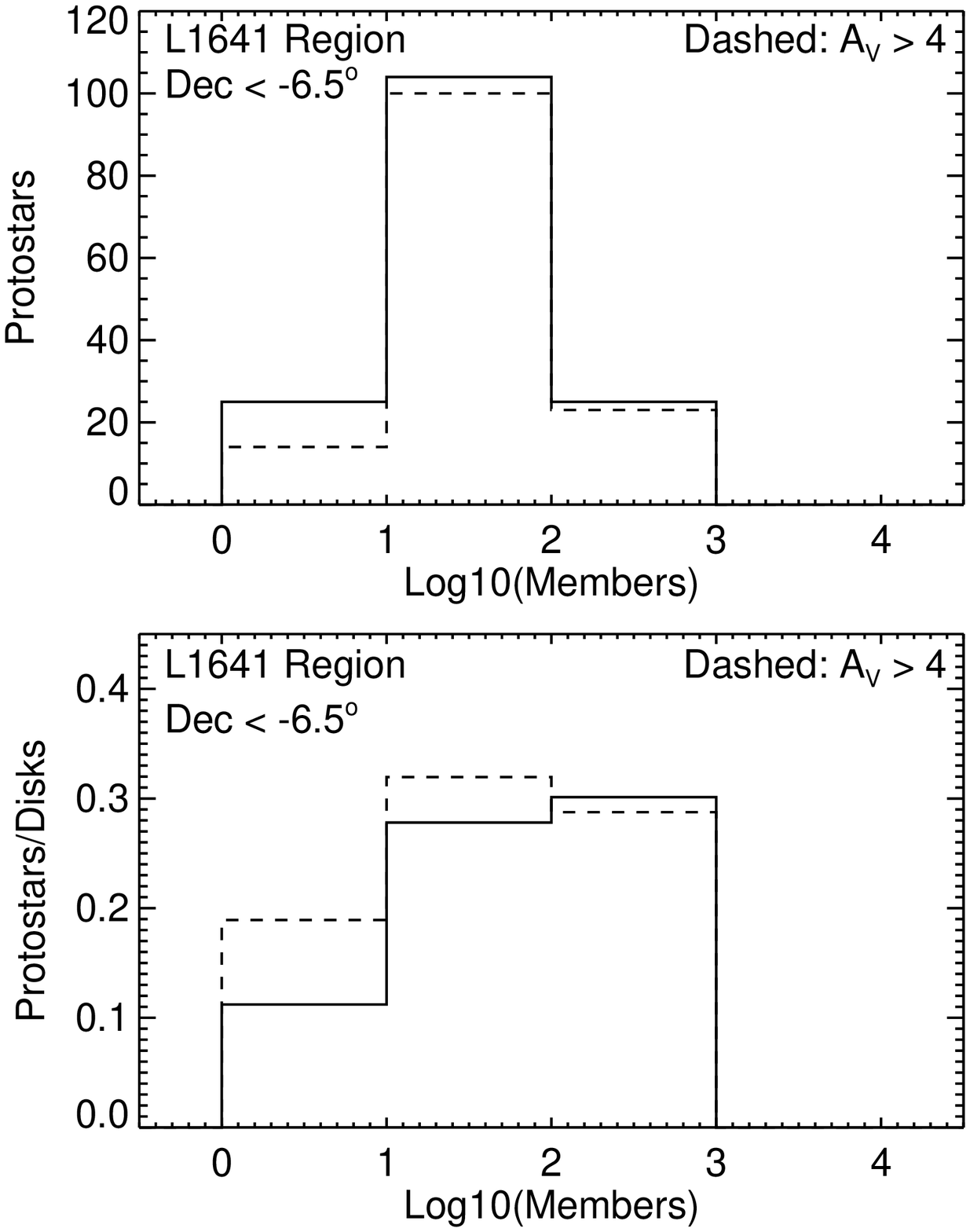}
\caption{{\bf Top Left:} the number of protostars vs. the number of members in the host assemblage; the four bins partition the distributed population (1-10 members), groups (10-100 members,  clusters (100-1000 members) and large clusters (100-1000 members).  The vertical dashed lines shows the approximate number of members at which the numbers of YSOs become incomplete. {\bf Bottom Left:} the protostar/disk ratio for the same bins.  There has been no correction for completeness, thus the numbers and ratios for the clusters are affected strongly by incompleteness. {\bf Top Right:} the number of protostars as a function of the number of members in the host assemblage for the L1641 region; this region suffers less from incompleteness than the other regions of the Orion clouds. The dashed lines show the number of protostars for the regions of the clouds with $A_V > 4$. {\bf Bottom Right:} the protostar/disk ratio vs. the number of members for the L1641 region.}
\label{fig:cluster_proto}
\end{figure}

\begin{figure}
\plotone{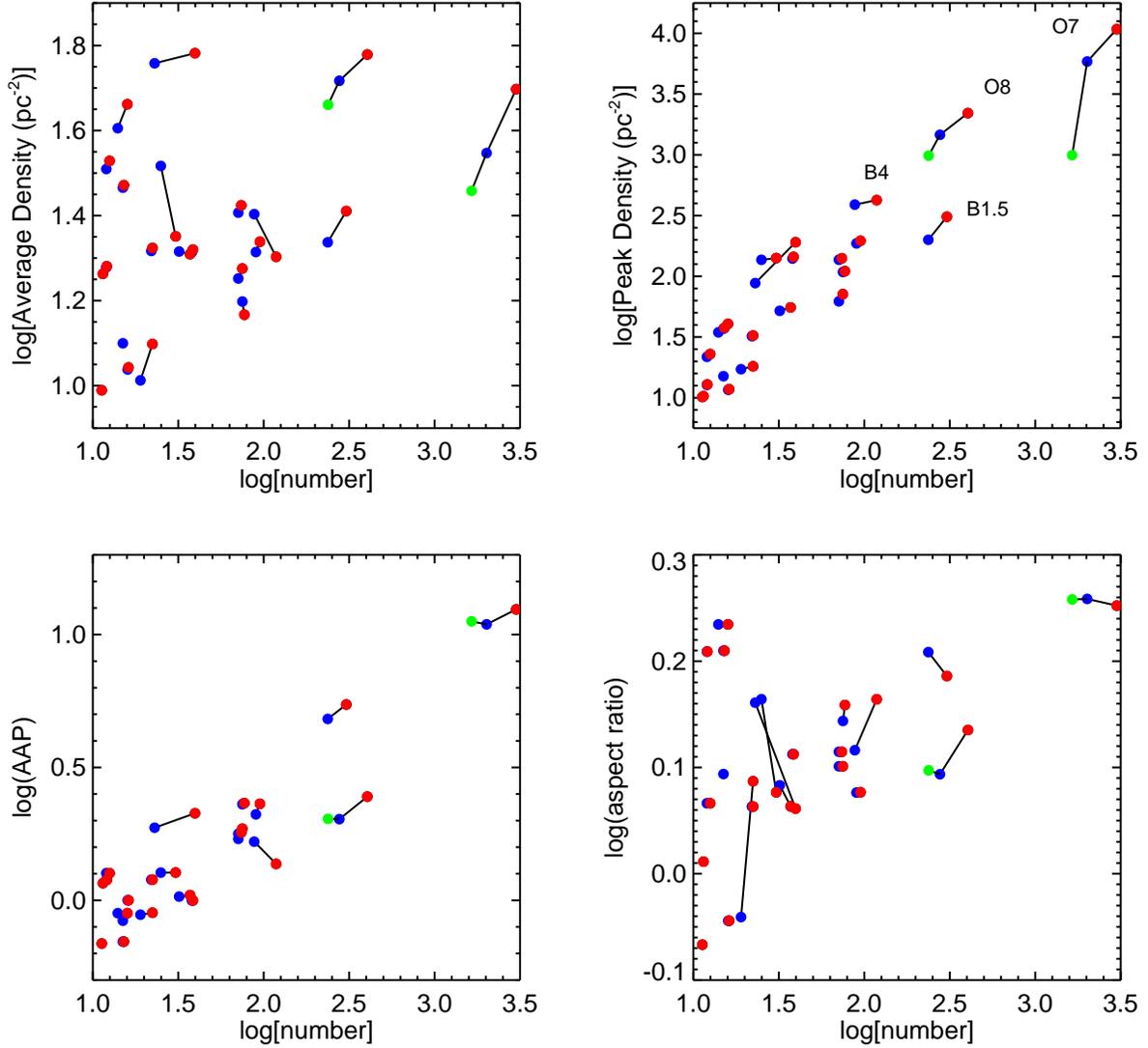}
\caption{The properties of the identified groups and clusters as a function of the number  of members. For each cluster, we show the uncorrected properties in blue  and the completeness corrected  properties in red; the values are linked together by a black line.  For the ONC and NGC~2024, the uncorrected properties includes the X-ray sources identified by {\it Chandra}.   For these clusters, we show their properties before its membership was augmented by the {\it Chandra} sources in green. }
\label{fig:cluster_radius_size}
\end{figure}

\begin{figure}
\plotone{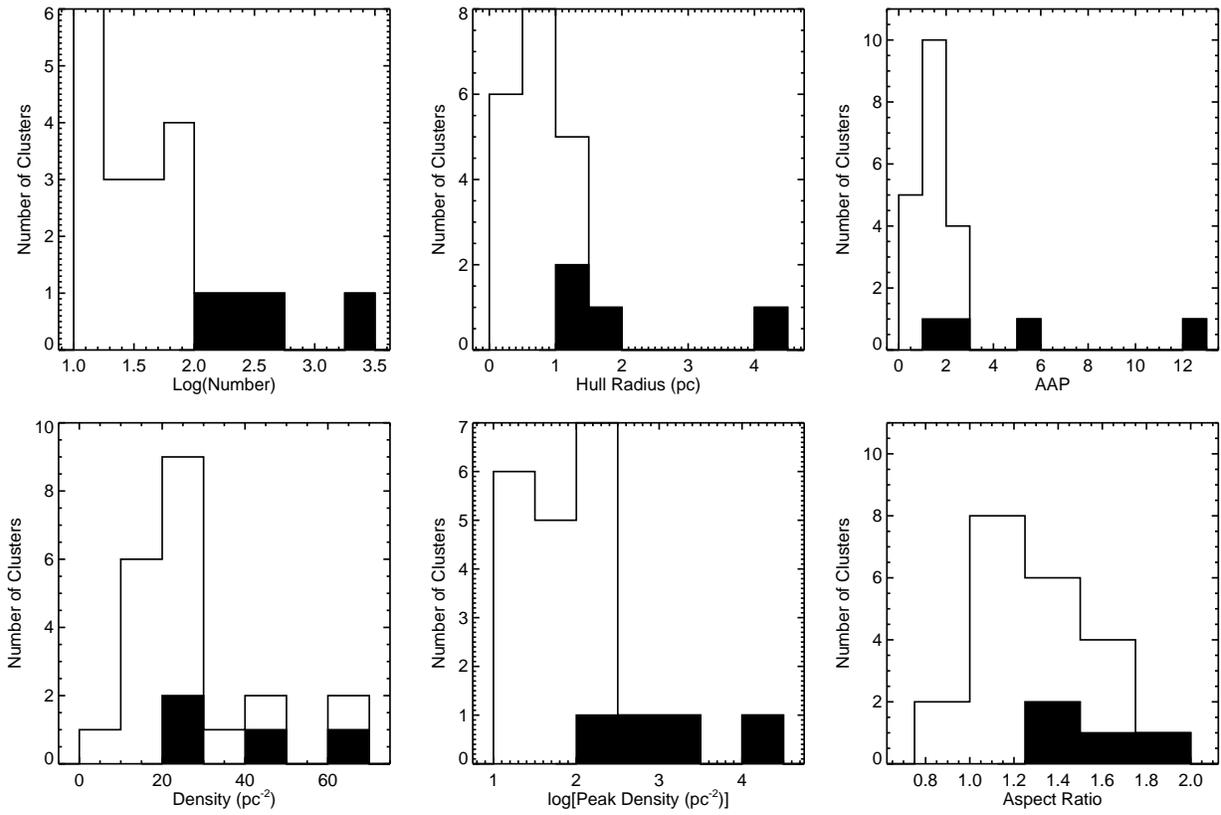}
\caption{Histograms of the cluster and group properties as defined in the text.  The black shaded histograms show the values for the four largest clusters, the combined shaded and unshaded histogram shows distribution from all clusters and groups.}
\label{fig:cluster_prop}
\end{figure}

\begin{figure}
\plotone{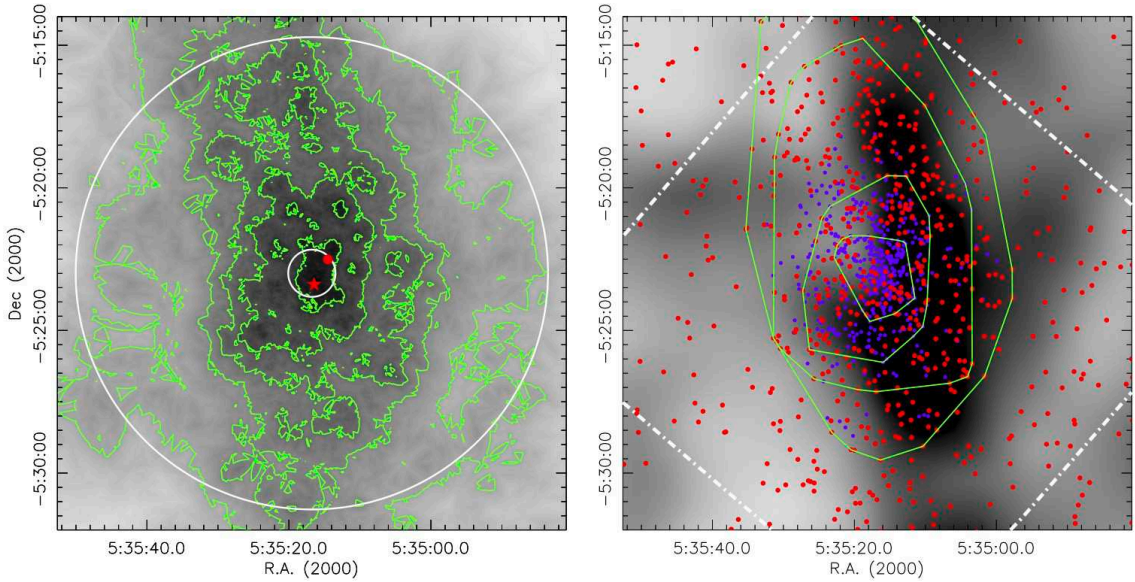}
\caption{Maps of the ONC cluster. {\bf Left:} the N$_{10}$ surface density map of the ONC.  The red star and dot indicate the position of $\theta^1$~C and BN, respectively.  The green countours are for surface densities of 100, 250, 500, 1000, 3000~pc$^{-2}$.  The circles are at radii of 0.1 and 1~pc from the central position of the cluster. {\bf Right:} distribution of YSOs overplotted on the A$_V$ map of the Orion~A cloud.  The red dots are {\it Spitzer} identified IR-excess sources while the blue dots are the X-ray identified YSOs from the COUP survey which were not identified by Spitzer.  The green lines give the convex hulls for threshold densities of 250, 500, 1000 and 2000~pc$^{-2}$. The dot--dash trapezoid gives the position of the COUP field.}
\label{fig:onc_map}
\end{figure}

\begin{figure}
\plotone{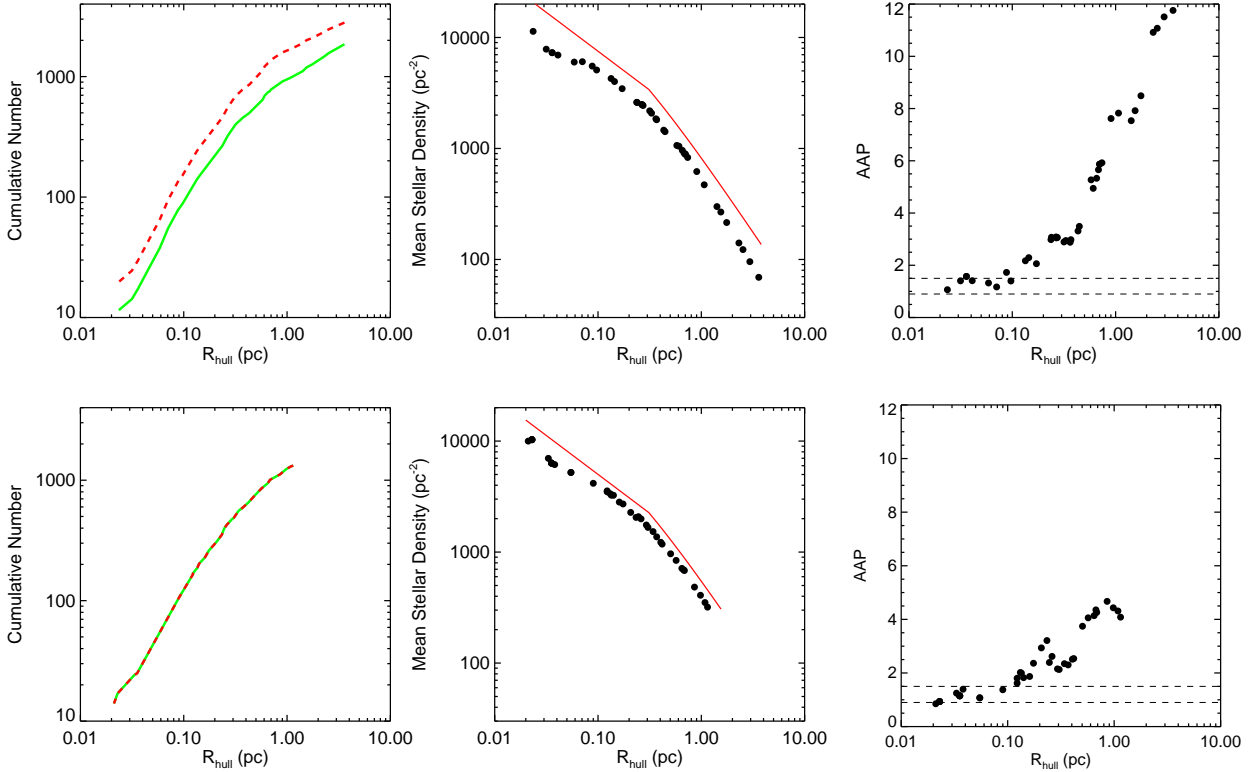}
\caption{The properties of the ONC cluster as a function $R_{hull}$. In the upper panels we show the results for the combined IR and COUP X-ray sample, in the bottom panels we show the same analysis for the COUP X-ray sample alone. The values of $R_{hull}$ were calculated for a series of threshold densities as described in the text; the properties are calculated for the YSOs that  fall within the corresponding convex hull.  The number of members, surface density and asymmetry of sources within the region defined by that threshold are plotted as a function of $R_{hull}$.   The left panels show the cumulative number of dusty YSOs within a given convex hull; the green line gives the uncorrected number and the red lines gives the weighting corrected number. Note that no corrections is applied to the X-ray data in the bottom row of panels.  The middle panels show the mean stellar density within the convex hulls for the corrected data on the top and the X-ray data on the bottom. For comparison, the red lines show YSO surface density $\propto R_{hull}^{-0.7}$  for $R_{hull} < 0.3$~pc and $\propto R_{hull}^{-1.4}$ for $R_{hull} > 0.3$~pc. The right panels give the AAP for each of the convex hulls.  Again, the data in the top panel are corrected for incompleteness.  This show that the ONC cluster has significant azimuthal asymmetries for $R_{hull} > 0.1$~pc.}
\label{fig:onc_prop}
\end{figure}

\begin{figure}
\plotone{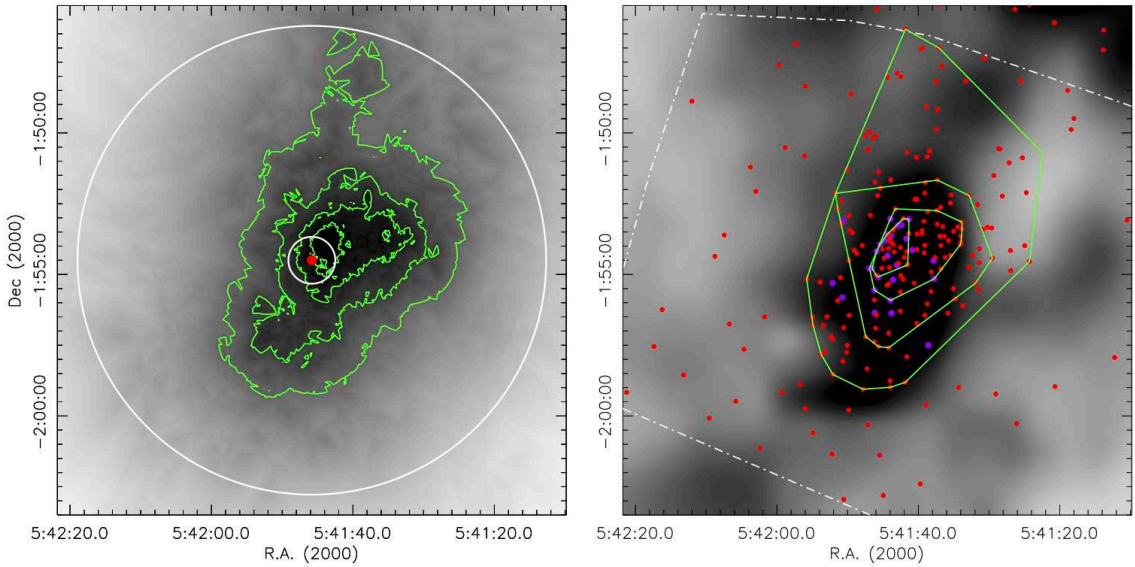}
\caption{Maps of the NGC2024 cluster.  {\bf Left panel:} the N$_{10}$ surface density plot for the clusters.  The red dot marks the position of IRS~2. IRS~2b, the likely exciting star of the NGC~2024 HII region, is located 5" to the north-west of IRS~2 \citep{2003A&A...404..249B}. The circles give radii of 0.1 and 1~pc centered on IRS~2. The green contours trace the 100, 250, 500,1000~pc$^{-2}$ levels. {\bf Right panel:} the extinction map of the region with the positions of the dusty YSO overlaid in red and the newly added {\it Chandra} X-ray sources in blue. The white dot/dashed line gives the approximate outline of the {\it Chandra} field.  The green lines show the convex hulls for threshold densities of 100, 250, 500 and 1000~pc$^2$.}
\label{fig:n2024_map}
\end{figure}

\begin{figure}
\plotone{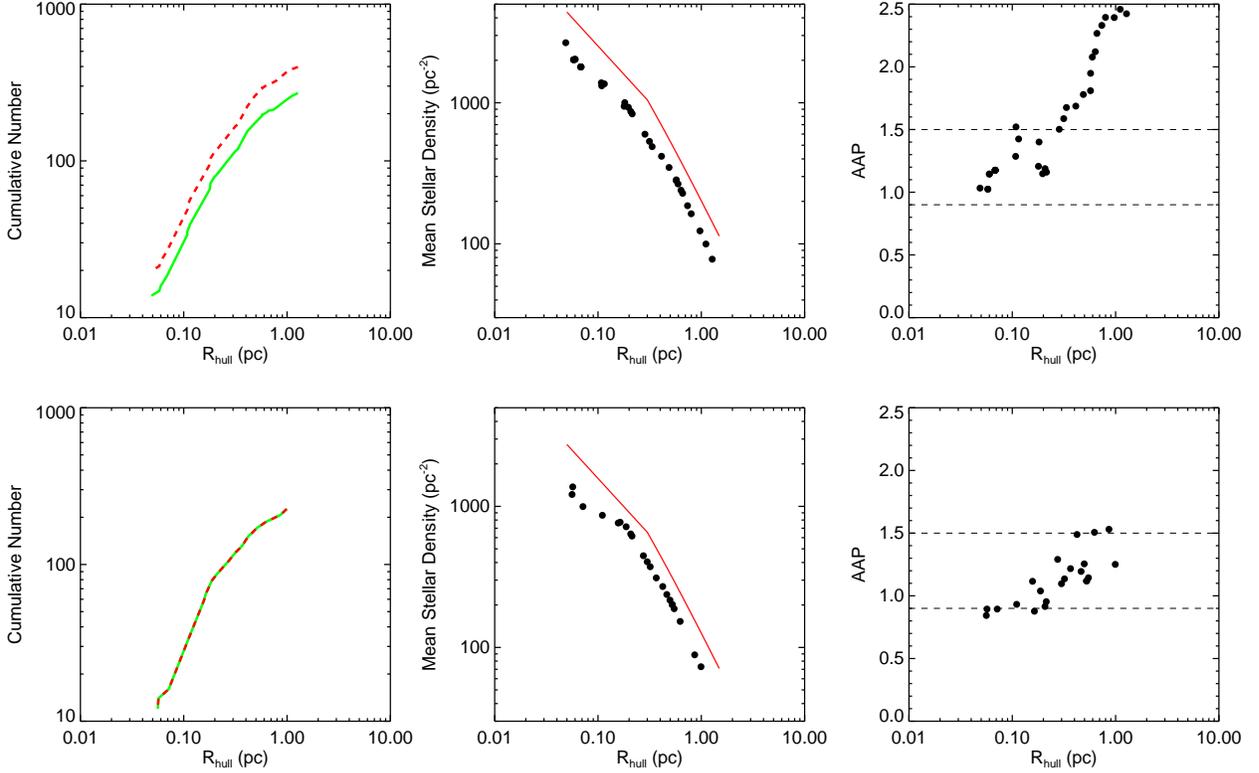}
\caption{The properties  of the NGC~2024 cluster as a function of $R_{hull}$  In the upper panels we show the results for the combined IR and {\it Chandra} X-ray sample, in the bottom panels we show the same analysis for the {\it Chandra} X-ray sample alone. The axes of the panels and the values plotted are the same as for Fig~20 with one difference: the red lines in the middle panel shows the profile for YSO surface density $\propto R_{hull}^{-0.8}$ for $R_{hull} < 0.3$~pc and $\propto R_{hull}^{-1.5}$ for $R_{hull} > 0.3$~pc.  Like the ONC, the AAP plots show that the NGC~2024 cluster has significant azimuthal asymmetries over much of the range of $R_{hull}$, particularly for the corrected IR and X-ray sample.}
\label{fig:n2024_prop}
\end{figure}

\begin{figure}
\plotone{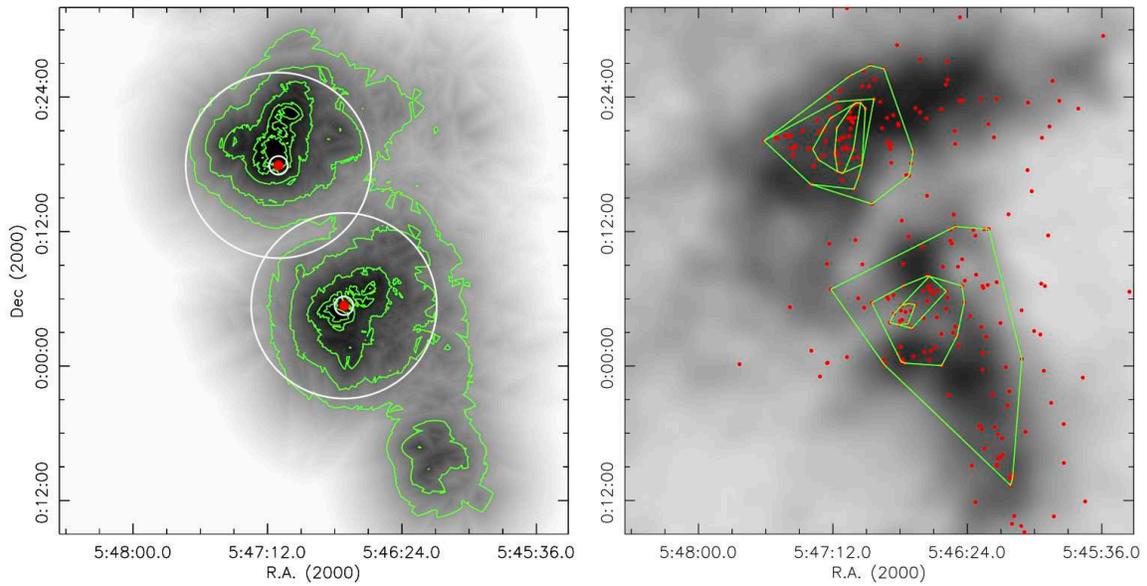}
\caption{Maps of the NGC2068/2071 cluster.  {\bf Left panel:} the N$_{10}$ surface density plot for the clusters.  The lower red dots mark the B1.5V star BD~+00~1177B in the NGC~2068 nebula while the upper red dot marks the B5 star V1380 Ori in the NGC~2071 Nebula.  The circles give radii of 0.1 and 1~pc centered on those stars. The green contours trace the 10, 20, 50, 100, 150~pc$^{-2}$ levels. {\bf Right panel:} the extinction map of the region with the positions of the dusty YSO overlaid.  The green lines show the convex hulls for threshold densities of 10, 20, 50, 100, 150 YSOs pc$^{-2}$.}
\label{fig:n2068_map}
\end{figure}

\begin{figure}
\plotone{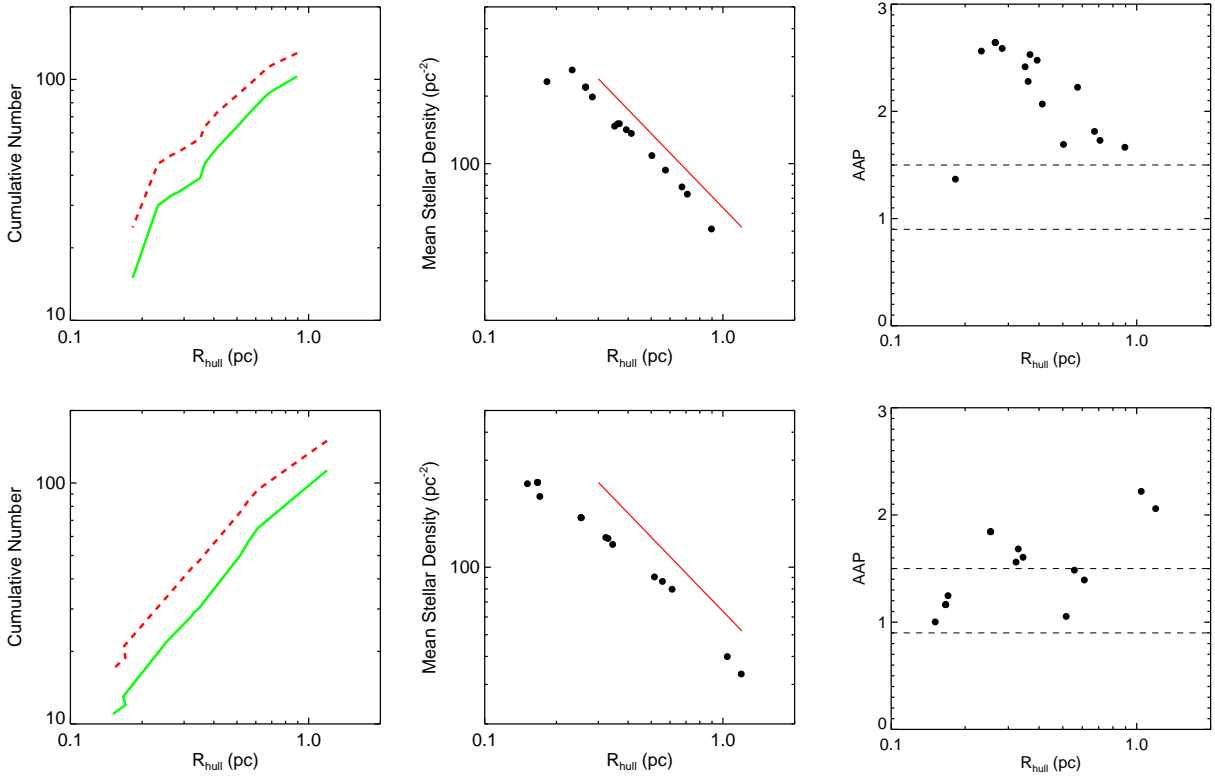}
\caption{The properties of the NGC~2068/2071s sub-clusters as a function of $R_{hull}$. The upper panels show the properties of the sub-cluster centered on NGC~2071 and the lower panels give the properties of the sub-cluster centered on NGC~2068.  The axes of the panels and the values plotted are the same as for Fig~20 with one difference: the red lines in the middle panels show YSO surface density $\propto R_{hull}^{-1.1}$. Like the ONC and NGC~2024, both sub-clusters show significant azimuthal asymmetries over most of the range of $R_{hull}$ with $AAP \ge 1.5$ .}
\label{fig:n2068_prop}
\end{figure}

\begin{figure}
\plotone{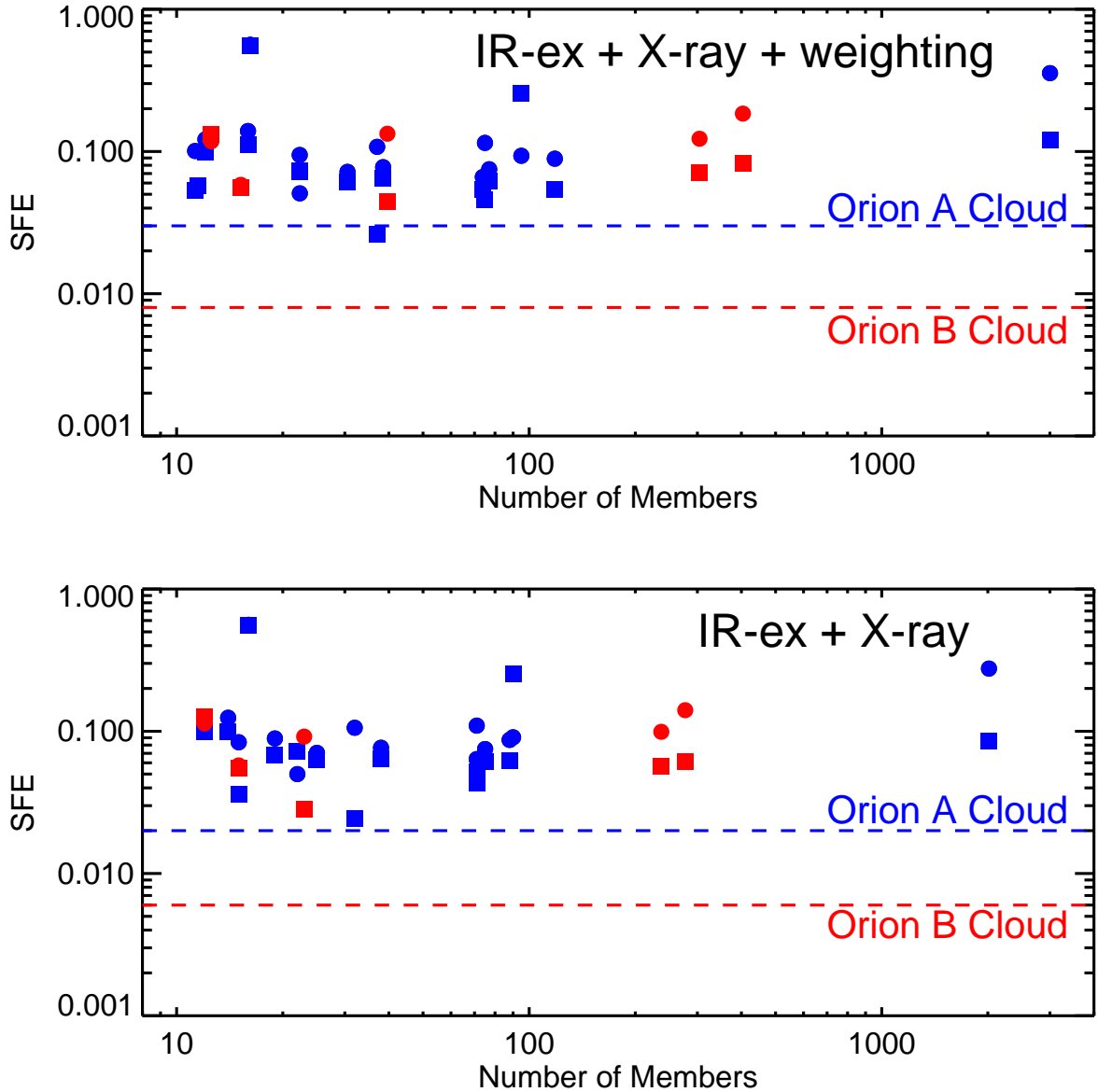}
\caption{The star formation efficiency of the groups, clusters and clouds in the Orion complex with the blue symbols denoting regions within Orion~A and the red symbols denoting regions within Orion~B.  The circles show the values determined for clusters and groups with the extinction map of \citet{2011ApJ...739...84G}  while the squares give the values determined from the $^{13}$CO maps from \citet{2013MNRAS.431.1296R}. The dashed lines give the SFE for the entire Orion A and B clouds using total cloud masses from \citet{2005A&A...430..523W}  and  \citet{2011A&A...535A..16L}. The upper panel uses the number dusty YSO corrected for incompleteness by both the application of the weighting correction and the inclusion of X-ray sources in the ONC and NGC~2024 clusters.  The lower panel uses  the number of detected dusty YSO augmented by the inclusion of X-ray sources and therefore gives systematically lower values.}.
\label{fig:sfe}
\end{figure}

\begin{figure}
\plotone{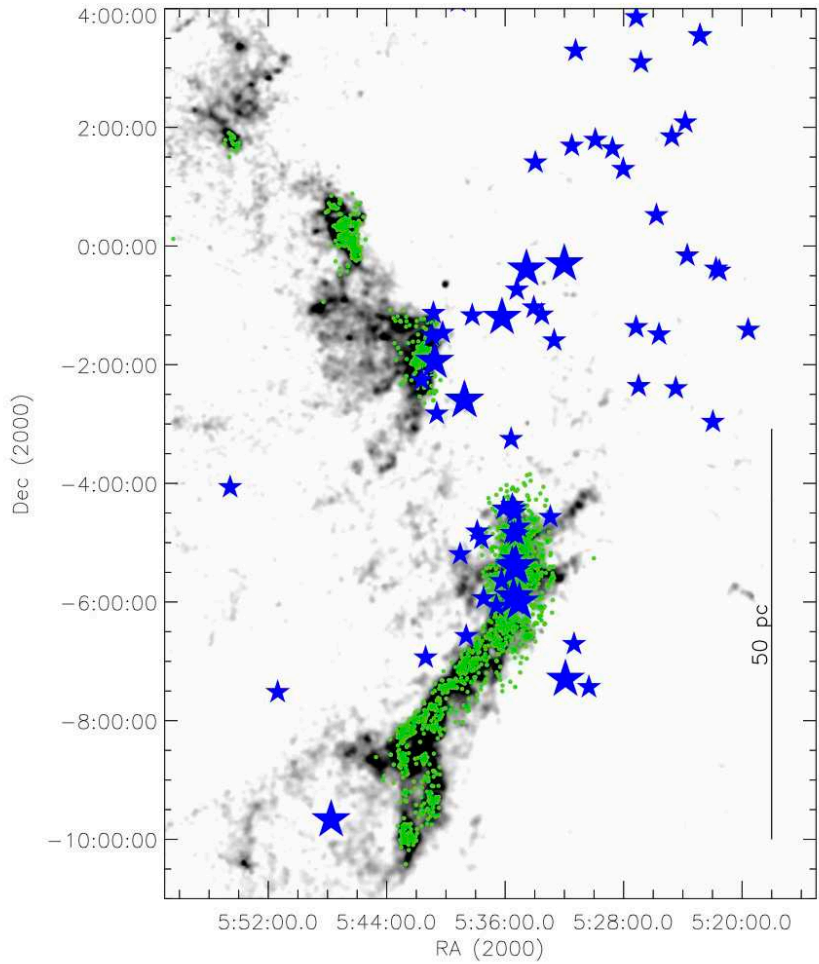}
\caption{The Orion OB1 Association.  The extinction map of the Orion molecular clouds \citep{2011ApJ...739...84G} is shown in grayscale.  The green dots are the dusty YSOs identified in this survey by their IR-excesses.  The blue stars are the OB stars identified by \citet{1994A&A...289..101B}.}.
\label{fig:ob}
\end{figure}

\begin{figure}
\plotone{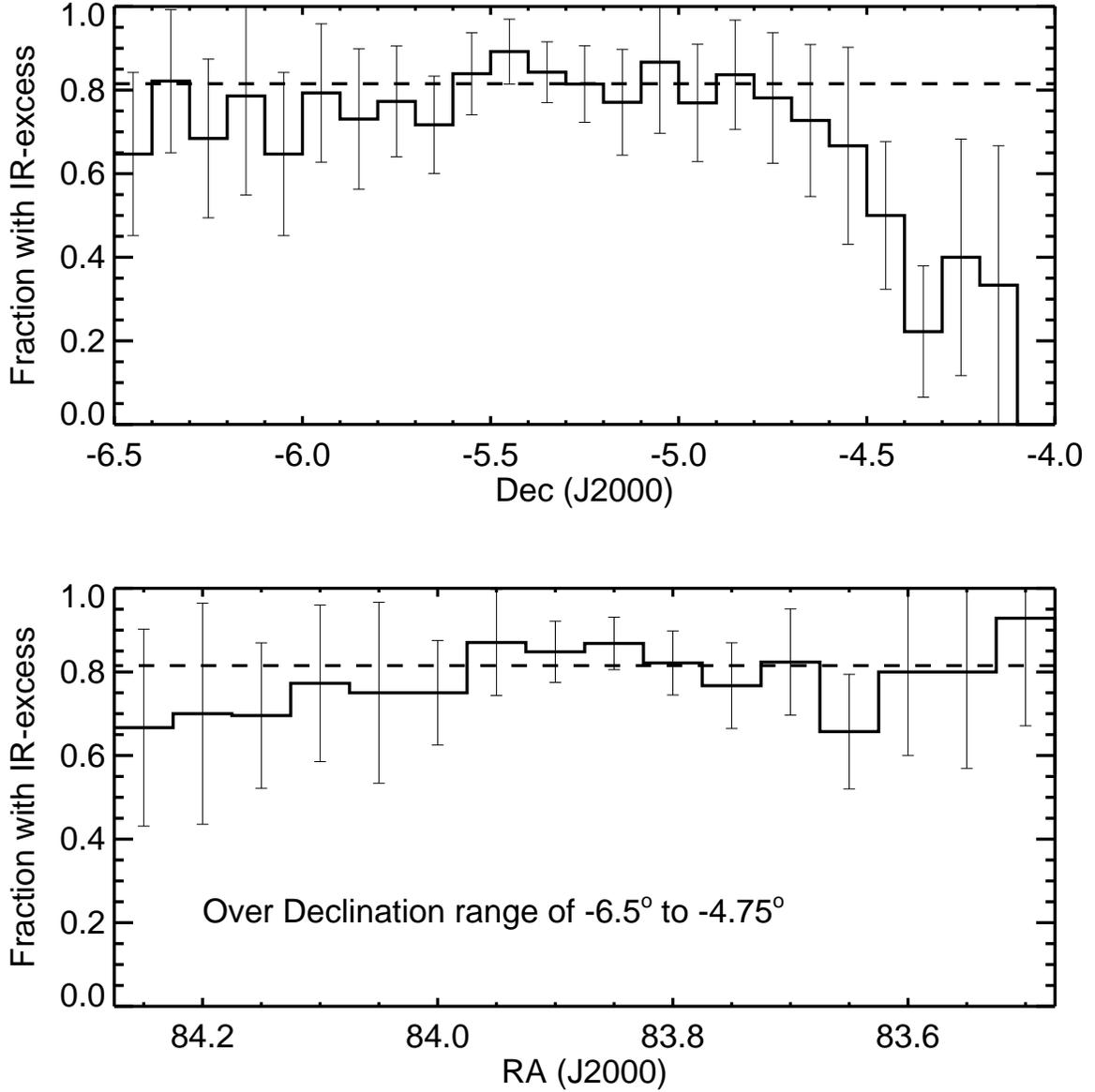}
\caption{The fraction of variables with IR-excesses as a function of right ascension ({\bf top}) and declination ({\bf bottom}). The center of the ONC is between $5.25^{o}$ and $5.5^{o}$; no decline in the fraction of disks is apparent at this location. The decline at $decl. \ge -4.5^{o}$ occurs at the edge of the Orion~A molecular cloud; the region with the low IR-excess fraction outside of the cloud is the more evolved NGC~1981 cluster \citep{2013ApJ...768...99P}.}
\label{fig:variex}
\end{figure}

\begin{figure}
\plotone{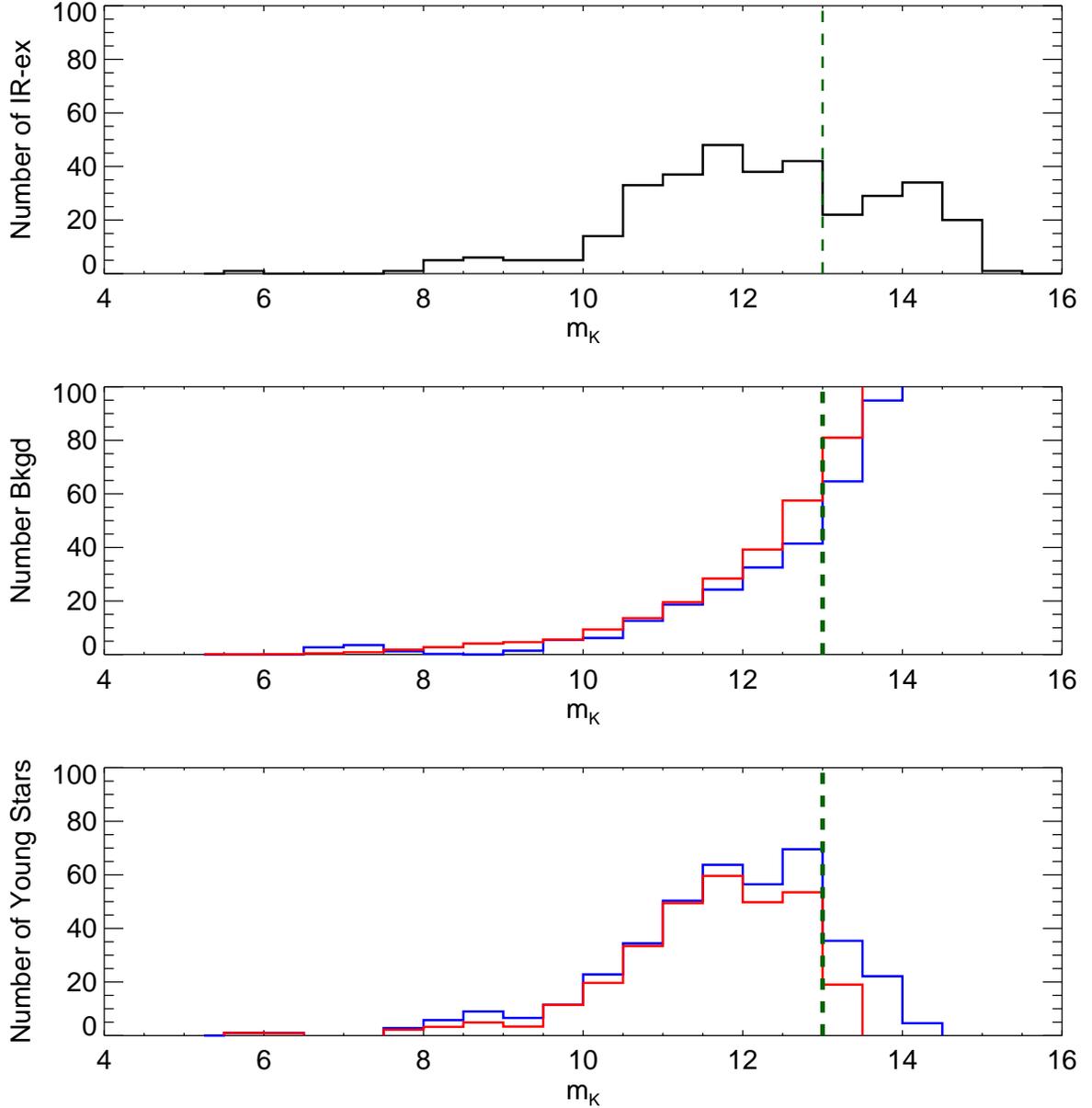}  
\caption{{\bf Top panel:} the histogram of $K_s$-band magnitudes for the IR-excess sources (i.e.~dusty YSOs) for the groups and clusters of L1641. {\bf Middle panel:} the histogram of $K_s$-band magnitudes for the estimated sky contamination toward the L1641 groups and clusters.  The blue histogram uses the {\it Spitzer} references fields near Orion A to measure the background contamination.  The red histogram uses two 1~sq.~deg. reference fields, one at $l=206^o$, $b = -19.3^o$ and the other at $l=216^o$, $b = -19.3^o$, which were observed by 2MASS but not Spitzer. {\bf Bottom panel:} the background subtracted histogram for all sources toward the L1641 groups and clusters which have sufficient photometry to search for IR-excesses.  The blue and red histograms in the bottom panel are those created with the blue and red histograms in the middle panel, respectively. The vertical dashed line is the 13~mag cutoff used in the disk fraction analysis.}
\label{fig:disk_l1641}
\end{figure}

\begin{figure}
\plotone{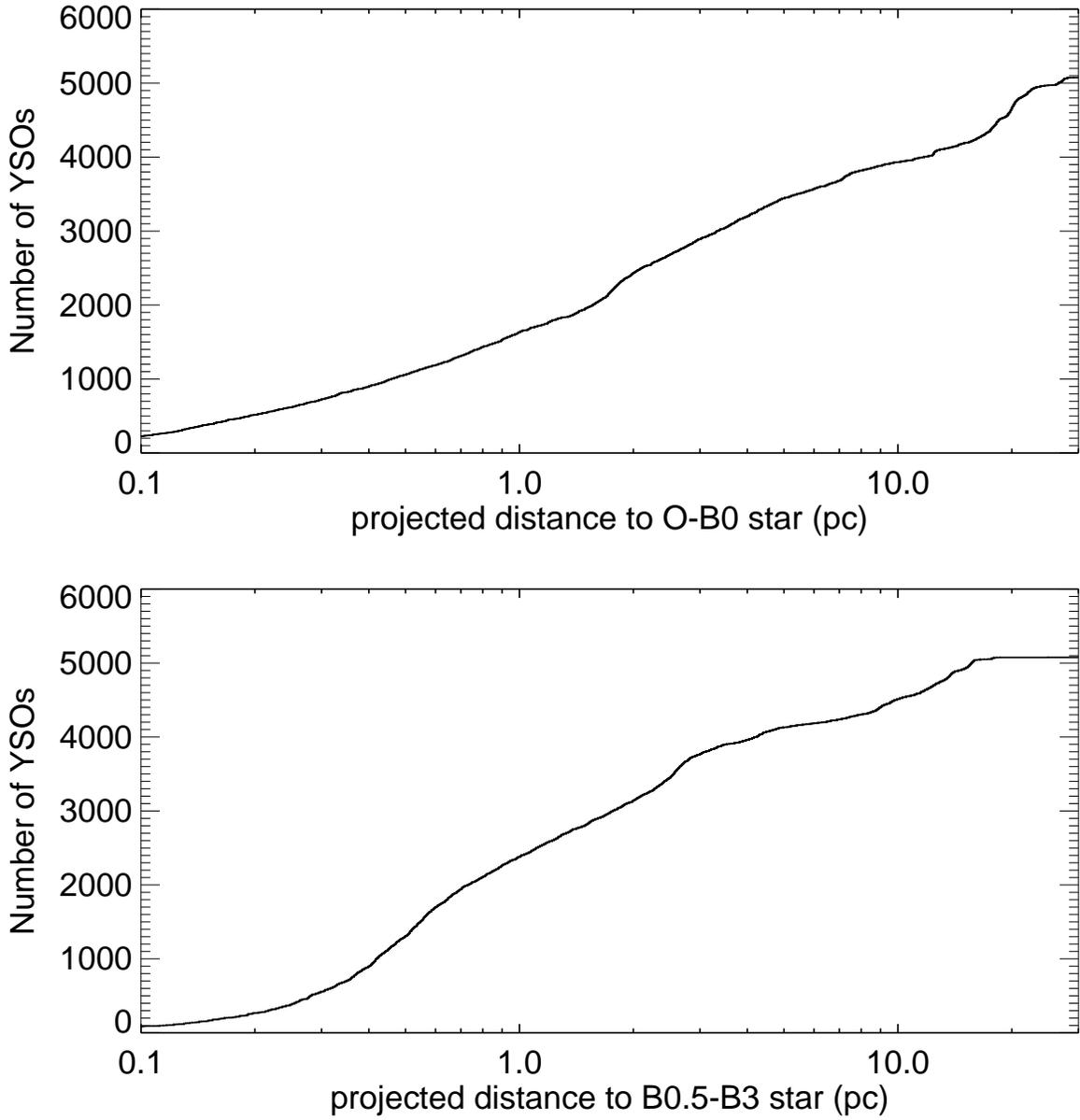}
\caption{The cumulative distribution of projected distances to the nearest OB stars for the YSOs identified in this paper.
{\bf Top:} distribution of projected distances to the nearest O-B0 star.  {\bf Bottom:} distribution of projected distances to the
nearest B0.5-B3 stars.}
\label{fig:dist_nearest_ob}
\end{figure}

\begin{figure}
\epsscale{1.}
\plotone{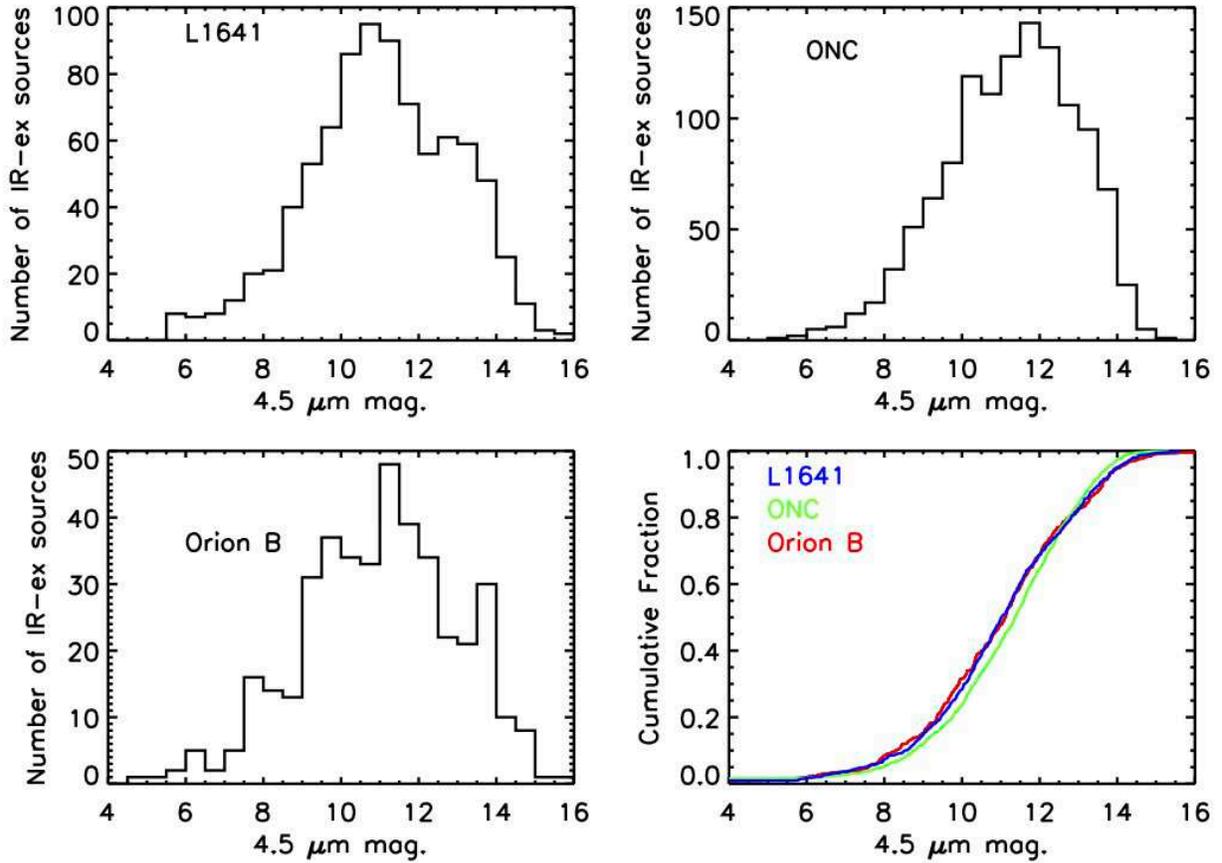}
\caption{{\bf Top panels and bottom left panel:} histograms of m$_{4.5}$ for sources with RMEDSQ deviations less than 30~DN.  These are the samples of  YSOs identified in regions with low nebulosity and point source  confusion.  The histograms are shown for three different fields: the  field containing the ONC and its surroundings, the field containing
 L1641, and the three fields covering the Orion~B cloud. {\bf  bottom right panel:} a comparison of the cumulative fraction of sources  with magnitudes brighter than m$_{4.5}$ for the three different fields. Note that  the L1641 and Orion~B sample are very similar, while the ONC is distinctly different with an excess of faint stars.}
\label{fig:lowmsqd}
\end{figure}

Figure~\ref{fig:frac} shows the fraction of YSOs for a given value of RMEDSQ.  These data were created with the following analysis. The first step was to adopt a representative sample of YSOs from a region with minimal confusion from nebulosity and point sources. We did this by selecting the YSOs below a limiting RMEDSQ value.  To assess the potential biases induced by the adopted value  of the limiting RMEDSQ value, we chose three values for the  RMEDSQ in the 8~$\mu$m band:  $< 30$, $20$ and $10$.   Furthermore, to examine the possible bias incurred by  the intrinsic properties of the YSO populations, we extracted YSOs in three distinct regions: the ONC field, the L1641 field,  and three fields covering Orion B.  The Orion~B fields  include the L1622 field, NGC2023/2024 field and the NGC2068/2071 field (see Figure 1 in Paper~I). This resulted  nine different baseline samples.  In Figure~\ref{fig:lowmsqd} we show the 4.5~$\mu$m distribution for the three different regions.   Note that there are visible differences between the distributions; in particular, the ONC shows a broader distribution than L1641 or Orion~B.   The distributions extracted from Orion~B and L1641 using the three different RMEDSQ limits are virtually  indistinguishable. KS tests were performed to compare all possible pairs of the 4.5~$\mu$m distributions from those two samples;  the lowest probability that these two distributions were drawn from a common parent distribution was $0.32$.  In comparison, the  ONC sample showed a very low probability of being drawn  from the same parent distributions as those from the Orion B and L1641  clouds, the  probability being less than $0.01$.  

We then randomly selected sources from each of the baseline YSO samples and added artificial YSOs to the IRAC
images drawn from the low RMEDSQ sample.  The artificial stars were placed to the north, east, south and west of known YSOs in all four IRAC bands; this formed a cross of artificial YSOs centered on each known YSO.  This procedure was repeated three times with offsets of $10''$,  $20"$ and $30"$ from the bonafide YSO.  The stars were then  recovered and their magnitudes  were extracted with the IDL routine PhotVis (???).    Finally, the stars were then run through the YSO identification  scheme to determine which of the stars would still be identified as YSOs.  For example, sources may no longer be considered YSOs if their uncertainties or colors no longer satisfied the limits and criteria in Paper~I. Since the 24~$\mu$m data contain large saturated regions, the detection limits at 24~$\mu$m cannot be simply  parameterized by RMEDSQ. For this reason, we did not include  the 24~$\mu$m data in this analysis.   This will only have a small effect on the results as only 5\% of the  IR-excess sources require detection  at 24~$\mu$m. We also did not perform the analysis for the 2MASS data; hence, for each of the baseline sources added to an image we adopted the 2MASS PSC magnitudes for that source. We note that 2MASS uncertainties did not show the same dependence on RMEDSQ as the IRAC data.  This is due to the much lower level of nebulosity in the near-IR data where the primary source of confusion is source confusion.  

To calculate the completeness we then followed the procedure outlined for the point source detection in Paper 1. The number of sources added and recovered were binned by their 8~$\mu$m $log(RMEDSQ)$ values and the fraction of sources recovered was calculated for each bin. This procedure was repeated using each of the nine different baseline samples.  We also repeated this procedure using the NGC~2024/2023 field to test whether the result was repeatable in different parts of Orion.  This field was chosen since it showed the next highest variations in RMEDSQ after the ONC field.   The results are  summarized in Figure~\ref{fig:frac}. The same overall trend is apparent for each of the baseline populations and the ONC and NGC~2024/2023 fields, but there is a dependence on the level of completeness depending on which population is used for our fiducial YSO sample in the artificial star test. In particular, the curves  using the ONC baseline population are systematically lower.  Given the good agreement between the points using the Orion~B and L1641 curves, we averaged the fractions from all RMEDSQ values in those two baseline populations  (leading to six iterations in total).  Finally, we calculated the uncertainties using the same approach as Paper~I and fit the equation

\begin{equation}
x = log(RMEDSQ),~f = \frac{1}{2}(1-\erf(\frac{x-a}{\sqrt{2}b}))
\label{eqn:logerf}
\end{equation}

\noindent
to the averaged fractions.  The fit is shown in Figure~\ref{fig:frac}, the resulting coefficients are $a=3.22 \pm 0.04$ and $b =1.05 \pm 0.05$. 

\acknowledgements
This work benefited from immeasurable from discussions with Fred Adams, John Bally, Cesar Briceno, Neal Evans, Gabor Furesz, Charlie Lada, Pavel Kroupa , Thomas Henning, Amy Stutz and Scott Wolk.  The analysis of the data would not be possible without the superb support we received from the staff of the Spitzer Science Center. This publication makes use of data products from the Two Micron All Sky Survey, which is a joint project of the University of Massachusetts and the Infrared Processing and Analysis. Center/California Institute of Technology, funded by the National Aeronautics and Space Administration and the National Science Foundation.  This work is based in part on observations made with the Spitzer Space Telescope, which is operated by the Jet Propulsion Laboratory, California Institute of Technology under a contract  with NASA.  It received support through that provided to the IRAC and MIPS instruments by NASA through contracts 960541 and 960785, respectively, issued by JPL.  Support for this work was also provided by NASA through awards  issued to STM and JLP by JPL/Caltech. This paper was mostly completed while STM was on sabbatical at the Max-Planck-Institut f\"ur Radioastronomie and further revised while continuing his sabbatical at the Max-Planck-Institut f\"ur Astronomie. STM thanks those institutes for their support.

\bibliography{ADS}

\newpage

\begin{table}[h]
\caption{Demographics of Dusty YSOs in {\it Spitzer} Orion Survey}
\vskip 0.1 in
\begin{tabular}{lccccc}
\tableline\tableline
  & $< 10$~YSOs & 10-100 YSOs & 100-1000 YSOs & $> 1000$~YSOs & Total \\
\tableline
\multicolumn{6}{c} {Orion A and B}\\
No X-ray\tablenotemark{a}   &  722  &  638    & 475   & 1646  &  3481  \\
No weight\tablenotemark{b}  &    722  &    638  &    514  &   2015 &  3889 \\     
Corr.\tablenotemark{c}       & 670   &    606  &    826   &  3002  &  5104  \\
\multicolumn{6}{c} {Orion A}\\
No X-ray\tablenotemark{a} 	  &    581   &  594 &  0  &  1646   &  2821 \\
No weight\tablenotemark{b}	 &     583  & 593  &  0  &  2015 & 3191 \\
Corr.\tablenotemark{c}  		& 540  &  539  & 118   & 3002 & 4199 \\
\multicolumn{6}{c} {Orion B}\\
No X-ray\tablenotemark{a}  & 135    & 50   &  475  &  0    &  660 \\
No weight\tablenotemark{b} & 134  & 50  &  514   &  0    &  698 \\
Corr.\tablenotemark{c} &         130  & 67  &  708   &  0   &  905 \\
\tableline
\tablenotetext{a}{Numbers from {\it Spitzer} IR sample without augmentation and corrections}
\tablenotetext{b}{Numbers augmented with X-ray data from {\it Chandra} data in the ONC and NGC~2024.}
\tablenotetext{c}{Numbers with augmentation from {\it Chandra} and with weighting corrections.}
\end{tabular}
\label{tab:demographics}
\end{table}

\newpage 

\includegraphics[scale = 0.75]{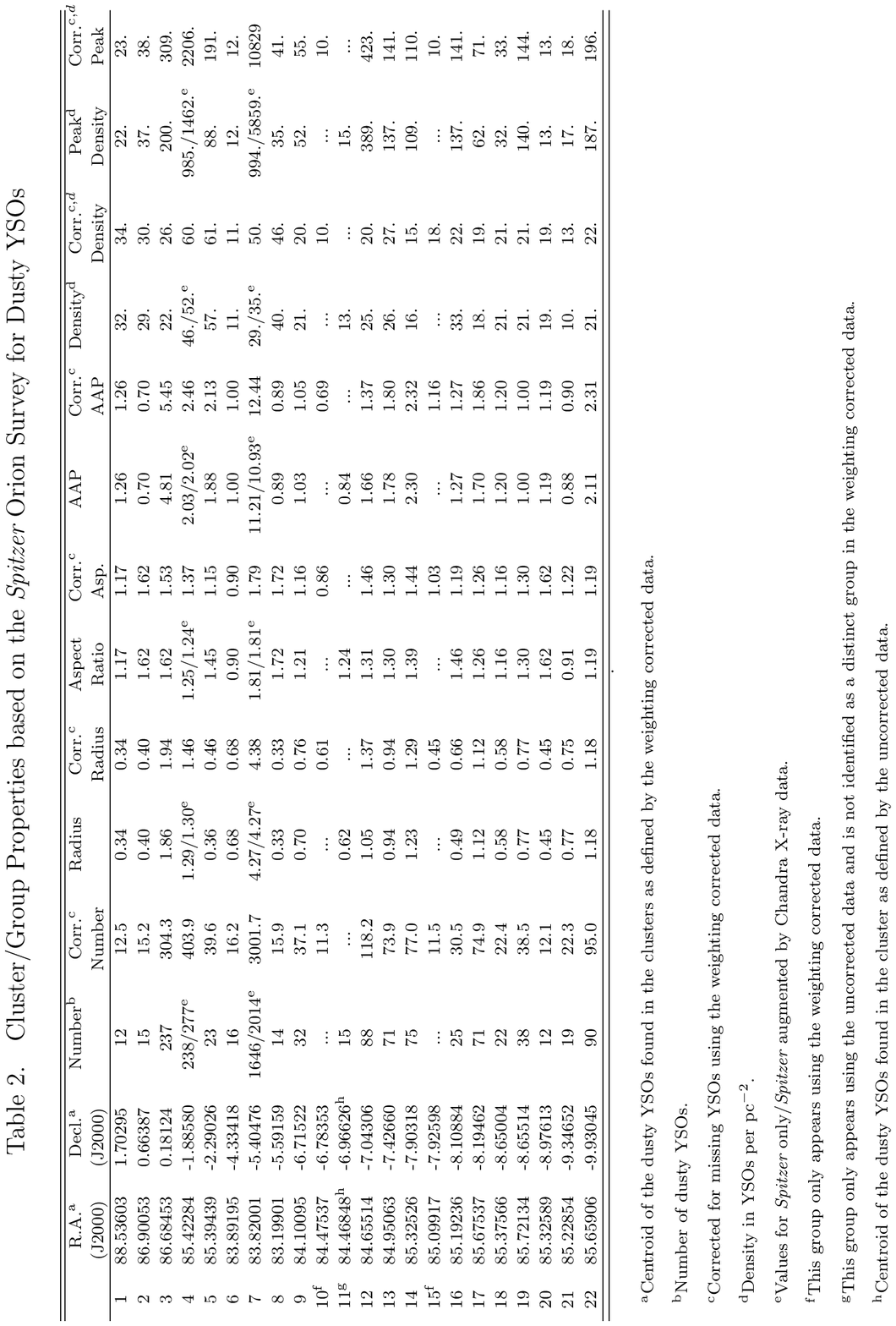}

\includegraphics[scale = 0.75]{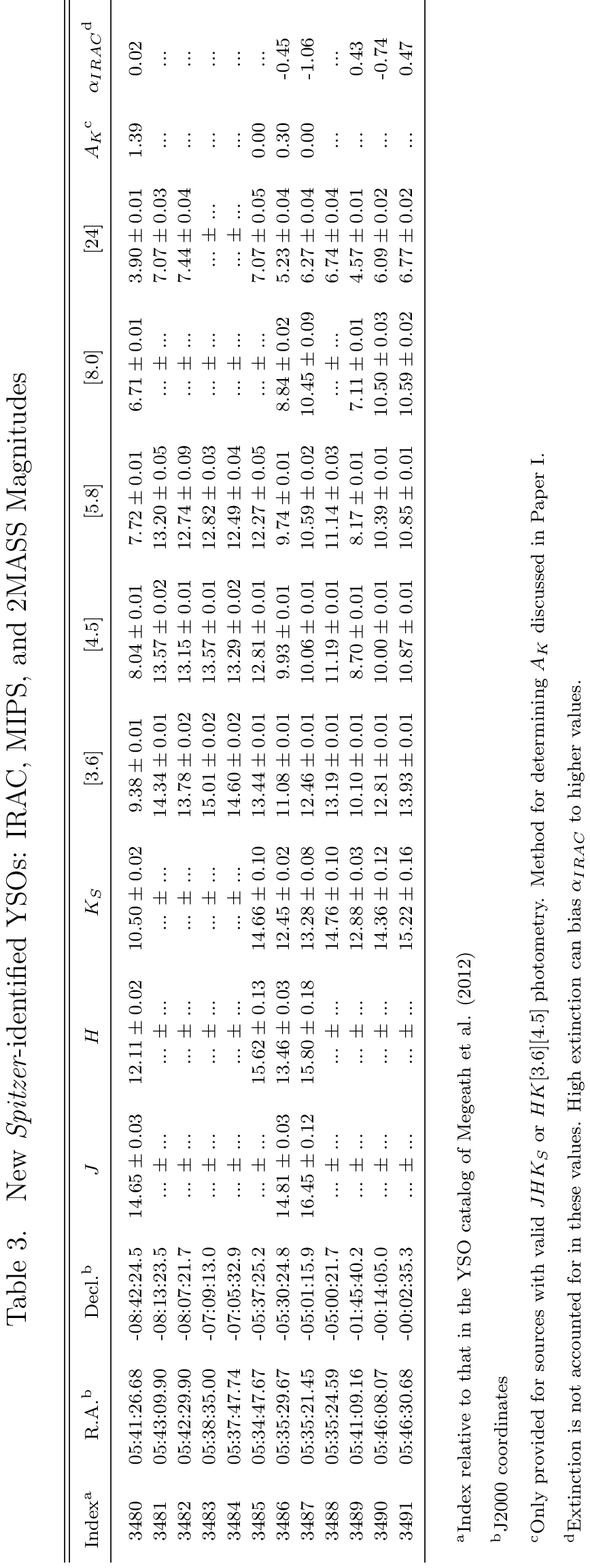}

\includegraphics[scale = 0.75]{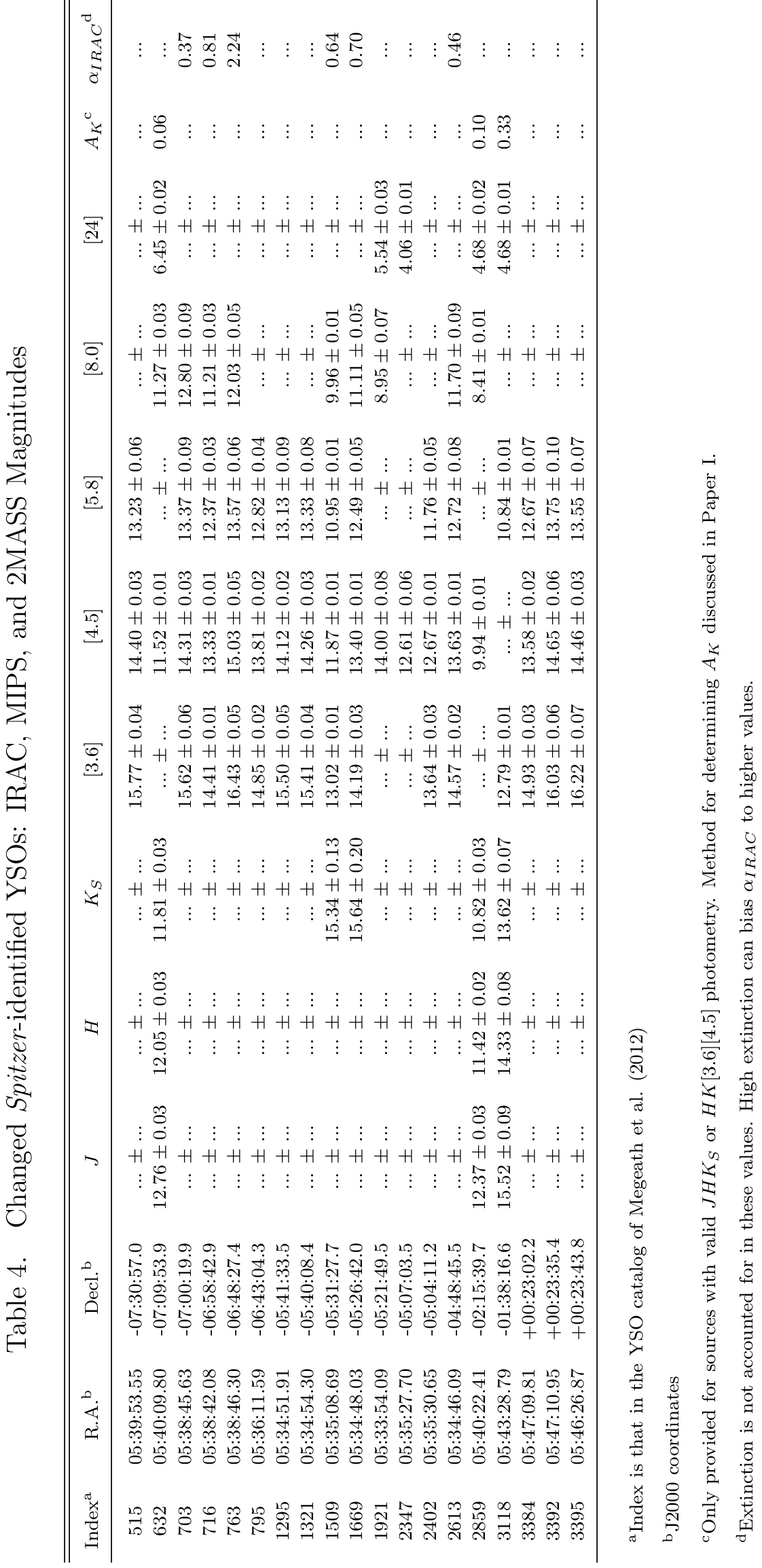}

\clearpage

\clearpage

\clearpage

\clearpage

\clearpage

\clearpage

\clearpage

\clearpage

\clearpage

\clearpage

\clearpage

\clearpage

\clearpage

\clearpage

\clearpage

\clearpage

\clearpage

\clearpage

\clearpage

\clearpage

\clearpage

\clearpage

\clearpage

\clearpage
\newpage

\clearpage
\newpage

\clearpage
\newpage

\clearpage

\clearpage

\clearpage

\clearpage

\end{document}